\documentclass[a4paper,12pt, oneside]{article} 
\usepackage{palatino}
\usepackage{amssymb}
\usepackage{amsmath}
\usepackage{amsfonts}
\usepackage{amsthm}
\usepackage[margin=1.1in]{geometry}
\usepackage[doublespacing]{setspace} % QJE requiere doble espacio para revisión
\usepackage{fancyhdr}
\usepackage{appendix} 
\usepackage{afterpage}
\usepackage[pdftex]{graphicx}
\usepackage{epstopdf}
\usepackage{comment}
\usepackage{float}
\usepackage{rotating}
\usepackage[para,online,flushleft]{threeparttable}
\usepackage{tabulary}
\usepackage{tabularx}
\usepackage{natbib}
\setcitestyle{chicago-authoryear}
\usepackage[utf8]{inputenc}
\usepackage[english]{babel}
\usepackage{graphicx}
\usepackage{latexsym}
\usepackage{amssymb}
\usepackage{amsmath}
\usepackage{footmisc}

\usepackage{bbm}
\usepackage{rotating}
\usepackage{amsxtra}
\usepackage[mathcal]{eucal}
\usepackage{enumerate}
\usepackage{rotating}
\usepackage[labelsep=period,justification=centering]{caption} 
\usepackage{subcaption}
\usepackage{lscape}
\usepackage{multirow}
\usepackage{paralist}
\usepackage{indentfirst}
\usepackage[affil-it]{authblk}
\usepackage[dvipsnames,svgnames]{xcolor}
\usepackage{setspace}
\usepackage{booktabs}
\usepackage{threeparttablex}
\setTableNoteFont{\fontsize{8}{9.6}}
\usepackage{titlesec}
\usepackage{lipsum}
\usepackage{bbm}
\usepackage{afterpage}
\usepackage{pdflscape}
\usepackage{url}
\usepackage{csquotes}
\usepackage{mathtools}
\usepackage{dsfont}
\usepackage{array,booktabs}
\usepackage{accents}

\usepackage {tikz}
\usepackage{textcomp}
\usetikzlibrary{positioning}
\usetikzlibrary{shapes,arrows}

\usepackage{caption}

\AtBeginEnvironment{table}{\singlespacing}
\AtBeginEnvironment{figure}{\singlespacing}
\newcommand{\figurenotes}[1]{
  \vspace{4pt} % Small gap between figure and notes
  \parbox{\linewidth}{\small \noindent \textit{Notes:} #1}
}

\usepackage{chngcntr}

\usepackage{mathptmx} % Times New Roman
\usepackage{sectsty}
\sectionfont{\large\centering\scshape\mdseries} % Small caps
\subsectionfont{\normalsize\mdseries\itshape}
\subsubsectionfont{\normalsize\mdseries\itshape}
\paragraphfont{\normalsize\mdseries\itshape}

\usepackage{etoolbox}

\AtBeginEnvironment{table}{
\singlespacing
}

\renewenvironment{abstract}
  {%
    \par\vspace{-2em}
    \noindent
    \begin{center}
    \begin{minipage}{\textwidth}
    \textbf{Abstract:}
  }
  {%
    \end{minipage}
    \end{center}
    \par\vspace{1em}
  }

\makeatletter
\renewcommand{\@maketitle}{%
  \newpage
  \null
  \vspace{-3.5em} % controls space above title
  \begin{center}%
    {\LARGE \@title \par}%
    \vskip 1em%
    {\large
      \lineskip .5em%
      \begin{tabular}[t]{c}%
        \@author
      \end{tabular}\par}%
    \vskip 1em%
    \vspace{-0.5cm}
    {\small \@date}%
  \end{center}%
  \par
  \vskip 2em%
}
\makeatother

\renewcommand{\thefigure}{\Roman{figure}}
\renewcommand{\thetable}{\Roman{table}}
\renewcommand{\thesection}{\Roman{section}}

\definecolor{blue}{rgb}{0,0.08,0.5}
\definecolor{red}{rgb}{.6,0,0}
\definecolor{green}{rgb}{0,0.376,0}
\usepackage[bookmarks, pdftitle={}, pdfauthor={}, colorlinks=true, linkcolor=red, citecolor=blue, urlcolor=blue]{hyperref}

\newcommand{\Expect}{{\rm I\kern-.3em E}} %fancy expectation symbol

\newcolumntype{Y}{>{\centering\arraybackslash}X}

\begin{document}
\vspace{-1cm}
\title{\textbf{Remote Work and Women’s Labor Supply: The New Gender Division at Home}\thanks{\begin{doublespace}We thank Michael Best, Sandra Black, Nicholas Bloom, Pierre Andre Chiappori, Jonathan Dingel, Tatiana Mocanu, Suresh Naidu, Cristóbal Otero, Sebastián Otero, Bernard Salanie, Alessandra Voena, and participants of the 2025 Remote Work Conference at Stanford University and seminars at Columbia University for their helpful comments.\end{doublespace} \vspace{-0.65cm} }}
%\thanks{Acknowledgments go here.} %--- insert inside the title brackets

\author{
  \begin{tabular}{c}
    Isabella Di Filippo, Bruno Escobar, and Juan Facal\thanks{\begin{doublespace} Di Filippo: Columbia University, \textbf{id2403@columbia.edu}. Escobar (Corresponding Author): Columbia University, \textbf{be2329@columbia.edu}. Facal: Columbia University, \textbf{jmf2285@columbia.edu} \end{doublespace} \vspace{-0.55cm}  }
  \end{tabular}
}

\date{\small Last updated: \today}

\maketitle

\vspace{0.5cm}

\begin{abstract}
\noindent
We study how increases in remote work opportunities for men affect their spouses' labor supply. Exploiting large variation in work-from-home (WFH) exposure across occupations before and after the COVID-19 pandemic, we find that increases in men's WFH exposure led to sizable improvements in their wives' intensive-margin labor market outcomes. Further analysis shows that these effects are driven by intra-household \textit{spillovers}, whereby the husband's presence at home relieves wives' childcare constraints and enables wives' greater labor market engagement.
%Evidence shows women are less likely to engage in primary childcare activities, while husbands working from home partially compensate by covering more childcare for their spouse.
%relieves wives'
%childcare constraints 
%housework time constraints and enables greater labor market engagement. 
%Women are less likely to engage in primary childcare activities, while husbands working from home partially compensate by covering more childcare for their spouse. 
These results highlight the role of household dynamics in shaping the labor market consequences of remote work and narrowing gender gaps.

\thispagestyle{empty} %don't number title page


\vspace{1cm}

%\vfill

%Workplace flexibility has the potential to reduce persistent gender disparities in labor market outcomes, particularly following parenthood. However, little is known about its causal impact on mothers’ labor supply, and even less about the indirect effects of fathers’ access to flexible work arrangements through intra-household spillovers. This paper examines how increased work-from-home (WFH) probability for husbands affects their wives’ labor market outcomes, focusing on U.S. households with children of child-rearing age. Exploiting quasi-experimental variation in the probability of WFH across occupations before and after the COVID-19 pandemic, we estimate the causal impact of exogenous shifts in husbands’ WFH exposure. Comparing otherwise similar households exposed to different degrees of this shock, we find that treated women are over 2 percentage points more likely to be employed and participate in the labor force, a 4\% increase relative to the pre-pandemic mean. Time-use data reveal that this effect is facilitated by a reallocation of household tasks: women reduce time spent on household production and shift toward secondary childcare (childcare while engaged in other activities), with partial compensatory increases in men’s involvement. These findings highlight the critical role of intra-household dynamics in shaping the labor market effects of workplace flexibility. \vfill

\end{abstract}

\pagebreak
\setcounter{page}{1} %start numbering on intro

\section{Introduction}

Over the past century, gender gaps in education, human capital, and labor market outcomes have narrowed substantially in developed countries \citep{olivetti2016evolution}. Nonetheless, a persistent gap remains in labor market trajectories following parenthood \citep{kleven_child_2024}. After childbirth, women frequently reduce their labor force participation or exit employment altogether \citep{kleven2022geography}, and those who remain employed often face large earnings penalties relative to men \citep{bertrand_dynamics_2010}. These gender disparities are largely driven by (i) the unequal division of childcare responsibilities within households, and (ii) the fact that many jobs - including high-paying white collar positions - have historically lacked flexibility options to allow women to balance both their careers and motherhood \citep{goldin_grand_2014}. 

This paper examines whether the expansion of remote work opportunities can shift intra-household dynamics and encourage greater labor market participation among women with children. Our motivation builds on the idea, emphasized by \citet{goldin_grand_2014}, that the \emph{last chapter} of gender convergence may depend on restructuring work to allow for greater flexibility. Although not synonymous with flexibility, remote work can relax physical and time constraints faced by parents by allowing job tasks to be performed outside the workplace. While previous literature has established that the advent of remote work has the potential to increase female labor supply overall \citep{heggeness2021telework, farooqi2023essays, harrington2023has}, we argue and show that access to remote work for the husband is itself a key enabler of this effect for mothers with young children. Specifically, a meaningful reallocation of childcare responsibilities -- one that sufficiently relaxes mothers' time constraints to expand their labor supply \citep{gottlieb2024gender, simintzi2025effect} -- is more likely to occur when the husband himself can work from home and take care of children, regardless of whether the wife's own job offers this option. %Workplace flexibility has been shown to help parents, especially mothers, accommodate both predictable and unexpected caregiving demands \citep{angelici_smart_2024}. Evidence also shows that women and parents of young children place a high value on flexible work arrangements \citep{mas_valuing_2017, aksoy_working_2022, bloom_how_2022}, and are more averse to commuting \citep{le2021gender}.
 %Despite this potential, causal evidence on how remote work options available to the household, even when limited to the husband, affect wives' labor supply remains scarce. This paper fills that gap by documenting whether and how the expansion of remote work for husbands affects wives' labor supply decisions and the household division of market work and childcare. {\color{red} [A thing we can add here is sth like `we believe looking at the husbands shock on women is the most relevant margin as for example i) he can take of children when home whereas if wife has WFH shock if the husband is not home himself then she would still be the one to take care of the children so her direct wfh as not effect on her intensive labor supply ii) they are the ones more likely to be emplyoed and in wfh fasible occupations wrt to the ones women have so the household shock must operate through them]}

%We focus on U.S. dual-parent households with young children, combining occupation-level data on remote work vacancies with rich household-level survey data from the ACS and other national datasets. Leveraging quasi-experimental variation in remote work exposure across occupations before and after the pandemic, we estimate whether large-scale increases in remote work availability for male spouses altered allocation of time between market work and childcare in their households. Particularly, we emphasize whether men's increased time at home enabled their wives to expand their labor supply.

The backdrop for our analysis is the COVID-19 pandemic, which triggered a rapid and persistent shift in remote work norms. In the U.S., only 7\% of full paid workdays were conducted from home in 2019. By June 2023, this share had quadrupled to 28\% \citep{barrero_evolution_2023}. The magnitude of this shift varied widely across occupations: while some sectors (e.g., computer science and professional services) adopted remote work at scale, others (e.g., transport, healthcare, education) largely returned to in-person arrangements. We illustrate this in \autoref{fig:evolution_WFH}, which shows very large, heterogeneous and hand-in-hand changes in both the \textit{supply} of job postings offering fully remote work positions and the \textit{demand} for remote work embedded in the uptake of fully remote work jobs in household survey data.\footnote{Data for supply of remote work jobs comes from the publicly available \citet{hansen2023remote} dataset on job posting, and for the demand comes from the ACS. We describe both datasets in more detail in Section \ref{sec:data}}

We use this variation and a sample of dual-parent households with young children from the American Community Survey (ACS) to study whether women’s labor market outcomes respond to changes in household exposure to remote work opportunities -- specifically through their husband’s occupation.
Our empirical strategy compares the labor market outcomes of wives with young children whose husbands experienced a large increase in remote work availability, measured as the change in the share of job vacancies offering WFH in their occupation, to those whose husbands experienced little or no such change. This design has the advantage of identifying a \emph{reduced-form household-level effect} of remote work expansion. By plausibly expanding the set of feasible work and care arrangements available to the household -- and specifically increasing the male spouse's time at home -- the shock allows couples to jointly reoptimize their labor and childcare supply decisions, creating scope for increases in women's labor supply.

Our first-stage estimates from the ACS data indicate that occupation-level increases in WFH probability after COVID translated into substantial increases in men's realized WFH uptake, rising by approximately 12 percentage points on average and nearly tripling relative to pre-pandemic levels. Using an analogous sample of dual-parent households from the American Time Use Survey (ATUS), we also find that exposed husbands experience an increase in the share of work time spent at home by 3.7 p.p. and reduce their commuting time by 17\% with respect to baseline. Importantly, men's overall labor supply and earnings have remained unchanged in response to the WFH shock. 

Our baseline estimates show their wives' labor market outcomes improved along the intensive margin: annual earnings increased by approximately 5\%, weekly hours worked rose by roughly half an hour, weeks worked increased by about 0.5 weeks, and the likelihood of part-time work declined by approximately 9\%. These findings have a meaningful impact in narrowing the labor gender gaps: a 5.9\% reduction in earnings gap, a 3\% in weekly work hours, a 1.2\% in weeks worked, and, a -1.4\% reduction in the odds ratio of women working part-time.

To validate our identification strategy, we conduct a broad set of robustness checks, including controls for differential pre-trends in education and broad occupation categories, tests for changes in sample composition around the pandemic, and alternative treatment definitions. Results remain consistent throughout, reinforcing that increased WFH options for men causally raised women's labor market engagement.

%First, we 
%also assess whether our results are driven by specific occupation groups through leave-one-out analyses, and 
%evaluate the sensitivity of our estimates to alternative fixed effect structures and demographic controls, with particular attention to differential trends across education categories \citep{goldin2022understanding} and major occupational groups. We then examine the possibility of changes in household composition or selection into treatment following the pandemic, including tests for differential fertility and marriage patterns and reweighting strategies that align pre- and post-COVID covariates distributions. Finally, we explore the robustness of our findings to varying definitions of the post-treatment period and sample restrictions. Across all these exercises, our main results remain consistent, reinforcing the interpretation that increased WFH options for men played a causal role in raising women’s labor market engagement. 

We find no effect on the probability of employment, consistent with the shock being insufficient to overcome the fixed costs of labor market entry, nor on hourly wages.\footnote{Both findings align with recent evidence on the effects of WFH on the motherhood penalty \citep{basso2026wfh, scott2025flex}.}

%{\color{red} [Here I would add a paragraph about how then we focus on decomposing this household reduced-form effect into a spillover and a correlated direct-effect component].}

Our main effects on women's labor market engagement could reflect spousal spillovers from men's increased time at home, or simply assortative matching: women whose husbands work from home more often may also, independently, work in occupations with greater WFH exposure themselves, which could directly increase their own labor supply. We present several pieces of evidence pointing to the former channel. Chief among them, when we condition on women with a pre-determined low likelihood of working from home,
%based on their college major or their own occupation's WFH exposure,
spousal shocks still deliver significant increases in wives' working hours and weeks worked across the year, closely mirroring our main findings. %Since these women's own labor supply is unlikely to be directly affected by remote work availability in their own occupation, this result suggests that our estimates are not simply picking up assortative matching in remote work access. %Consistent with this interpretation, a decomposition analysis that separates spousal spillovers, direct effects of women's own WFH exposure, and assortative matching shows that matching accounts for most of the changes in women's own WFH and commuting, while spillovers -- rather than matching or direct effects -- are the main force behind the effects on women's labor supply.

We then evaluate whether shifts in the household's allocation of childcare responsibilities and other home production activities can account for the labor market effects on women. To do so, we rely mainly on the ATUS and the childcare topical modules from the Survey of Income and Program Participation (SIPP). We present novel evidence of intra-household reallocation of childcare activities within the couple: women reduce time spent on childcare as a primary activity by 2.4 p.p., and husbands partially offset this reduction with a 1 p.p. increase in childcare performed while working. This amounts to a 11\% reduction of the gap between women's and men's time allocated to childcare. Treated husbands are also more likely to care for children while their wives are working. In addition, treated households are less likely to enroll children in daycare and they spend less on childcare services. We also find evidence ruling out other potential channels behind our main results: women's increased labor supply does not come at the expense of their own leisure and personal time, nor is it driven by differential migration to better geographical labor markets.
%Our findings reveal that wives of men that saw large increases in work from home went on to experience a large increase in their share of day spent working, with no significant change in overall household production. Women themselves are also much more likely to work remotely, with approximately a 75\% rise relative to the pre -treatment mean, suggesting that remote work enables labor supply for both spouses. 

%We also document a 10 percentage-point increase in women’s own WFH incidence —approximately a 75\% rise relative to the pre -treatment mean.

Overall, our findings suggest that increases in husbands' access to remote work can influence their spouses' labor supply decisions along the intensive margin. We find novel evidence that the effects stem mostly from intra-household spillovers. Time-use and childcare patterns point to substitution across spouses as the mechanism: as men become more available at home, they take on a larger share of childcare, relaxing women's childcare time constraint and allowing them to expand their market engagement. %This implies almost a 20\% reduction in the baseline gap of the time women and men allocate to childcare.
%This offers a new insight on how structural changes to working arrangements, such as a massive increase in remote work options, can alter the childcare and housework technology and expand households' feasibility set, enabling Pareto-improving scenarios.
%{\color{red} [shall we delete this if we remove the model?] Our findings can also be rationalized through the lens of a household model where remote work options change the childcare technology and expands households' feasibility set.} 
%This casts light on the fact that remote work options can change the childcare technology and expands households' feasibility set.

%\subsection{Related Literature}
%not only wage gap
\paragraph{Related Literature}
%{\color{red} Let's remember to add paragraph about added-worker papers two referees mention we should speak to, as well as papers about intensive margin being the relevant one!! Potentially we can combine these 2 arguments together into one literature motivating paragraph. I think apart from my professor's paper \citep{basso2026wfh}, we should check these papers he mentions: . Prepandemic US studies show that WFH feasibility reduces the motherhood employment penalty but may induce sorting into lower-paying firms (Harrington and Kahn, 2025; Jack et al., 2025); European and Latin American evidence associates remote-work feasibility with narrower gender gaps (Arntz et al., 2022; Crescenzi et al., 2025; Zarate, 2025); pandemic-era evidence is more mixed, likely reflecting contemporaneous child4 care disruptions (Dunatchik et al., 2021; Farooqi, 2023; Goldin, 2022; Heggeness and Suri, 2021; Pabilonia and Vernon, 2022; Song, 2025).}

Our work contributes to the growing literature on how working arrangements affect labor market dynamics. While existing research has primarily focused on the direct effects of flexible work arrangements on individual outcomes -- highlighting their growing value for both workers and firms \citep{bloom_does_2015,mas_valuing_2017,bloom_how_2022,angelici_smart_2024,choudhury2024hybrid} -- women and parents in particular are known to place high value on flexibility \citep{mas_alternative_2020,aksoy_working_2022}. Yet little work has studied how changes in one spouse's work arrangements reshape decision-making and labor supply for the other spouse within the household.\footnote{We are most closely related to recent  work studying the relationship between remote work access and household collective labor supply \citep{buckman2025gendered, scott2025flex, su2025help}.} We contribute to this emerging strand by leveraging a large-scale, quasi-experimental shift in remote work feasibility to study how increased WFH access for men affects household labor supply and home production.  While \citet{harrington2023has} and \citet{gallen2025childcare} examine how increases in remote work opportunities affect mothers' own employment, earnings, and childcare demand, we show that husbands' access to remote work is fundamental in relaxing wives' childcare time constraints, thereby enabling changes in mothers' intensive margin of labor supply -- and is the main driver behind the increase in mothers' intensive labor supply after the pandemic. %%Our paper shows that husbands' access to remote work relaxes wives' childcare time constraints, thereby enabling changes in women's intensive margin of labor supply, and is an important driver behind the increase in women's intensive labor supply after the pandemic.

Another body of literature has investigated the effects of parental leave policies as potential tools to increase mothers' labor supply. However, this literature has found that such policies have limited, and sometimes even negative, effects on women's long-term labor market outcomes \citep{ginja_parental_2020, canaan_parental_2022, ginja_employer_2023, kleven_child_2024}.\footnote{At best, fathers' access to temporal flexibility has been shown to improve maternal postpartum health, with little impact on female labor supply \citep{persson_when_2024}.} How, then, do we reconcile these findings with our own evidence of sizable and persistent effects from increased paternal presence at home due to remote work? We argue that the scale and perceived permanence of the WFH shock fundamentally differ from traditional parental leave policies, and our findings suggest that this distinction matters for outcomes. Rather than offering short-term accommodations around childbirth, the pandemic-driven shift toward remote work restructured household production possibilities, giving families a broader set of tools to jointly optimize labor supply and caregiving responsibilities over the longer run. This points to the importance of structural, rather than short-term, changes in work arrangements for driving lasting shifts in women's labor supply -- a distinction that this literature has not, to our knowledge, previously isolated.

Finally, we contribute to the extensive literature on gender inequality in the labor market \citep{pol2,hum1,hum2,kleven_child_2024}. \citet{goldin_grand_2014} emphasizes that closing gender gaps requires structural changes in the labor market that enhance flexibility. More recently, \citet{goldin2022understanding} has argued that the rise of WFH should eventually benefit the labor supply of college-educated mothers. By exploiting the long-run effect of COVID-19 on the availability of remote work options, we are among the first papers to show that WFH can help close gender gaps in labor market outcomes, even in developed countries.\footnote{For example, \citet{jalota2024works} find in an RCT in India that increased remote work options double female labor force participation among data-entry workers. \citet{zarate2025child} use pseudo-panels from several Latin American countries to study how changes in remote work mitigated the child penalty. \citet{basso2026wfh} leverages administrative data and WFH contracts to study the impact of WFH on the motherhood penalty in Italy.}
%It has been difficult to disentangle net employment effects in the immediate aftermath of the Covid pandemic due to a shortage in the supply of childcare services \citep{heggeness2021telework}. 

\paragraph{Outline}
The remainder of the paper is structured as follows. Section~\ref{sec:data} describes our data sources and sample construction. Section~\ref{sec:empirics} presents our empirical strategy and identification assumptions. Section~\ref{sec:results} reports our main results on women's labor market outcomes. Section~\ref{section:robustness} presents robustness checks and placebo tests. Section~\ref{sec:mechanisms} uses time-use and childcare data to examine the intra-household mechanisms behind our results. Section~\ref{sec:conclusions} concludes.

%%%%%%%%%%%
%% Data %%%
%%%%%%%%%%%
\section{Data} \label{sec:data}

\subsection{Data Sources and Sample}

We draw on several public-use U.S. household surveys. Our primary results use the American Community Survey (ACS, 2013--2024), which provides detailed demographics and labor market information on all  household members, including the presence and age of children, and occupational codes for both husbands and wives.\footnote{Occupation codes at the 4-digit level are based on the 2010 Census Occupation Classification (\textsc{OCC2010}). According to ACS documentation, respondents are asked to report their primary occupation, defined as the one from which they earn the most income; if uncertain, respondents are instructed to report the occupation at which they spend the most time. Unemployed individuals are asked to report their most recent occupation, and when more than one occupation is listed, the first one recorded is used. In our baseline sample, 81.9\% of unemployed men report a non-missing occupation, compared to only 50.6\% of unemployed women, consistent with women being more likely to have never been employed. %We do not restrict the sample to employed husbands, and we show robustness of our results to controlling for the husband's employment status in \autoref{tab:acs_baseline_annual_spemp_annual_control}.
} The bulk of our analysis about labor market outcomes comes from the ACS data's retrospective measures of employment and labour supply. This usually cover the preceding 12 months, and include indicators for employment, usual hours worked per week, number of weeks worked, part-time status, wages and total salary income (earnings).\footnote{Throughout the paper, we define intensive margin outcomes such as wages conditional on employment.} Other outcomes, such as work from home uptake and commuting time, are measured at the week prior to interviews. We also provide complementary evidence from the Current Population Survey (CPS, 2013--2026) and the Survey of Income and Program Participation (SIPP, 2014--2024). To investigate underlying mechanisms, we use time-use data from the American Time Use Survey (ATUS, 2013--2024). To measure changes in WFH exposure at the occupation level, we use job postings data on remote and hybrid work arrangements \citep{hansen2023remote}, complemented by the WFH feasibility index of \citet{dingel_how_2020}. Appendix~\ref{app:data_construction} describes each data source in further detail.

\paragraph{Sample Restrictions}

We restrict our analysis across datasets to married dual-parent households in which both spouses reside together and have at least one child aged 0--8, excluding households with more than four children; we further restrict the sample to women aged 20--50 and men aged 18--64. %\footnote{We adopt standard age restrictions for women, as is common in the literature \citep{kleven2024eitc}. We additionally impose age restrictions on men to ensure they are of typical labor-force age.}

\subsection{Treatment Definition} \label{subsection:treatment_defn}

We use job postings data from \citet{hansen2023remote} to construct occupation-level measures of the change in WFH probability before and after the COVID-19 pandemic, computed as the difference in the share of hybrid and remote-work job postings between the post-pandemic (2023--2024) and pre-pandemic (2019) periods. Appendix \autoref{fig:shock_distribution} shows the distribution of this measure. %We convert occupation codes from the SOC 3-digit classification to ACS occupation codes and 
We classify occupations with a WFH probability change at or above the median as treated, and those below as control; these measures are merged to household survey data based on the husband's occupation at the time of the survey.\footnote{Ideally, we would match individuals to their pre-pandemic occupation to avoid concerns about endogenous selection into more WFH-intensive occupations on the basis of unobservables. Unfortunately, the cross-sectional nature of our dataset does not allow us to do so.} 

Appendix \autoref{tab:top_occ} lists the 15 most common occupations in the treated and control groups. As expected, occupations that experienced the largest increases in WFH probability after COVID-19 include those in the tech sector, law, and management, such as computer and software engineers, lawyers, and various types of managers. In contrast, occupations with minimal changes in WFH feasibility are concentrated in sales, construction, and retail, as well as in sectors like healthcare (e.g., physicians and surgeons) and education (e.g., teachers), where in-person presence remains a key job requirement. Data from the 2019 Occupational Employment Statistics (OES) (Appendix \autoref{tab:occ_oes}) and ACS (Appendix \autoref{tab:occ_acs}) show that treated occupations are associated with higher wages, both on average and at any percentile of the distribution, and tend to employ more educated workers with longer schooling and longer work hours.\footnote{Somewhat surprisingly, women appear slightly more represented in treated occupations in the ACS sample, although the difference is small (i.e., 1 percentage point) and not economically significant.}

Using the flexibility indexes from \citet{goldin_grand_2014} -- updated with the latest O*NET data -- we find little correlation between our treatment definition and these flexibility measures (see Appendix Figures \ref{fig:onet2025_five} and  \ref{fig:onet2025_seven}), easing concerns about pre-existing occupational differences. As shown in Appendix \autoref{tab:desc_stat_onet}, treated occupations are somewhat more amenable to remote work (e.g., less physical proximity, more irregular schedules, less structured work, and greater freedom to determine tasks and priorities), but are not necessarily more ``flexible'' in the sense of \citet{goldin_grand_2014}. Indeed, \citet{goldin_grand_2014}'s indexes are statistically higher for treated occupations\footnote{Higher index values indicate lower worker substitutability and lower job flexibility.} and are designed to capture temporal flexibility and the degree to which workers are substitutable -- a distinct concept from remote work suitability.\footnote{ \citet{mas_valuing_2017} show that workers value different dimensions of work arrangements quite differently. In a field experiment with a national call center, they find that workers place little value on scheduling flexibility -- such as the ability to choose days, times, or hours worked -- but do value working from home substantially, with a willingness to accept up to an 8\% wage cut for the option.} This distinction underscores that while remote work and flexibility are related, the two concepts are not equivalent.

\subsection{Descriptive Statistics}

Appendix Tables \ref{tab:desc_labor} and \ref{tab:desc_hh} present descriptive statistics for baseline labor market outcomes and household-level covariates, respectively. On average, treated households are more likely to have both spouses employed, with both husbands and wives working at higher rates than in control households. Treated households also report a higher probability of both spouses working from home and slightly longer commuting times relative to control households, with husbands commuting longer than wives on average. As expected, treated households are more likely to have both spouses highly educated and less likely to have attained only a high school degree or less. Treated wives are also slightly older, on average, than their counterparts in the control group. In terms of family composition, treated households tend to have fewer children and younger children on average. Additionally, labor market outcomes for women in the treated group exhibit larger average changes post-COVID compared to women in the control group, though this pattern is purely descriptive. In all regressions, we control for the education levels of both spouses, the wife's age, and the number of children in the household.

Appendix \autoref{tab:desc_atus_out} reports descriptive statistics for the time-use outcomes. Men devote roughly twice as much time to market work as women. In contrast, women allocate a substantially larger share of their workday to working from home and spend only about half as much time commuting. Women also devote approximately twice as much time to household production as men. Childcare is the most time-intensive household activity for both spouses and exhibits a large gender disparity. At baseline, women spend about one-third of the day caring for children, compared with one-sixth for men. For both genders, childcare performed as a primary activity accounts for roughly one-third of total childcare time. Time devoted to leisure and personal care -- such as sleeping and grooming -- is similar across spouses. Although treated husbands spend more time working from home at baseline, the treatment and control groups are otherwise similar in terms of housework, personal care, and leisure activities.

%%%%%%%%%%%%%%%%%%%%%%%%%%%%%%%%%%%%%%%%%%%%%%%%%%%%%%%%%%%%%%%%%%%%%%%%%%%%%%%%%%%%%%%%%%%%%%
%%% EMPIRICAL STRATEGY %%%%
%%%%%%%%%%%%%%%%%%%%%%%%%%%%%%%%%%%%%%%%%%%%%%%%%%%%%%%%%%%%%%%%%%%%%%%%%%%%%%%%%%%%%%%%%%%%%%

\section{Empirical Strategy}\label{sec:empirics}

\paragraph{Baseline Difference-in-Differences}

We exploit variation in the change in WFH probability across occupations induced by the COVID-19 pandemic. Our identification strategy compares the labor market outcomes of women whose husbands were employed in occupations that experienced larger increases in WFH probability to those whose husbands experienced small or no changes, before and after the pandemic. As explained in the previous section, change in WFH exposure is collapsed at the occupation level and reflects the difference in demand-driven WFH probabilities pre- and post-COVID.

Our baseline specification is:
\begin{equation}\label{eq:1}
y_{ist}
= Post_t \times (\beta  + \delta During_t) \times \text{High-}\Delta\text{WFH}_{\,o(h(i))}
+  X'_{it}\psi
+ \gamma_{st}
+ \Theta_{o(h(i))}
+ \varepsilon_{ist}.
\end{equation}

where \( y_{ist} \) denotes a labor market outcome for woman \( i \), observed in state \( s \) and time \( t \).\footnote{Time is defined at the month-year level in CPS, ATUS, and SIPP. In ACS, where month of interview is unavailable, time is defined at the year level.} 
\( \text{High-}\Delta\text{WFH}_{\,o(h(i))} \) is a dummy equal to one if the husband’s occupation experienced an above-median increase in WFH probability between the pre- and post-pandemic periods. \( During_t \) and \( Post_t \) are time dummies for the years 2020--2021 and for 2020 onwards, respectively.

The coefficient \( \beta \) captures the average post-pandemic change in outcomes for women whose husbands are employed in high-\(\Delta\)WFH occupations relative to those in low-\(\Delta\)WFH occupations --- that is, the difference-in-differences effect after 2020. The coefficient \( \delta \) captures any additional differential effect during the COVID-19 period (2020--2021), relative to the post-2022 period.\footnote{In the ACS sample, where we use annual data from 2013 to 2024, we define the during-COVID period as 2020--2021 and the post-COVID period as 2022 onward. In the ATUS (a subsample of the CPS) and main CPS samples, where we have monthly data from 2013 to 2024, we define the during-COVID period as March 2020 to February 2022, and the post-COVID period as March 2022 onward.}

The control vector \( X_{i,t} \) includes the woman’s age, education, and number of children, as well as the husband's education. All regressions include state-by-time fixed effects \( \gamma_{st} \) to account for local labor market shocks and policy variation, and occupation-of-husband fixed effects \( \Theta_{o(h(i))} \) to absorb time-invariant differences across husband’s occupations.

%When leveraging the panel structure of one of our datasets, the SIPP, we estimate a panel version of Equation~\eqref{eq:1} by including household fixed effects. As a result, we remove any control or fixed effect that is time-invariant, differencing out all unobserved and observed factors that do not vary over time within households and/or individuals.

%When conducting heterogeneity analysis by age of children or other relevant categories, we interact the treatment term with the categorical variables of interest, omitting the base category, and similarly interact controls and fixed effects to allow for differential trends across groups.

%We also conduct industry-level specifications, where we construct a treatment indicator based on whether the husband's Census 2012 industry ($n(h(i))$) experienced an above-median increase in WFH probability. In these regressions, we replace the occupation-level treatment term in Equation~\eqref{eq:1} with the industry-level indicator interacted with the post and during-COVID periods. As with the main specification, we include state-by-time and spouse occupation and/or industry fixed effects, along with the standard set of demographic controls.

\paragraph{Identification Assumption}
Under a standard parallel trends assumption --- namely, that absent the pandemic, the labor market outcomes of women whose husbands were employed in occupations experiencing large increases in WFH probability would have followed trajectories similar to those of women whose husbands were employed in occupations with minimal WFH changes --- the coefficient $\beta$ identifies the causal effect of interest. Specifically, $\beta$ captures the reduced-form \emph{household-level effect} of the husband's exogenous shift in WFH probability, induced by the pandemic, on the wife's labor market outcomes.\footnote{As discussed more extensively in Appendix Section~\ref{app:metrics} and Section~\ref{sec:results}, our baseline empirical strategy necessarily combines the spillover effect arising from the husband's WFH exposure with the direct effect operating through the wife's own WFH exposure, which is correlated with the husband's by virtue of assortative matching.}

\paragraph{Dynamic Difference-in-Differences} To assess pre-trends and the dynamic evolution of treatment effects, we estimate a dynamic version of Equation~\eqref{eq:1}. Specifically, we estimate the following specification:

\begin{equation}
\label{eq:2}
\begin{aligned}
y_{ist}
= \sum_{t\neq 2019} \beta_t \, \text{High-}\Delta\text{WFH}_{\,o(h(i))}\times D_t
+ X'_{it}\psi
+ \gamma_{st}
+ \Theta_{o(h(i))}
+ \varepsilon_{ist}.
\end{aligned}
\end{equation}

where 2019 serves as the omitted (base) year. The coefficients \( \beta_t \) capture the differential evolution of outcomes for women whose husbands were employed in occupations with large WFH shifts, relative to those in occupations with smaller or no such changes, for each year over the period 2012--2023 (excluding 2019).\footnote{As discussed in the data section, the effective coverage period for outcomes referring to the previous 12 months spans 2012--2023, whereas for outcomes measured at the time of the interview it spans 2013--2024.} All notation follows as in Equation~\eqref{eq:1}.

\section{Main Results}\label{sec:results}

\subsection{WFH Shock Exposure and Husbands' WFH Uptake}

Our identification strategy exploits occupation-level WFH shocks to husbands, which plausibly increase their time availability for household production and, in turn, raise wives' labor supply. Consequently, our first-stage analysis verifies that large shocks to occupation-level remote work options available to fathers of young children in our ACS sample led to a sizable increase in their own uptake of remote work.\footnote{We use as measure of remote work in ACS the following question: \emph{How did you  usually get to work LAST WEEK. Mark (X) ONE box for the method of transportation used for most of the distance}. As discussed in \citet{buckman2025measuring}, this question design yields a measure for the incidence of fully remote work, and aligns with comparable measures from the Survey on Working Arrangements SWAA and other surveys. However, we note that ACS captures only fully remote work; nowadays hybrid options are more prevalent.}  

\begin{figure}[H]
    \caption{Dynamic Effects of Husbands' WFH Shock on Own Remote Work Uptake}\label{fig:acs_dynamic_DiD_spwfh_only}    
    \includegraphics[width=0.9\linewidth]{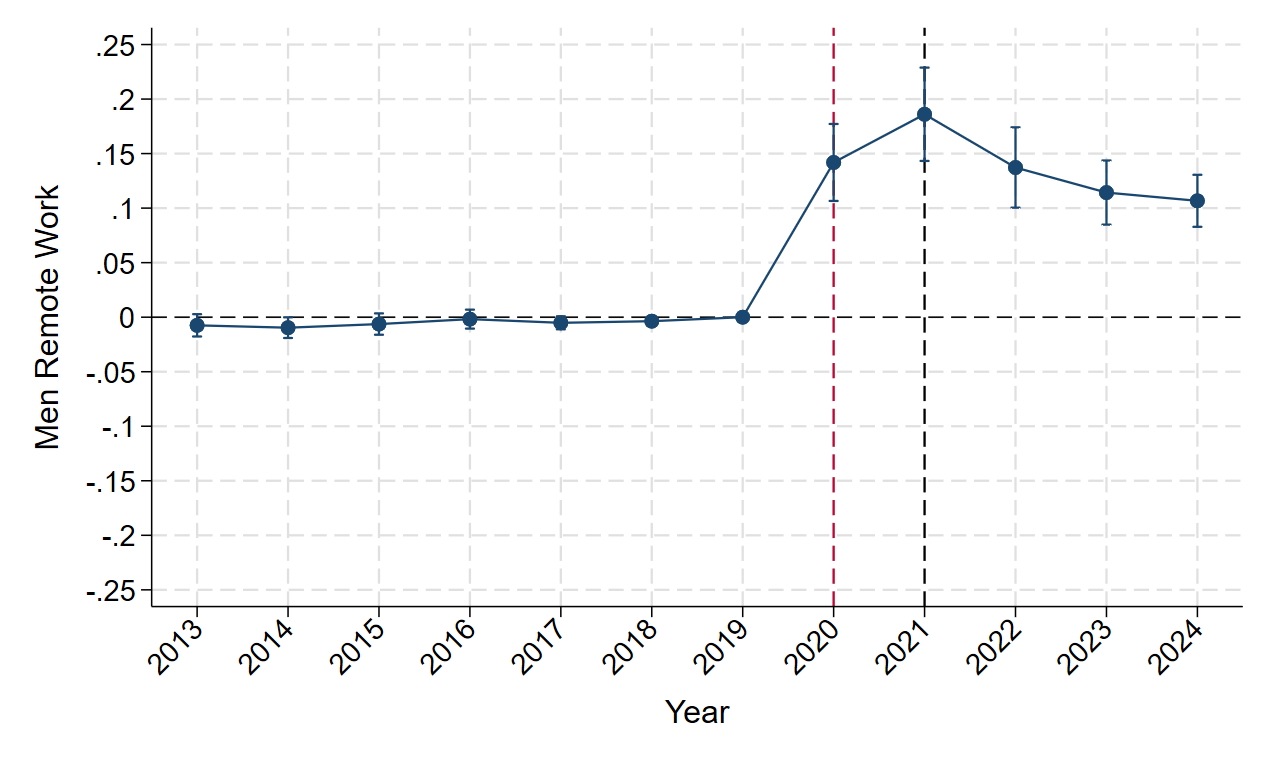}
    \centering
    \begin{minipage}{\textwidth} \small
         \textit{Notes:} This figure plots yearly coefficients from the dynamic difference-in-differences specification in Equation~\eqref{eq:2}, estimated on the 2013--2024 ACS sample of dual-parent households with children of child-rearing age. The outcome is the husband's self-reported work-from-home (WFH) status (see \autoref{tab:acs_first_stage_combined} for the outcome definition). The omitted year is 2019. All regressions include state-by-time and husband's-occupation fixed effects, and control for the wife's age, number of children, and education levels of both spouses. Confidence intervals are shown at the 95\% level and are based on standard errors clustered at the husband's occupation level.
    \end{minipage}
\end{figure}

In \autoref{fig:evolution_WFH}, we already documented large and heterogeneous changes in the demand and supply of remote work options in the US labor market across occupations. We now test this finding more rigorously using our baseline dynamic specification. The dynamic DiD in \autoref{fig:acs_dynamic_DiD_spwfh_only} shows that men in treated occupations increased their uptake of remote work by as much as 20 p.p. relative to the control group and to the pre-COVID baseline year of 2019. Although the effects are smaller (around 10 p.p.) by 2024 --- coinciding with a growing trend of firms mandating returns to the office --- the shock has had lasting and economically meaningful effects on households. With a pre-pandemic average WFH rate of 3\%, rates for men in our sample have tripled. %This increase broadly mirrors the large rise in the share of remote work job vacancies for highly exposed occupations documented in \autoref{fig:evolution_WFH}. 

\autoref{tab:acs_first_stage_combined} reports estimates of spouses' WFH uptake using the baseline three-period DiD specification in equation \eqref{eq:1}. On average, the fifth column shows men's WFH uptake increased by approximately 12 p.p. in the post-pandemic period. In the final column, we examine commuting time as share of the day and find that men in more exposed occupations reduce their commuting time by roughly 17\% of their pre-treatment average.

%To delve deeper in how households reorganized their market and non-market division of tasks, we exploit the ATUS and ACS time-use variables. 
Using our time-use diary data -- whose sample size is an order of magnitude smaller than that of the ACS -- we also find that  men exposed to more remote work options experience a substantial reallocation of their work location. Appendix \autoref{fig:dynamic_laborprod_men_ATUS} shows that they increase the share of time spent working from home by 4.3 p.p., with a roughly offsetting reduction in time spent working outside the home.

Interestingly, both the ACS and ATUS show there is no change in men's overall working time. In the first four columns of \autoref{tab:acs_first_stage_combined} we find no change in men's earnings, hours worked on a week or part-time status; if anything there is a significant reduction in weeks worked across a year. Panel (a) of Appendix \autoref{fig:dynamic_laborprod_men_ATUS} shows no change in the overall share of day spent working. The change in men's labor supply, thus, seems to amount entirely to a relocation of their workplace.

\subsection{Effects on Wives' Labor Market Outcomes}
\label{section:baseline_results}

\paragraph{Effects on Wives' Labor Supply}

We hypothesize that expanded WFH opportunities for husbands relax household production constraints --- both temporal and physical --- thereby facilitating wives' labor market participation while raising children. \autoref{tab:acs_baseline_annual} presents estimates from Equation~\eqref{eq:1} in the ACS data and shows that an exogenous increase in remote work availability for husbands leads to a significant expansion in wives' labor supply at the intensive margin.\footnote{As discussed in Section~\ref{sec:data}, intensive margin outcomes are observed only for individuals with positive earnings and should therefore be interpreted as conditional on employment.} Specifically, wives' total annual earnings increase by approximately 5\%, weekly hours rise by around 0.5 hours, and the number of weeks worked increases by 0.5. Consistent with these intensive-margin responses, the probability of working part-time declines by approximately 2.5 percentage points, corresponding to a 9\% reduction relative to the pre-treatment mean. We find qualitatively similar results when splitting occupations by quartiles of treatment exposure, with effects concentrated among the third and fourth quartiles, as shown in Appendix \autoref{tab:acs_intensity}.

\autoref{fig:ddid_main} plots the annual coefficients from the dynamic DiD specification in Equation~\eqref{eq:2} for our main labor market outcomes. Pre-treatment coefficients are comparatively flat across all outcomes, and they exhibit sharp increases in 2020 that persist over time, consistent with the lasting effects on WFH uptake documented in the first stage analysis.\footnote{Using the dynamic DiD specification in the ATUS data, we also corroborate that women experience an increase in the time allocated to market work, consistent with the increase in hours worked documented in the ACS findings. As shown in Appendix \autoref{fig:atus_dynamic_DiD_working_share_wives}, a spousal WFH shock leads to women increasing the share of their day spent working after the pandemic by 5.8 percentage points, i.e., equivalent to 85 additional minutes.}  Overall, the evidence broadly supports the parallel trends assumption and indicates that treatment effects strengthen in the post-COVID period.\footnote{To assess whether these effects reflect improvements among treated households or deterioration among control households, \autoref{fig:ddid_main_by_group} plots adjusted labor market trends separately for the treated and control groups. Earnings increase only for the treated group, while remaining flat for the control group. Hours worked and weeks worked increase --- and part-time employment declines --- for both groups, with effects of larger magnitude among treated households.} \footnote{Importantly, the intensive margin effects are robust to controlling for linear and non-parametric trends in the husband's education category and broad occupation group, as discussed further in Section~\ref{section:robustness} -- see e.g. Appendix Tables \ref{tab:acs_appendix_trends_spedlevel} and \ref{tab:acs_appendix_trends_broadspocc}.}

One might ask whether the husband's WFH shock was sufficiently large to induce wives to enter employment, or whether, conditional on employment, wives sorted into higher-paying jobs. We find no evidence of either margin. As shown in Appendix \autoref{fig:ddid_emp_wages}, we cannot rule out pre-existing trends in employment, nor do we find a significant effect on wages. When controlling non-parametrically for trends in the husband's education level or broad occupation category, the employment estimates are small and statistically indistinguishable from zero. The null effect on employment is consistent with the husband's WFH shock being insufficient to overcome the fixed costs of labor market entry for women. The null effect on wages, in turn, aligns with the evidence in Section~\ref{sec:data} showing that WFH feasibility has little correlation with \citet{goldin_grand_2014}'s measure of temporal flexibility, suggesting that the shock did not necessarily alter the way firms structure and remunerate jobs.\footnote{As \citet{goldin_grand_2014} argues, reducing the gender wage gap requires firms to stop disproportionately rewarding long and inflexible hours --- a change in labor demand that lies beyond the scope of the husband's WFH shock.} These results are also consistent with recent evidence on WFH and the child penalty, which similarly finds no effects on employment or wages \citep{basso2026wfh, scott2025flex}.

Since employment and wages do not respond to the shock, the remaining analysis focuses on the intensive margin outcomes presented in \autoref{tab:acs_baseline_annual} --- hours worked, weeks worked, earnings, and part-time status --- where the effects are both economically meaningful and precisely identified.

\paragraph{Prevalence of Spousal Work from Home} To better characterize the role of spousal availability at home on wives' labor market supply, we estimate a modified 2SLS that combines our first stage with the baseline specification. Specifically, our second stage is the same as the main specification (\ref{eq:1}), but replacing the exogenous spousal occupation-level WFH shock, $High\,\Delta WFH_{o(h(i))}$, with the spouse's observed WFH uptake, $WFH_{h(i),t}$. For the first stage, we instrument observed uptake with the leave-one-out average fully remote work rate in the spouse's occupation using ACS data where we leave out the spouse's own WFH uptake from the occupation-year average, $\overline{WFH}_{o(h(i)),-i,t}$.

The second-stage results are reported in Appendix \autoref{tab:acs_baseline_2sls_loo} and are consistent with the baseline estimates. If anything, the effects are two to three times as large, which is intuitive: rather than averaging across households where husbands are only more likely to work remotely, we are capturing the effect of the husband being fully available at home on wives' labor supply.\footnote{Since we are identifying a LATE, compliers are households in which husbands are even more likely to work from home full-time when the WFH rate in their occupation has increased substantially.}

%In \autoref{tab:acs_baseline_ind_occ}, we include both the occupation- and industry-level WFH shocks, as well as their interaction, to disentangle the relative importance of each source of variation. The results indicate that the occupation-based measure primarily drives the effect. The coefficient on industry exposure is positive but not statistically significant, while the interaction term is near zero, suggesting no meaningful complementarity between the two dimensions of exposure.

\paragraph{The Role of Spousal Education} Since our empirical strategy exploits major structural changes in labor market occupations, understanding the role of the spouses' education level in mediating our effects is at the core of our analysis. We take a first look at this dimension in Appendix \autoref{fig:acs_het_edlevel}, where we re-estimate the baseline specification separately by spouses’ education levels. From panel A, we see that  effects among college-educated women are more precise and closer to our baseline results. This is not surprising on one hand as most women in our sample have college degrees. Moreover, this is consistent with the idea that college-educated women may be more likely to hold professional jobs in which they can more flexibly increase participation, hours, and earnings. Interestingly, as shown in panel B, effects are no larger in households with college educated male spouses than those where men have only high school completed. This suggests a limited role of assortative matching in driving our results.  %Furthermore, \autoref{tab:acs_het_reledlevel} examines how wives' treatment effect vary by their relative years of education with respect to their husbands, a stable proxy for women’s bargaining power \citep{chiappori2002b,browning1998}. The estimates reveal that the treatment effects are increasing in the level of the relative years of education at diminishing rate.

\paragraph{Presence and Age of Children}

Our argument hinges on the idea that having children imposes constraints on household production that restrict wives' labor supply, and that these constraints are relaxed as WFH increases the husband's availability at home. To validate this mechanism, Appendix \autoref{tab:acs_placebo_allchild} shows that the spousal WFH shock has no effect on wives' labor supply in households without children (Panel A) or in households where all members are already adults (Panel C). Effects are broadly similar for households with teenage children relative to our baseline results, suggesting that the age profile of children up to adulthood does not mediate effect sizes. We further confirm in Appendix \autoref{fig:acs_dynamic_DiD_ageyc} that effect sizes are statistically indistinguishable across households with children aged 1 to 8.
This is consistent with the well-documented finding that child-related penalties on women's labor market trajectories persist well beyond childbirth and the most intensive years of childcare \citep{waldfogel1998understanding, kleven2019children, cortes2023children, goldin2022understanding}.
%Indeed, in \autoref{fig:acs_dynamic_DiD_ageyc} we confirm that effect sizes for women labor supply are statistically indistinguishable among households with children in the ages between 1 and 8.

\paragraph{Are Men Affected By Spousal Shocks?} Our preferred specification uses husbands' changes in their occupation WFH rates following COVID-19 as an exogenous shock that impacts their wives' labor market decisions. However, it could be that when their wives' face an increased availability of remote work options, men also respond in their labor supply. While increasing women's availability at home is not bound to change division of home production, it may still increase the intensive margin of men's labor supply. To test for potential spillover effects in the opposite direction, Appendix \autoref{tab:acs_men_annual} inverts the baseline specification and estimates the impact of wives’ WFH shocks on husbands’ labor market outcomes using ACS data. 
We find no evidence of increases in men’s earnings, part-time status or amount of weeks worked. If anything, weekly work hours experience a small decrease of 0.6\% relative to the pre-treatment average. This pattern is consistent with the fact that almost all men in our sample are already employed in full-time positions, so an increase in spousal WFH flexibility does not shift a binding constraint for them. 
%This is in line with the well-documented finding that men's own-wage labor supply elasticity is close to zero \citep{keane2011labor}, leaving little scope for either own or cross-margin responses.\footnote{Although in a different setting and literature, the household labor supply insurance literature finds small cross-elasticities of labor supply for men: \citet{blundell2016consumption} estimate that husbands' hours respond minimally to permanent shocks in their wives' wages, consistent with men being already at or near full labor supply.}
%This pattern is consistent with men moving into better-paying jobs when their spouses experience increased work flexibility. Importantly, we find no evidence of negative income effects for husbands, alleviating concerns that declines in husbands’ labor supply or earnings might drive the observed increases in wives’ labor supply.

\paragraph{Closing Labor Gender Gaps}

Our main effects have direct implications for narrowing the gender gap in labor market outcomes. To quantify this, we conduct a simple accounting exercise following \cite{kleven2019children}. In Appendix \ref{app:gender_gap_accounting}, we report the step-by-step construction of these measures.

As Appendix \autoref{tab:acs_gender_gap} highlights, husband's WFH exposure implied a 5.9\% reduction in the earnings gap, a 3\% reduction in
the gap in weekly hours worked, and a 1.2\% reduction in the gap in weeks worked, while the gap in the probability of women working part-time relative to men decreases by 1.4\%.\footnote{The negative value arises due to the fact women are more likely to work part-time than men, so a negative change in this variable implies that the gap is narrowing.} Increases in husbands' access to remote work, even when the shock is not directly targeted at women, can therefore generate a meaningful narrowing of the labor market gender gap.

\subsection{Disentangling Effects}\label{subsec:disentangle}
As discussed in Section~\ref{sec:empirics}, our baseline coefficients combine three channels: the intra-household response generated by men’s WFH exposure (\emph{spillover effects}) and two components operating through own WFH exposure - one that is uncorrelated with their spouse's shock (\emph{direct effect}) and another correlated one (\emph{matching effect}). Indeed, there is reason to believe that the latter two channels might be relevant in explaining labor supply gains. Using our dynamic DiD specification in the ACS data, we show in Appendix \autoref{fig:markettimeuseI}  that the male spouses' WFH shock led to a significant and lasting increase in women working from home and a significant reduction in the time spent commuting.

As a first step toward isolating the component of the effects on wives that operates exclusively through the husband's WFH shock, we augment Equation~\eqref{eq:1} with wife's occupation fixed effects. In this specification, $\beta$ captures the effect of the husband's exposure to high-WFH occupations, holding constant the wife's exposure through her own occupation. Appendix \autoref{tab:acs_intensive_with_WomenOccFE} shows that economically meaningful effects persist, with point estimates closely mirroring those in \autoref{tab:acs_baseline_annual}. This strategy does not, however, allow us to separately quantify the relative contribution of each channel.
%\footnote{By construction, this specification conditions on employed wives, so the estimates pertain to intensive margins only.}
%\footnote{Consistent to the main findings, there is still no effect on hourly wages.} 
%Second, we estimate the effect of women's WFH shocks on their own labor market outcomes. \autoref{tab:acs_ownshock_annual} indicates that women do experience an increase in their intensive margin outcomes, mainly in the number of weeks worked.
%\footnote{Due to survey design, we are unable to estimate extensive-margin own-elasticities. Occupation is only reported for individuals who are currently employed or in the labor force, which makes occupation-based WFH exposure mechanically collinear with participation and employment status. As a result, it is not possible to separately identify the causal effect of own WFH exposure on the decision to enter the labor force or become employed.}  

 To overcome this, we develop an empirical strategy that separately estimates coefficients for each of the three channels. To do this, in addition to men's occupation which captures WFH options for men, we need a variable that can strongly predict WFH options for women on its own. For example,  \citet{harrington2023has} leverage women’s college major as a pre-determined proxy for WFH probability. For each major, we compute pre- and post-pandemic changes in WFH uptake and define $\text{High-}\Delta\text{WFH}_{e(i)}$ as above the median.\footnote{Women without a reported major form a separate bin. This allows us to retain our full sample from the main analysis which includes women without college degree.} We then estimate:
\begin{equation}\label{eq:3}
\begin{aligned}
y_{ist} =\;& 
Post_t \times \big( \beta_1 + \delta_1 During_t  \big)  \times \text{High-}\Delta\text{WFH}_{\,o(h(i))} 
+ Post_t \times \big( \beta_2 + \delta_2 During_t \big) \times \text{High-}\Delta\text{WFH}_{\,e(i)}  \\
&\quad + Post_t \times \big( \beta_3 +\delta_3 During_t \big) \times \text{High-}\Delta\text{WFH}_{\,o(h(i))} 
                 \times \text{High-}\Delta\text{WFH}_{\,e(i)}  \\
&\quad+ X'_{it}\psi
+ \gamma_{st}
+ \Theta_{o(h(i)),e(i)}
+ \varepsilon_{ist}.
\end{aligned}
\end{equation}

In this setup, $\beta_1$ identifies intra-household \emph{spillovers} as it estimates the labor market effect of WFH spousal shocks for women who are ex-ante unlikely to WFH. $\beta_2$ captures the direct effect as it capture labor market effect of WFH shocks for women whose husbands are unlikely to work from home. $\beta_3$ captures the incremental effect of having both spouses exposed to the WFH shock, i.e. \emph{matching effects} of the male spouses' shock on women's labor supply.

%there exist pure \emph{spillovers effects} explaining part of the increase in the labor supply of women.  %On average, wives whose husbands are in highly exposed occupations are 1.4 p.p.\ more likely to be employed in the last 12 months and work roughly 30 additional minutes. %Among high-exposed majors, the joint effect of both channels imply a 3.9 p.p.\ employment increase, a 3.1 pp reduction of part-time and more than an additional hour of work. 

\autoref{tab:acs_edfield_estimates} shows that treated households in which women hold low-exposure majors experience sizable effects, as reflected in the first-row coefficient ($\beta_1$). Earnings increase by 3.2\%, while hours and weeks worked, and part-time status mirror the results documented \autoref{tab:acs_baseline_annual}. On the other hand, the direct effect estimate ($\beta_2)$, determined by women holding high-exposed majors with spouses on low-exposed occupations, presents no detectable effects on earnings or hours worked. We still observe a significant increase of 0.4 weeks worked and reduction of 1.8 percent on part-time status. The matching effect estimate ($\beta_3$) -- reported as the last coefficient -- shows almost no detectable effects on women's labor outcomes, besides an imprecise 10\% significance on earnings.

\begin{figure}[H]
    \caption{Dynamic Effects of Husbands' WFH Shock on Wives’ Labor Market Outcomes} \label{fig:ddid_main}
    \begin{tabular}{cc}
        (a) Log Earnings & (b) Usual Weekly Work Hours \\
        \includegraphics[width=.5\linewidth]{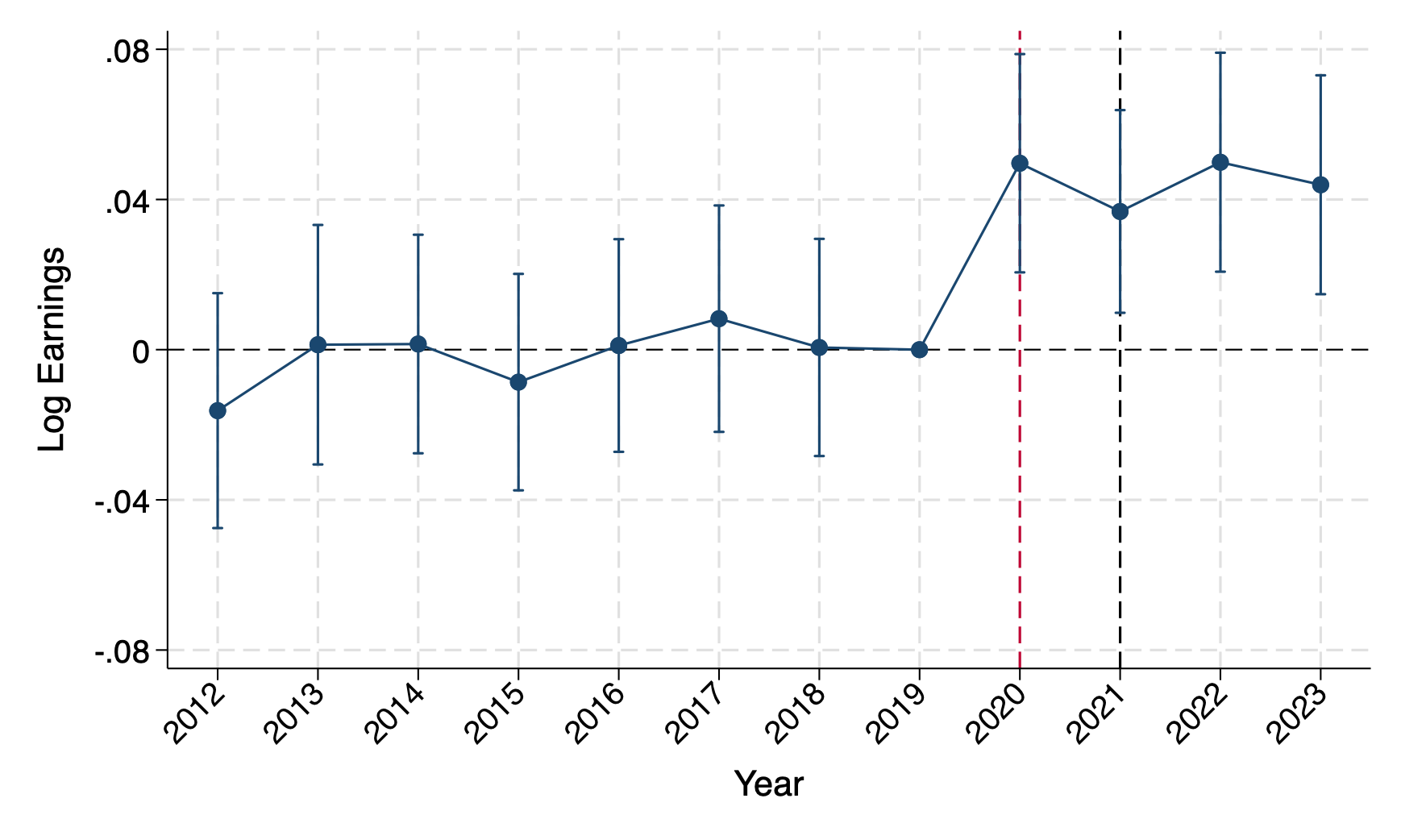} & \includegraphics[width=.5\linewidth]{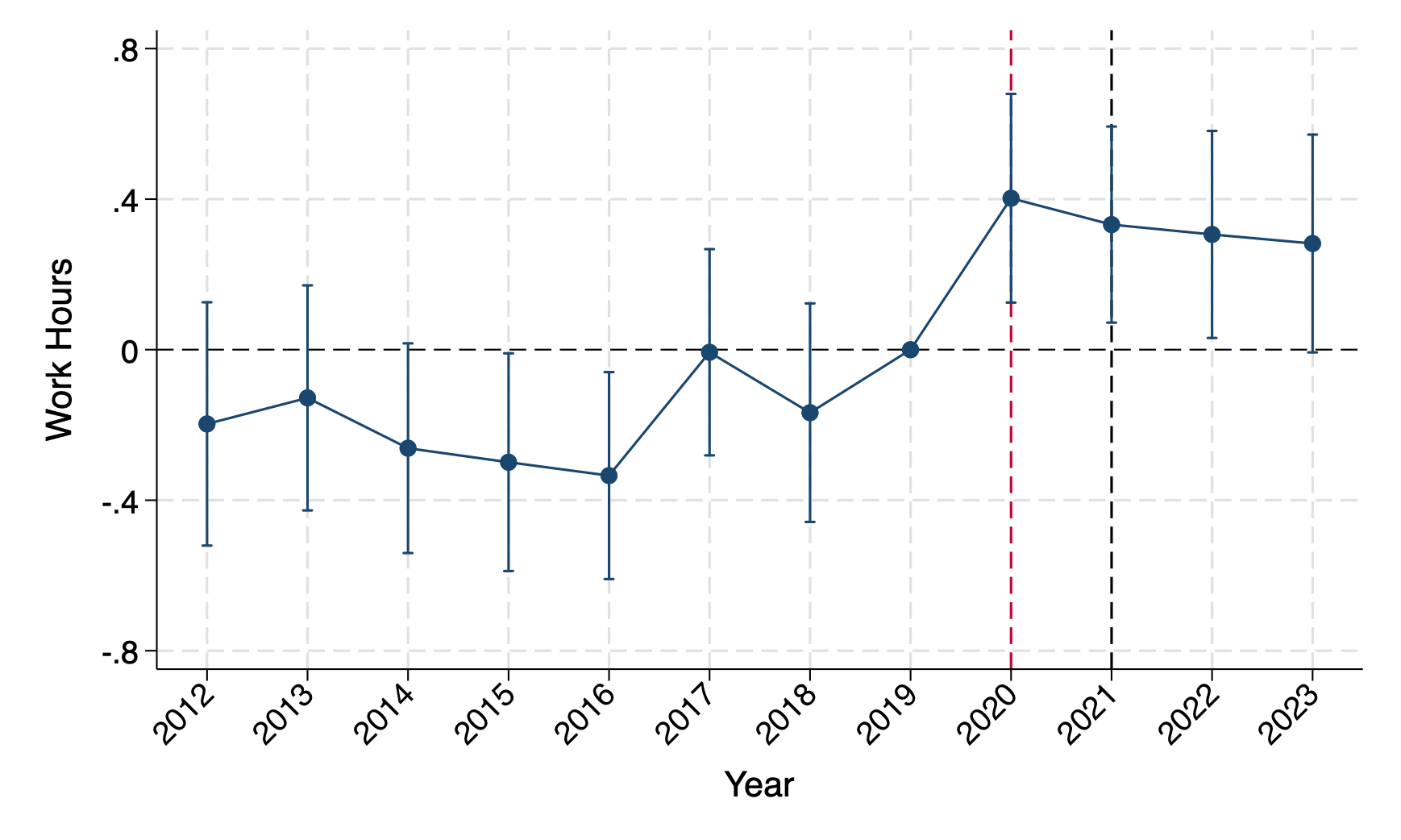} \\
         (c) Weeks Worked & (d) Part-time Status \\
        {\includegraphics[width=.5\linewidth]{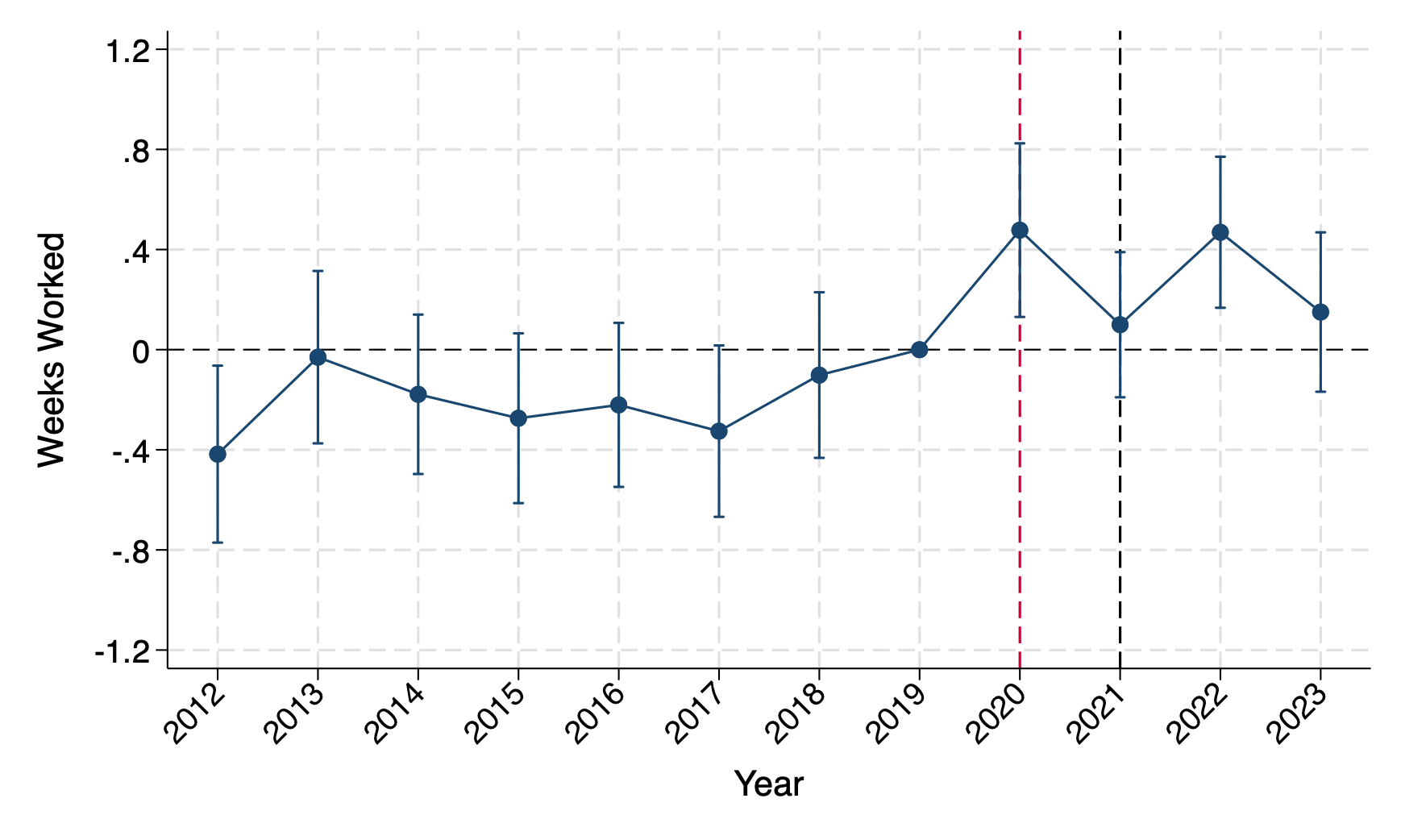}}
        & \includegraphics[width=.5\linewidth]{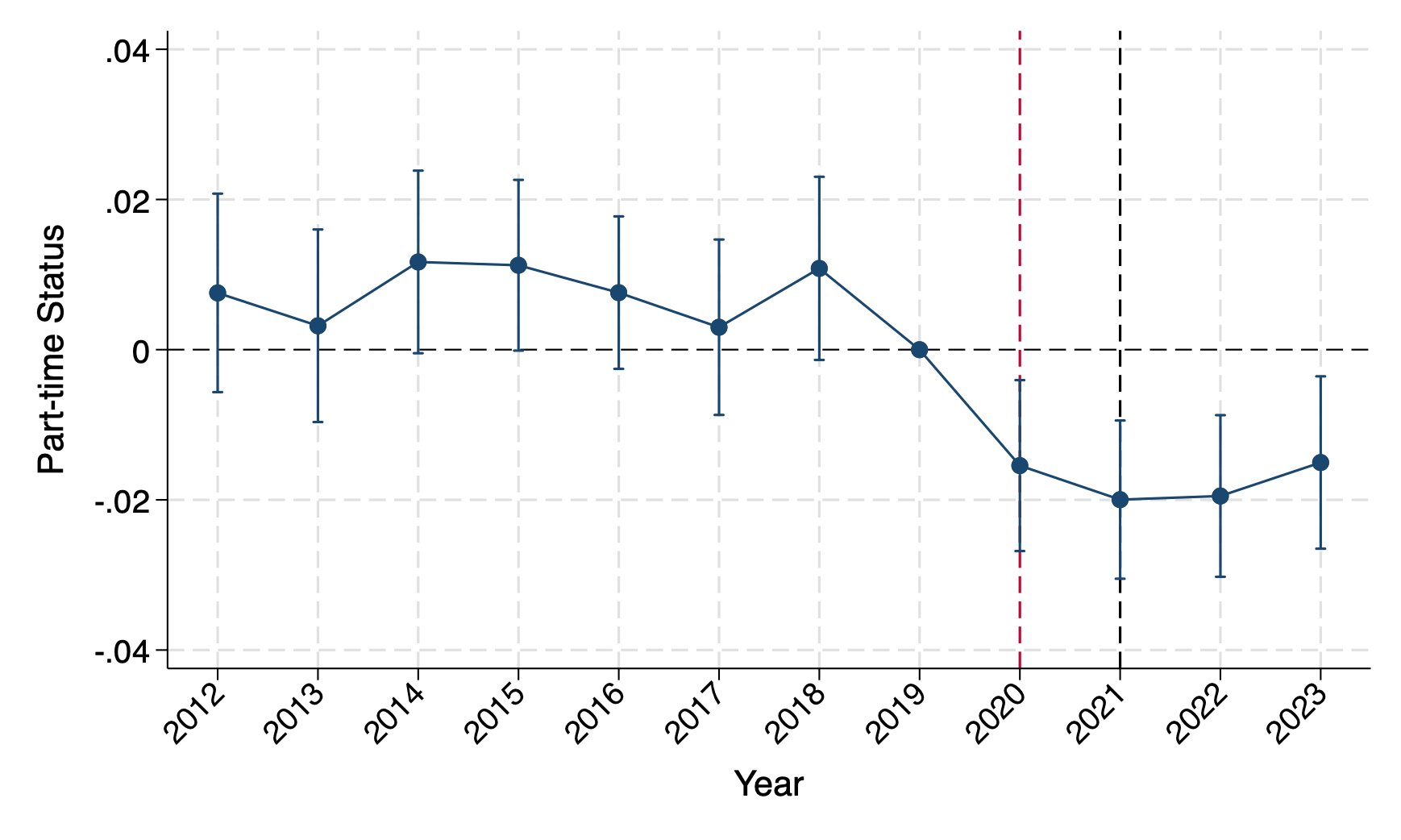} 
        
    \end{tabular}
    \vspace{0.5em}
    \centering
    \vspace{0.5em}
    \begin{minipage}{0.95\linewidth} \small
     \textit{Notes}: Each panel plots yearly coefficients from the dynamic difference-in-differences specification in Equation~\eqref{eq:2}, estimated on the 2013--2024 ACS sample of dual-parent households with children of child-rearing age. The outcomes are the wife's log annual earnings (panel a), usual weekly hours worked (panel b), weeks worked (panel c), and part-time status (panel d), all measured over the preceding 12 months and conditional on employment. The omitted (base) year is 2019. All regressions include state-by-time and husband's-occupation fixed effects, and control for the wife's age, number of children, and education levels of both spouses. Confidence intervals are shown at the 95\% level and are based on standard errors clustered at the husband's occupation level.
    \end{minipage}
\end{figure}

\begin{figure}[H]
    \caption{Event Study for Decomposition of Spillover, Direct and Matching Effects }\label{fig:acs_dynamic_DiD_edfield}
    \begin{tabular}{cc}
        (a) Log Earnings  &  (b) Weekly Work Hours \\
        \includegraphics[width=0.45\linewidth]{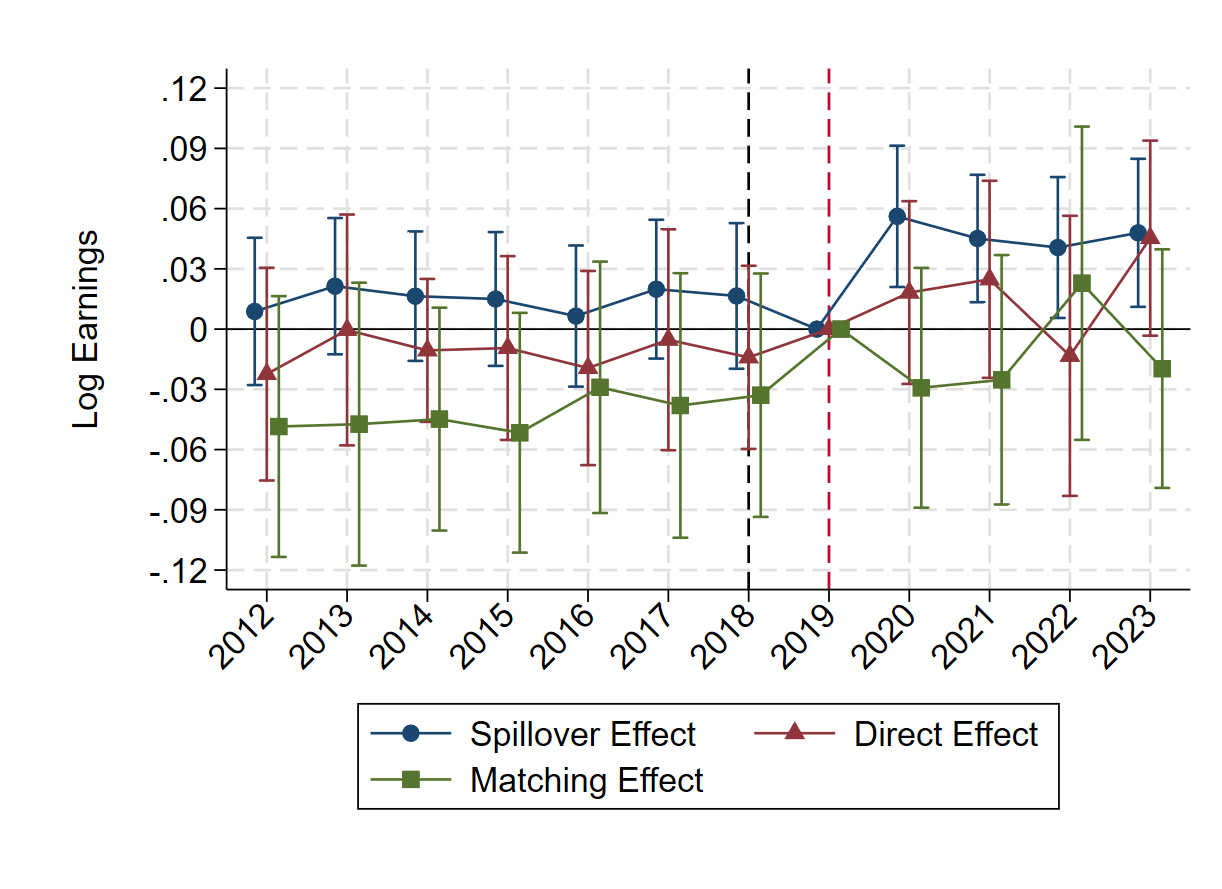} & \includegraphics[width=0.45\linewidth]{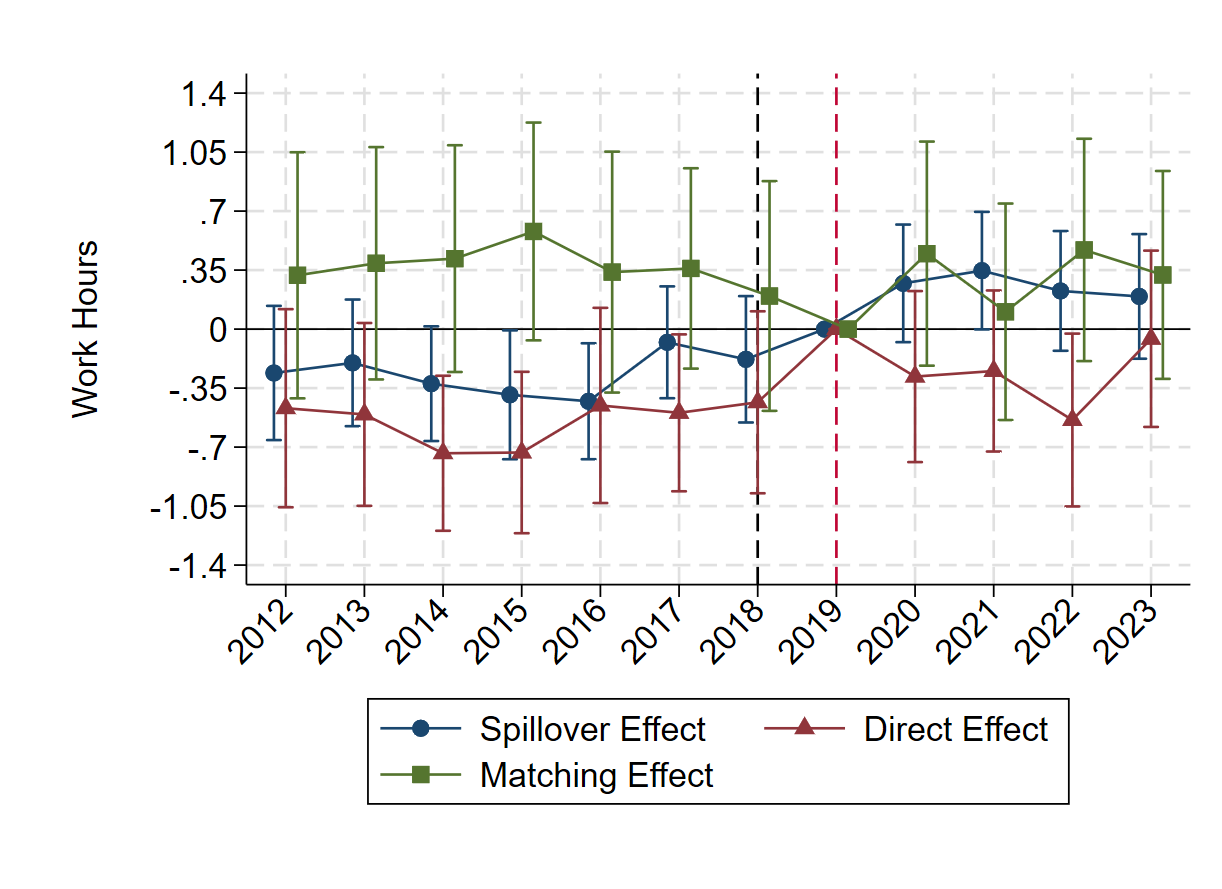} \\
        (c) Weeks Worked & (d) Part-time Status \\ \includegraphics[width=0.45\linewidth]{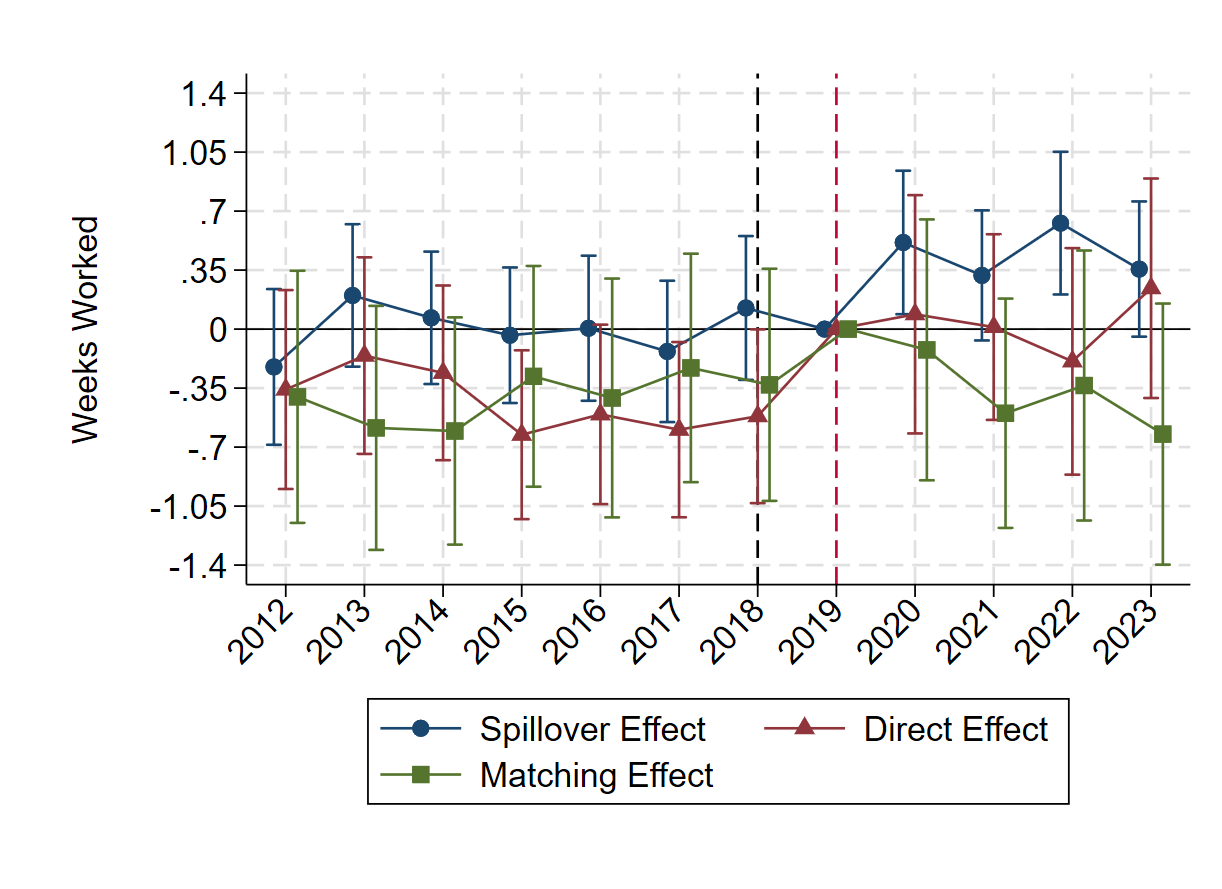} &         \includegraphics[width=0.45\linewidth]{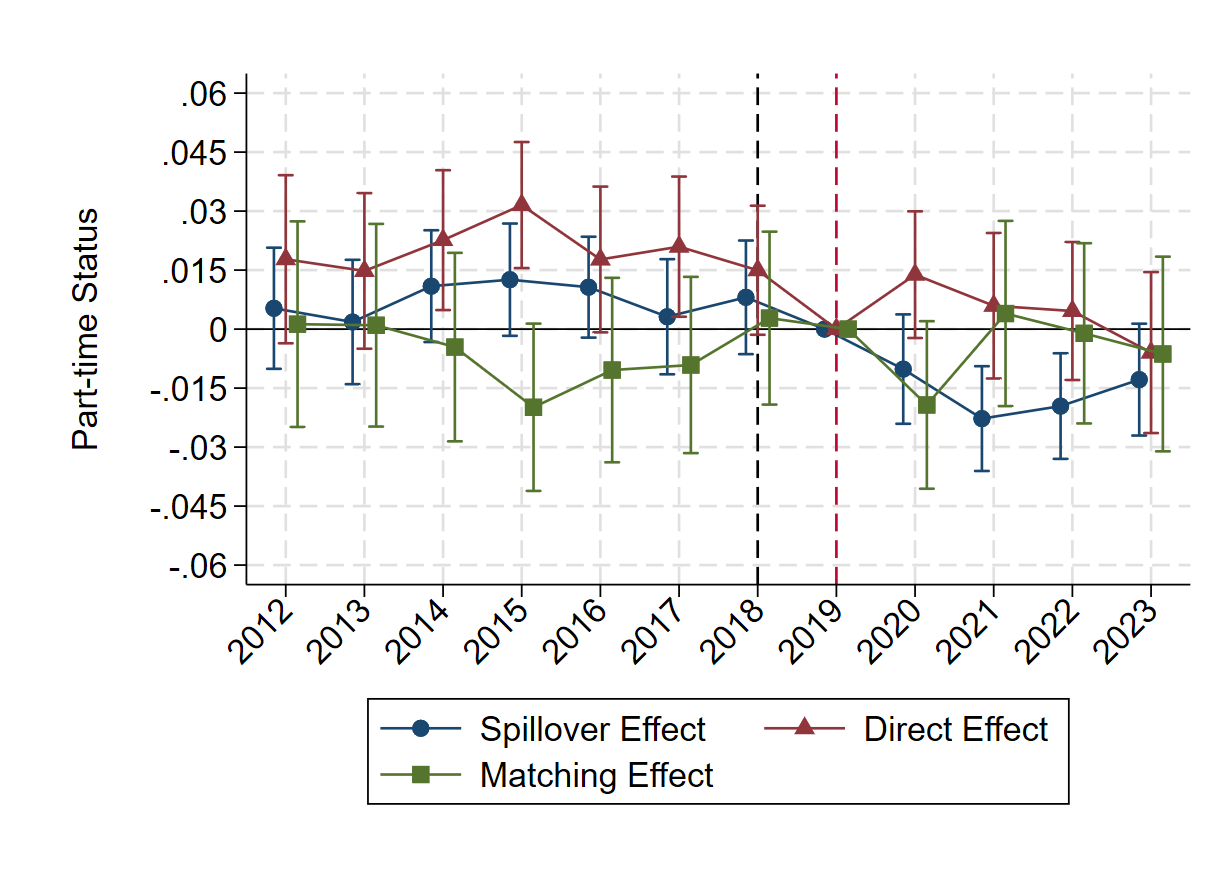} \\
        (e) WFH Uptake & (f) Commuting Time \\ \includegraphics[width=0.45\linewidth]{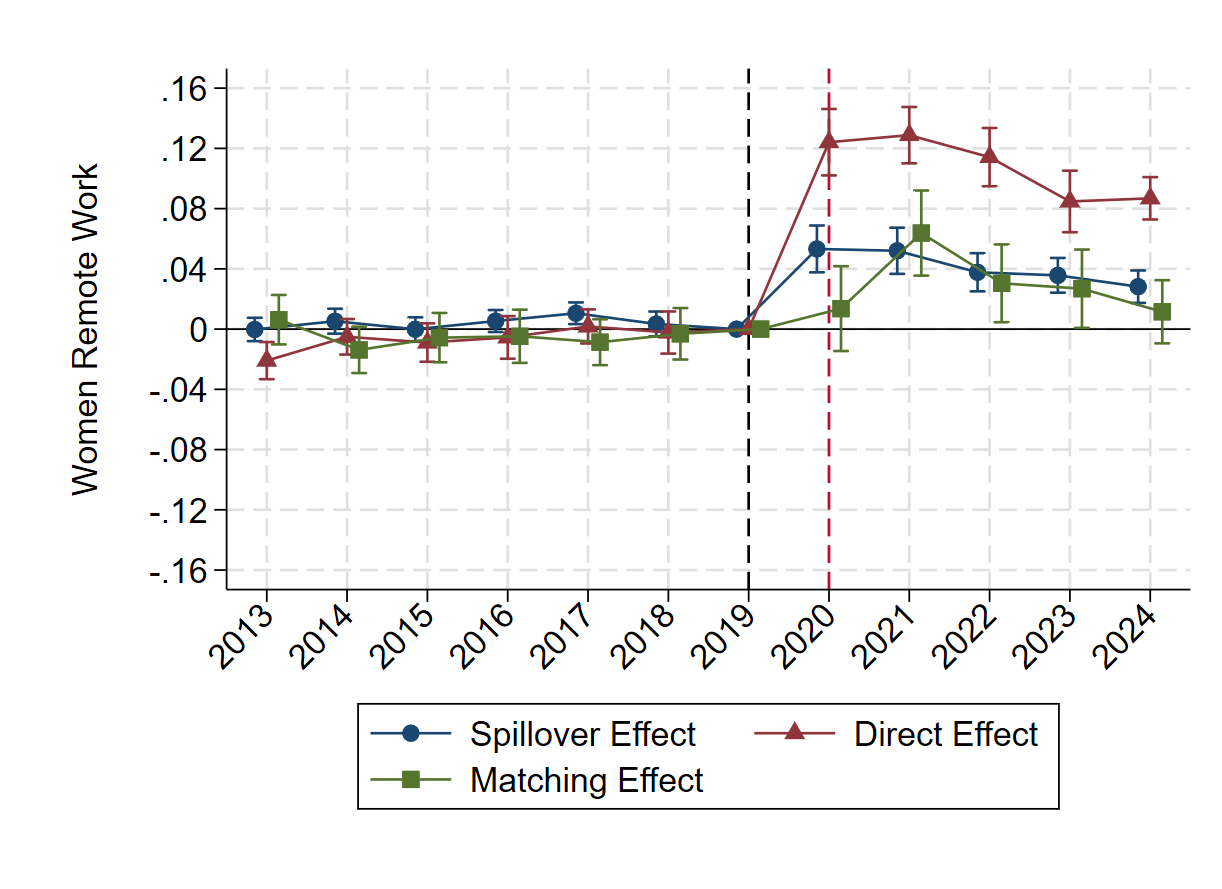} &         \includegraphics[width=0.45\linewidth]{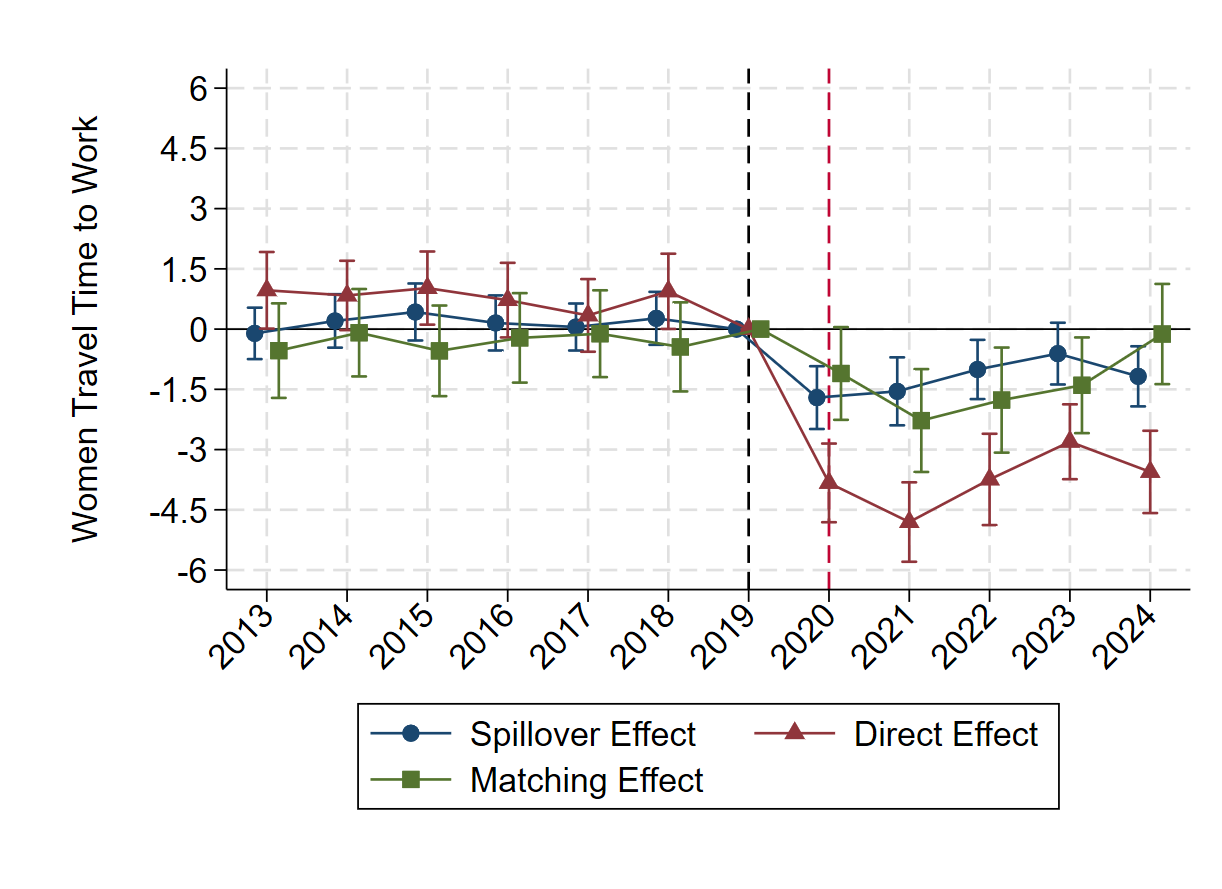}
        \end{tabular}
    \centering
    \vspace{0.5em}
    \begin{minipage}{\textwidth} \small
    \textit{Notes}: This figure plots coefficients from a dynamic version of the spillover decomposition specification in Equation~\eqref{eq:3} (the event-study analogue of \autoref{tab:acs_edfield_estimates}), estimated on the 2013--2024 ACS sample of dual-parent households with children of child-rearing age. For each yearly interaction, we plot three coefficients with 95\% confidence intervals: the husband's occupation-level WFH treatment (blue circles, \emph{spillover} effect), the wife's college-major-level WFH treatment (red triangles, \emph{direct} effect), and their interaction (green squares, \emph{matching} effect); see Subsection~\ref{subsection:treatment_defn} for the treatment definitions. The omitted year is 2019. All regressions include state-by-time, husband's-occupation, and wife's-college-major fixed effects, and control for the wife's age, number of children, and education levels of both spouses. Standard errors are clustered at the husband's occupation level.
    \end{minipage}
\end{figure}

These results are robust to using other strong predictors of women's likelihood to WFH, such as women's concurrent occupation when observable, to separate the three channels.\footnote{In  Appendix \autoref{tab:acs_sp_own_occ_estimates}, we estimate a modified version of the spillover decomposition equation (\ref{eq:3}) where we replace exposure at women's college major level, $Exposed_{e(i)}$, with exposure at women's own occupation level, $Exposed_{o(i)}$. Results are largely similar to the college degree-based decomposition of Appendix \autoref{tab:acs_edfield_estimates}. Since the college degree decomposition leads to no loss of sample size, it is our preferred specification.} We also develop a dynamic DiD version of the decomposition specification in $\eqref{eq:3}$ to visually inspect pre-trends in \autoref{fig:acs_dynamic_DiD_edfield}.\footnote{Specifically, we estimate an extended version of the baseline event study in \autoref{fig:ddid_main} and further add yearly coefficients for the treatment at the level of women's college degree major - $\text{High-}\Delta\text{WFH}_{\,e(i)} $ - and its interaction with the treatment at the level of the male spouses' occupation.} The plots show a significant and lasting intra-household \textit{spillover} effect -- the blue lines -- of the spousal WFH shock on women's earnings, hours and weeks worked for women who  are not more likely to work from home based on their college majors. Moreover, the pre-trends for the spillover effect estimates are non-significant and strikingly resemble the shape and trends of the main results in \autoref{fig:ddid_main}.

These results strongly suggest that the bulk of the gains in women's labor supply found in Section  \ref{section:baseline_results} stems from women who didn't increase much their uptake of WFH, and that changes in household arrangements \textit{arising} from the male spouse's increased availability at home are a key enabler of women's labor supply. Indeed, the last two columns in \autoref{tab:acs_edfield_estimates} show that the large changes in women's WFH uptake and work commuting shown at the beginning of Section \ref{subsec:disentangle} stem largely from matching effects ($\beta_3$). In Section \ref{sec:mechanisms}, we explore more directly the extent of changes in household production -- specifically looking at the role of childcare responsibilities.
\section{Robustness and Validation} \label{section:robustness}

\paragraph{Specification and Sample Robustness}
Our results are robust to differential trends across education categories and broad occupation groups.\footnote{Education categories are below high school, high school, and above high school; broad occupation groups are high-skill/professional, services and sales, and blue-collar/manual. Results are also robust to trends in the wife's education level (not reported).} Appendix Tables \ref{tab:acs_appendix_trends_spedlevel} and \ref{tab:acs_appendix_trends_broadspocc} report, for each control variable, the baseline estimate (Panel A), estimates with variable-by-year fixed effects (Panel B), linear time trends (Panel C), and a linear pre-trend adjustment following \citet{jakobsen2020wealth}\footnote{Following \citet{jakobsen2020wealth}, we estimate the pre-trend slope $\hat\theta$ from pre-2020 data by regressing $y_{ist}$ on $\text{High-}\Delta\text{WFH}_{o(h(i))} \cdot t$ (plus baseline controls and fixed effects) for $t<2020$, then subtract the extrapolated trend from the outcome in all periods and re-run the baseline event study. We repeat this using the husband's education, the wife's education, and the husband's broad occupation group in place of the treatment dummy. The identifying assumption is that post-event confounder behavior can be inferred from its pre-event linear trend \citep{freyaldenhoven2019pre, roth2022pretest}.} (Panel D). Point estimates are stable across all specifications, with only minor variation in magnitude under the broad occupation group controls.

Estimates are also insensitive to the fixed effect structure: Appendix \autoref{tab:acs_estimates_change_FE} varies occupation group definitions, geographic aggregation, and the level of interaction with state fixed effects, and coefficients remain close to our baseline (Panel E). Results are further unaffected by controlling for the husband's employment status (see Appendix \autoref{tab:acs_baseline_annual_spemp_annual_control}), by alternative demographic controls (see Appendix Tables \ref{tab:acs_change_controls1} and \ref{tab:acs_change_controls2}), by excluding self-employed individuals or trimming top-income households (see Appendix Tables \ref{tab:acs_baseline_estimates_no_self_emp}, \ref{tab:acs_trim_income_estimates}), and by dropping major occupation groups one at a time (see Appendix \autoref{tab:acsleaveoneoutoccupations2d}).

\paragraph{Group Composition and Selection}
A concern with our cross-sectional design is that estimates may reflect changes in assortative matching rather than intra-household adjustment. Appendix  \autoref{fig:assortative_matching} shows that while husbands' and wives' WFH exposure are positively correlated, this correlation is stable across the pre- and post-COVID periods, easing this concern. We also find no effect of the husband's WFH shock on the wife's probability of marrying or divorcing in the past 12 months (see Appendix \autoref{tab:acs_marriage_divorce}); a negative effect on childbearing is statistically significant but economically small (about 3\% of the control group mean). Consistent with this, Appendix  \autoref{fig:desc_fertility} and Appendix \autoref{tab:acs_fertility} show that fertility differences across husbands' WFH-exposure quartiles are small and, if anything, work against upward selection into our treated group.

To further address compositional change, we reweight the post-COVID sample to match the pre-COVID distribution of spouse occupation, wife's age, education, and state, using \citet{dinardo1995labor} and inverse-propensity-score weights.\footnote{We estimate a logit model where the outcome equals one if the observation is from a pre-COVID year and zero otherwise, using spouse occupation, wife's age and education, and state fixed effects as predictors; let $p$ denote the resulting predicted probability that an observation belongs to the pre-COVID period. We construct \citet{dinardo1995labor} weights as $p/(1-p)$ and inverse-propensity weights as $1/p$, both scaled by ACS sampling weights. Common support is verified in Appendix \autoref{fig:overlap}.} Reweighted estimates in Appendix \autoref{tab:acs_reweighting} are consistent with our baseline results.

\paragraph{Treatment Definition Robustness}
Appendix \autoref{tab:changeshockdef} shows that our results are unaffected by alternative treatment definitions: smoothing WFH job-postings shares with a three-month moving average (Panel B), alternative classification schemes and change measures (Panels C, E--G), and the \citet{dingel_how_2020} teleworkability index (Panels D, H), which captures inherent WFH feasibility rather than employer-side WFH supply.\footnote{We define treatment under this alternative measure as an indicator equal to one if the husband's occupation is at or above the median of the \citet{dingel_how_2020} teleworkability distribution, which averages O*NET-based task scores on the feasibility of remote work within each occupation. Notably, it classifies teachers as WFH-capable even though post-pandemic in-person instruction has largely persisted, driven by parental demand \citep{barrero_evolution_2023}.} Appendix Figures \ref{tab:changeshockdef_cont} and \ref{tab:changeshockdef_acs} further show quantitatively similar results using a continuous WFH-change measure and an alternative treatment based on husbands' self-reported WFH rates in the ACS.

\paragraph{Placebo Tests}
We artificially assign the COVID shock to 2016 and re-estimate our baseline specification on the pre-COVID ACS subsample (pre-period 2012--2015, post-period 2016--2019). Because the actual shock postdates this window and was unanticipated, we expect no treatment effect. Appendix \autoref{tab:acs_timing_placebo_baseline_annual} confirms this: we find no effect on earnings, hours, or part-time status, and only a small effect on weeks worked relative to our baseline estimates.

\paragraph{Evidence from CPS}
We replicate our analysis in the 2013--2026 CPS sample (Section~\ref{sec:data}). Despite a smaller sample, Appendix \autoref{tab:cps_baseline_estimates_combined} closely mirrors our ACS results: wives in higher-exposure households report more hours and weeks worked, are about 2 percentage points less likely to work part-time, and earn over 9 percent higher wages.

%%%%%%%%%%%%%%%%%%%%%%%%%%%%%%%%%%%%%%%%%%%%%%%%%%%%%%%%%%%%%%%%%%%%%%%%%%%%%%%%%%%%%%%%%%%%%%
%%% MECHANISMS %%%%
%%%%%%%%%%%%%%%%%%%%%%%%%%%%%%%%%%%%%%%%%%%%%%%%%%%%%%%%%%%%%%%%%%%%%%%%%%%%%%%%%%%%%%%%%%%%%%

\section{Household Specialization} \label{sec:mechanisms}
%BE SURE WE HAVE INFO ON WHEN POST TREATMENT

%% phrase it on how women can work more/enter labor supply when husband stays at home. what happened? we explore possible intrahousehold mechanisms but looking at how time use in childcare changes across partners. 

The remote work shock may have altered how households coordinate home production decisions, besides the market ones. Understanding these changes in household specialization and time use is central not only to explaining women’s increased labor market engagement, but also to interpreting the overall effects of the WFH shock on households. 

%should we have a benchmark??

%Understanding the household mechanisms that enabled women to increase their labor supply when their husbands became more likely to work from home is central to interpreting the overall effects of the WFH shock. 
In this section, we test whether the large labor effects we observe for women can be explained by changes in the household production function, particularly in the burden and organization of childcare duties.% As discussed in Section~\ref{sec:results}, these changes may arise from two forces: (i) spillovers from husbands’ increased presence at home and (ii) women’s own access to work from home through assortative matching.

\subsection{Childcare Production}

\begin{figure}[H]
    \caption{Dynamic Effects of Husbands' WFH Shock on Couples' Childcare Time-Use}\label{fig:dynamicmechI}
    \begin{tabular}{cc}
        %Primary Childcare
        \multicolumn{2}{c}{\emph{ (a) Primary Childcare (ATUS)}} \\
        Wives & Husbands \\
        \includegraphics[width=0.5\linewidth]{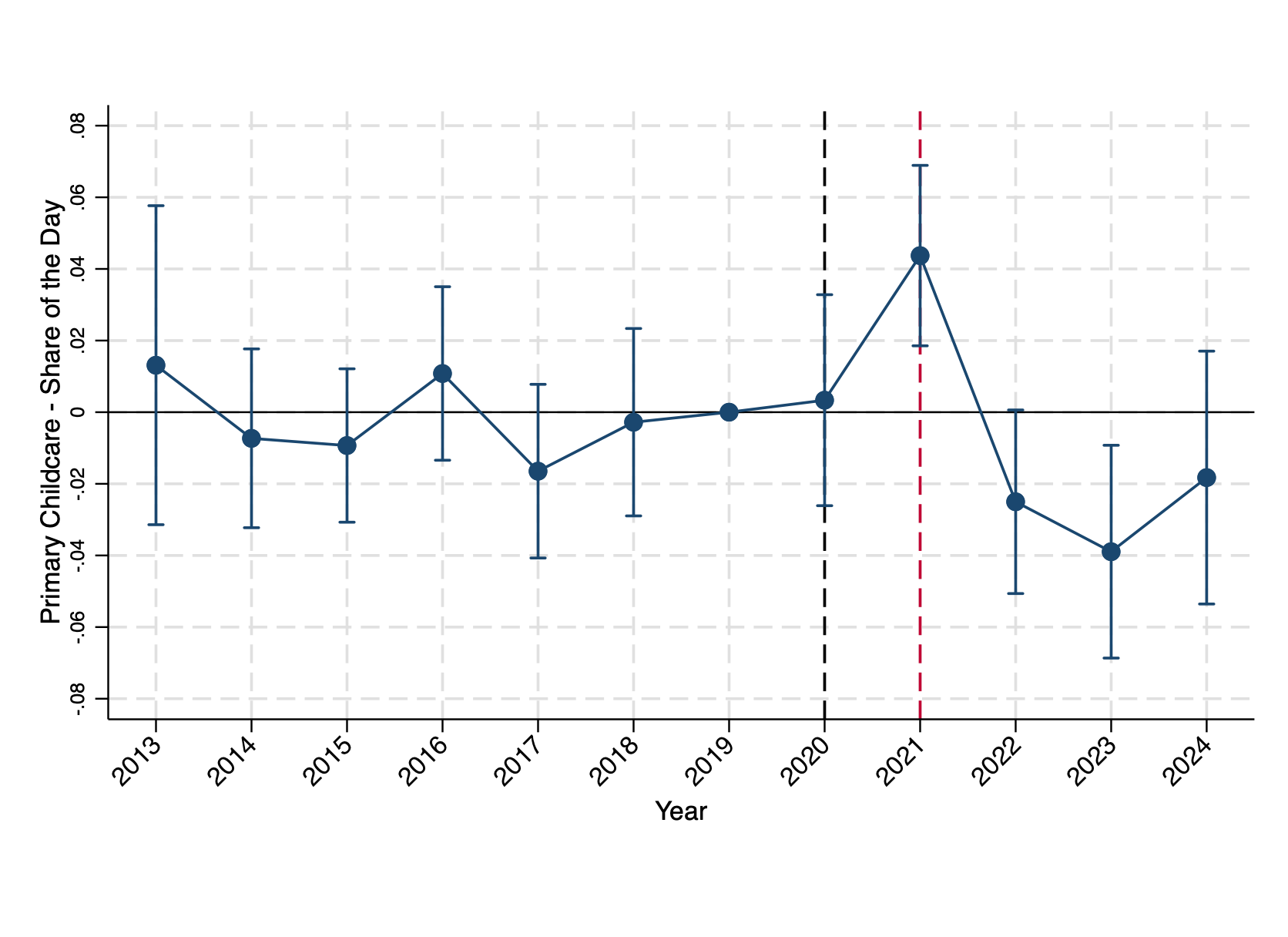} &
        \includegraphics[width=0.5\linewidth]{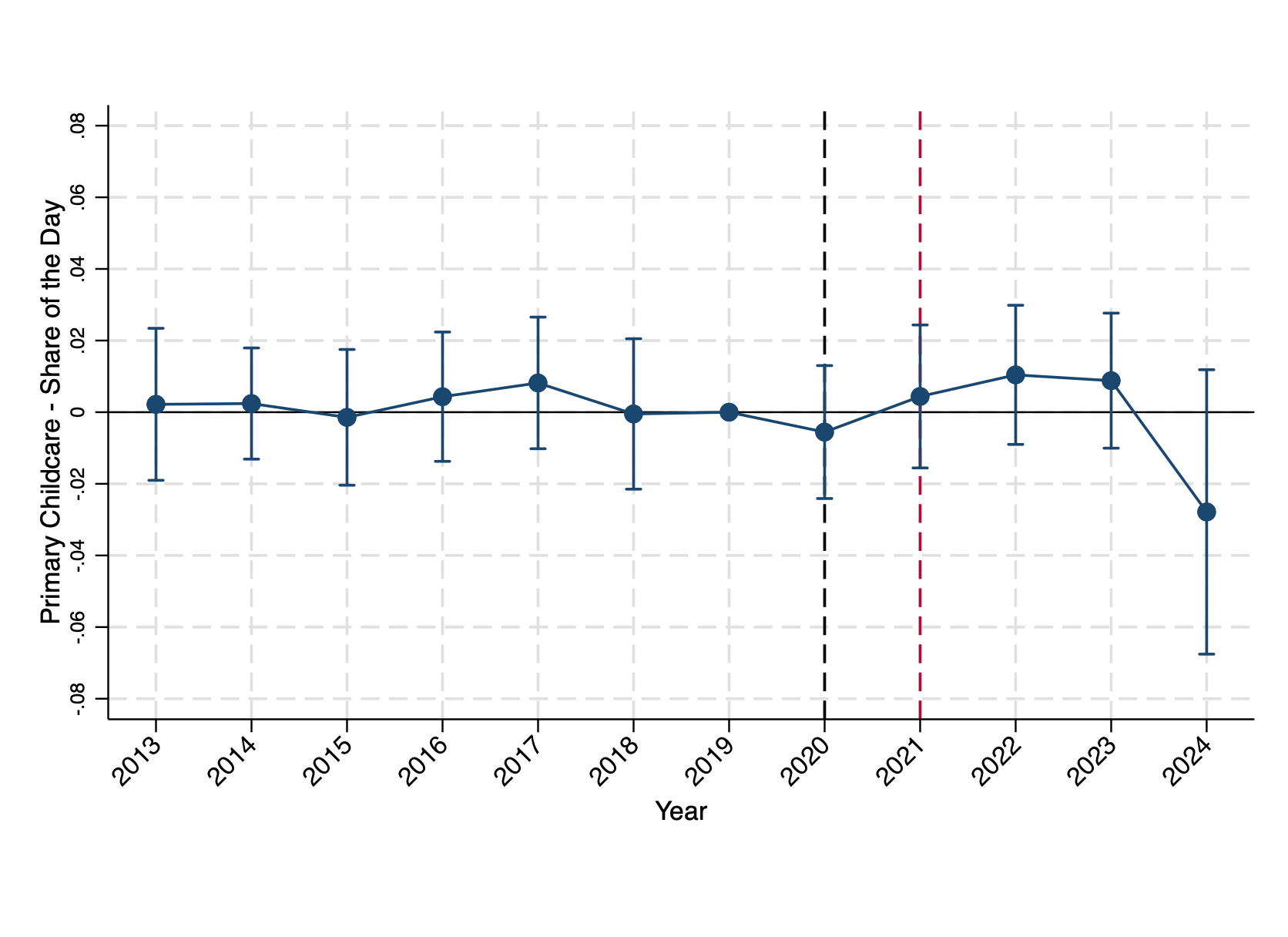}
        \\
        %Child Care while Working
        \multicolumn{2}{c}{\emph{ (b) Childcare while Working (ATUS)}} \\
        Female Spouse & Male Spouse \\
        \includegraphics[width=0.5\linewidth]{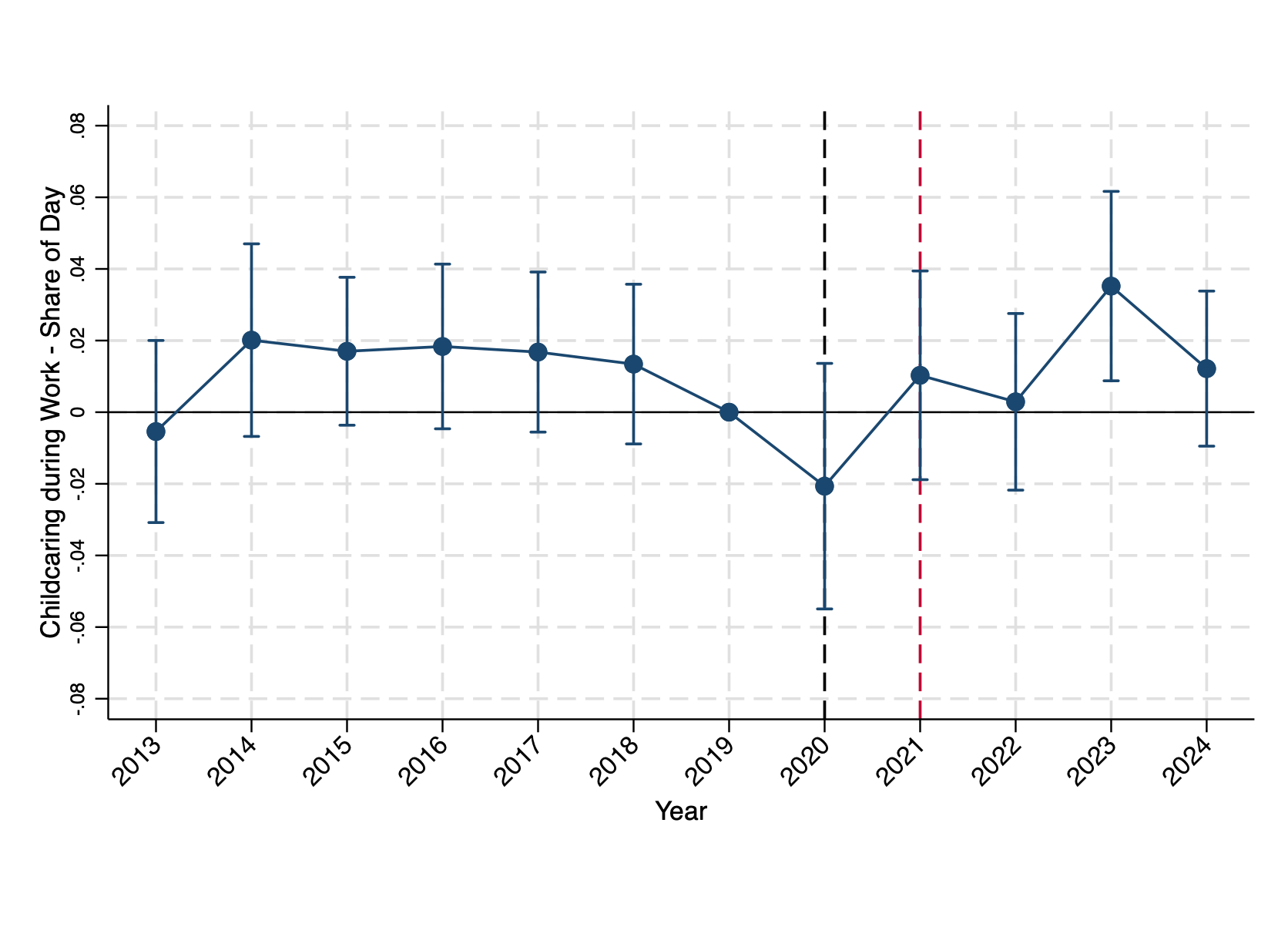} &
        \includegraphics[width=0.5\linewidth]{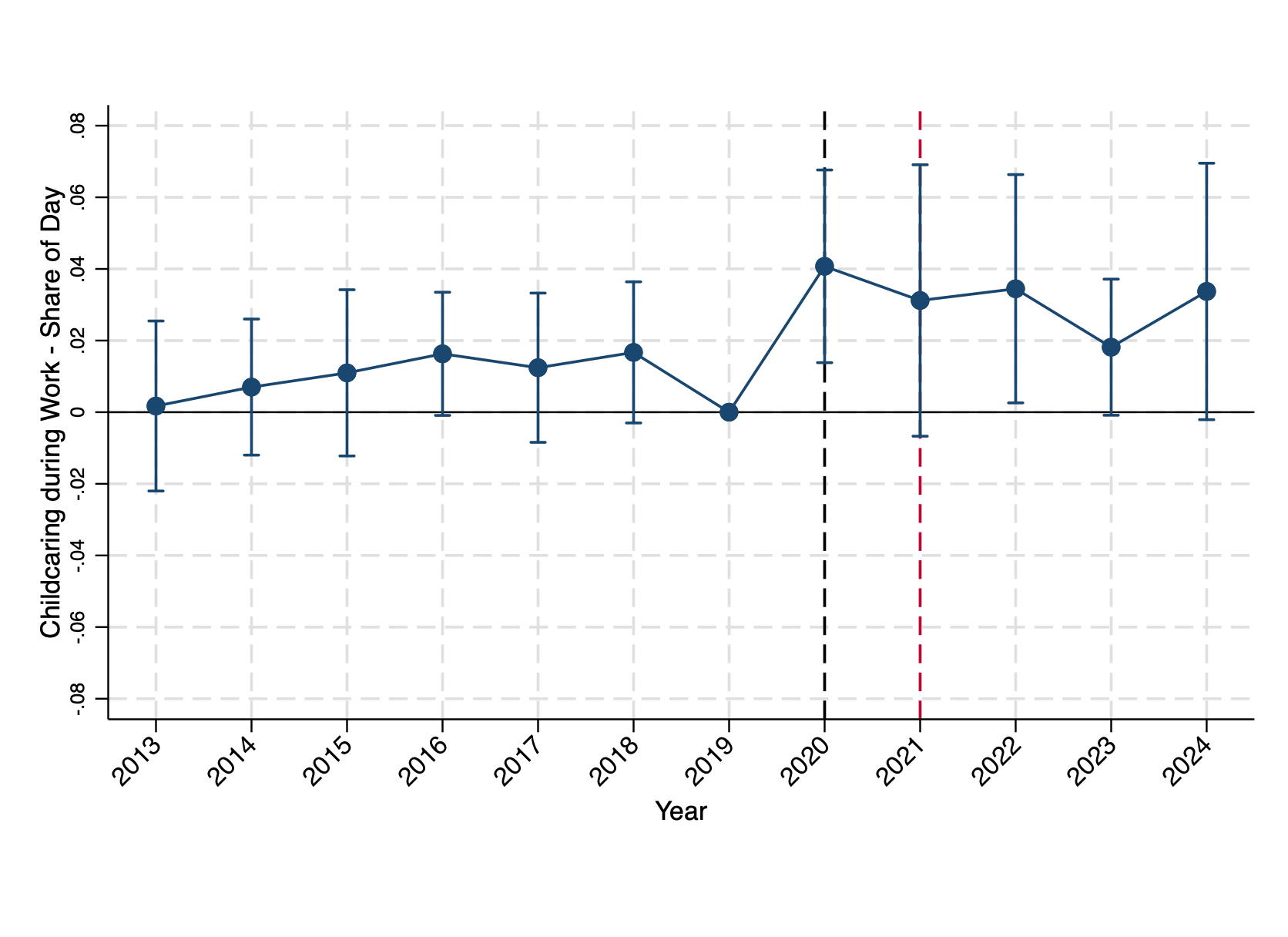}
        \\
    \end{tabular}
     \vspace{0.5em}
     \centering
    \vspace{0.5em}
    \begin{minipage}{0.99\linewidth} \small
    \textit{Notes}: This figure plots yearly coefficients from the dynamic difference-in-differences specification in Equation~\eqref{eq:2}, estimated on the 2013--2024 ATUS sample of dual-parent households with children of child-rearing age. Each dependent variable is expressed as the share of the 24-hour day spent in the activity. Panel (a) plots the effect on primary childcare time (childcare reported as the respondent's main activity) for the wife (left) and husband (right); panel (b) plots the effect on childcare performed while working (a secondary activity) for the wife (left) and husband (right). Because each ATUS respondent reports time use for only one household member, wife- and husband-side estimates come from separate men-respondent and women-respondent samples (see Subsection~\ref{subsection:treatment_defn} and Appendix~\ref{app:data_construction}). The omitted year is 2019. All regressions include state-by-time and husband's-occupation fixed effects, and control for the wife's age, number of children, and education levels of both spouses. Confidence intervals are shown at the 95\% level and standard errors are clustered at the husband's occupation level.
    \end{minipage}
\end{figure}

\paragraph{Childcare and Household Production Time Use}
%now comment on childcare and household activities!!
The change in the joint labor market choices discussed in Section \ref{sec:results} came alongside new patterns in the collective household production, as our dynamic DiD in \autoref{fig:dynamicmechI} documents. In particular, treated households exhibit meaningful shifts in the allocation of the most prominent household activity for the couple: childcare.
%Women reduce time spent on primary childcare, while men partially offset this by increasing childcare performed while working.    %the total time spent caring for children does not change significantly, there is a clear shift in how childcare is provided. 
Women in treated households reduce time spent on childcare as a primary activity by 2.4 p.p., an approximately 22\% decrease relative to the pre-treatment average, while husbands partially offset this by increasing time spent on childcare while working by 1 p.p. -- a 100\% increase relative to the pre-treatment average of 14 minutes.

Notably, these are the only two margins of childcare production within the household that do not shift. As \autoref{tab:mechatus1} shows, using our three-period specification \eqref{eq:1}, women do not reduce time spent on childcare as a secondary activity, implying that secondary childcare's share of their total childcare time rises by 6 p.p. Men, in turn, do not increase either their primary childcare time or their secondary childcare time when not working.\footnote{In \autoref{tab:mechatus1}, the average effect for husbands' childcare while working is barely insignificant. Still, \autoref{fig:dynamicmechI} documents a persistent, stable, and significant increase over the treatment years.} These findings imply a more equal distribution in the time spent caring for children between spouses, with the gender gap narrowing by 11\%.\footnote{To compute the gap in total childcare time, we follow the same strategy as described in \autoref{app:gender_gap_accounting}, but with men's fitted averages in the numerator and women's in the denominator. We invert the ratio here because, unlike in the labor market, women on average spend more time caring for children than men; using men/women as the ratio preserves the convention that a narrowing gap corresponds to greater equality between spouses.}

%While they are not reporting to be spending significant more time child-caring while working, although they are spending more time WFH. Additionally, there is no reduction in secondary childcare, as \autoref{tab:mechatus1} shows. In other words, women mainly cut back on primary childcare. As a result, secondary childcare becomes more prominent: its share of total childcare time rises by 8.4 p.p. relative to a pre-treatment mean of 64.5 percent.

%Husbands, on average, do not exhibit significant changes in total time spent on overall childcare. However, as shown in \autoref{fig:dynamicmechI}, their time spent on childcaring while working increases significantly by 1.5 percentage points. This is a 200\% increase relative to baseline. This aligns with the fact that men are more likely to work from home due to their shock. These patterns point that 

Besides childcare, men do not experience any change in time spent on housework, while women decrease their housework time by 2 p.p. This decline is concentrated in specific activities: time spent cooking falls by 0.9 p.p. and time spent shopping by 0.5 p.p.
%Still, these are less relevant activities in terms of their duration, relative to child caring.
%Childcare, being the most time consuming house chore for households with children as Appendix \autoref{tab:desc_atus_out} reflects, remains the one where we observe the most pronounced reallocation within the couple.
%, since it can be more substitutable and proportionally more important in time to be able to detect significant changes, compared to other household tasks. %This might explain the lack of precise effects on the other less prominent household activities. 

In Appendix \autoref{tab:mechplacebotiming}, we conduct a standard placebo test by assigning the onset of the post-treatment period to March 2016. Across all outcomes we find no effects, which alleviates concerns on differential pre-trends. Moreover, across a battery of previous exhibits using ATUS data (see e.g. \autoref{fig:dynamicmechI}, and Appendix Figures \ref{fig:dynamic_laborprod_men_ATUS} and \ref{fig:atus_dynamic_DiD_working_share_wives}), we do not find any significant pre-trends.

\paragraph{Other Sources of Childcare} Using the SIPP childcare topical modules, we are able to provide complementary evidence on how households change their sources of childcare  and hiring of daycare following shocks on spousal WFH options. We estimate our baseline cross-section specification on the 2014-2024 SIPP sample using a battery of childcare related outcomes, and present some results Appendix in \autoref{tab:sipp_mechanism_chcare_cross}. Our results show that in households shocked by the occupation-level WFH treatment, where the child's mother is working, the spouse is almost 7 p.p. more likely to be involved in childcare, roughly a 14\% increase over the baseline. On the other hand, the woman herself is not any more likely to be involved in childcare. %This pair of results are most consistent with the \textit{substitution hypothesis}.

\begin{figure}[H]
    \caption{Effect of Husband's WFH Shock on Daycare Hiring (SIPP)}\label{fig:daycare}
    \begin{tabular}{c}
        (a) Hired Daycare  \\
        \includegraphics[width=0.8\linewidth]{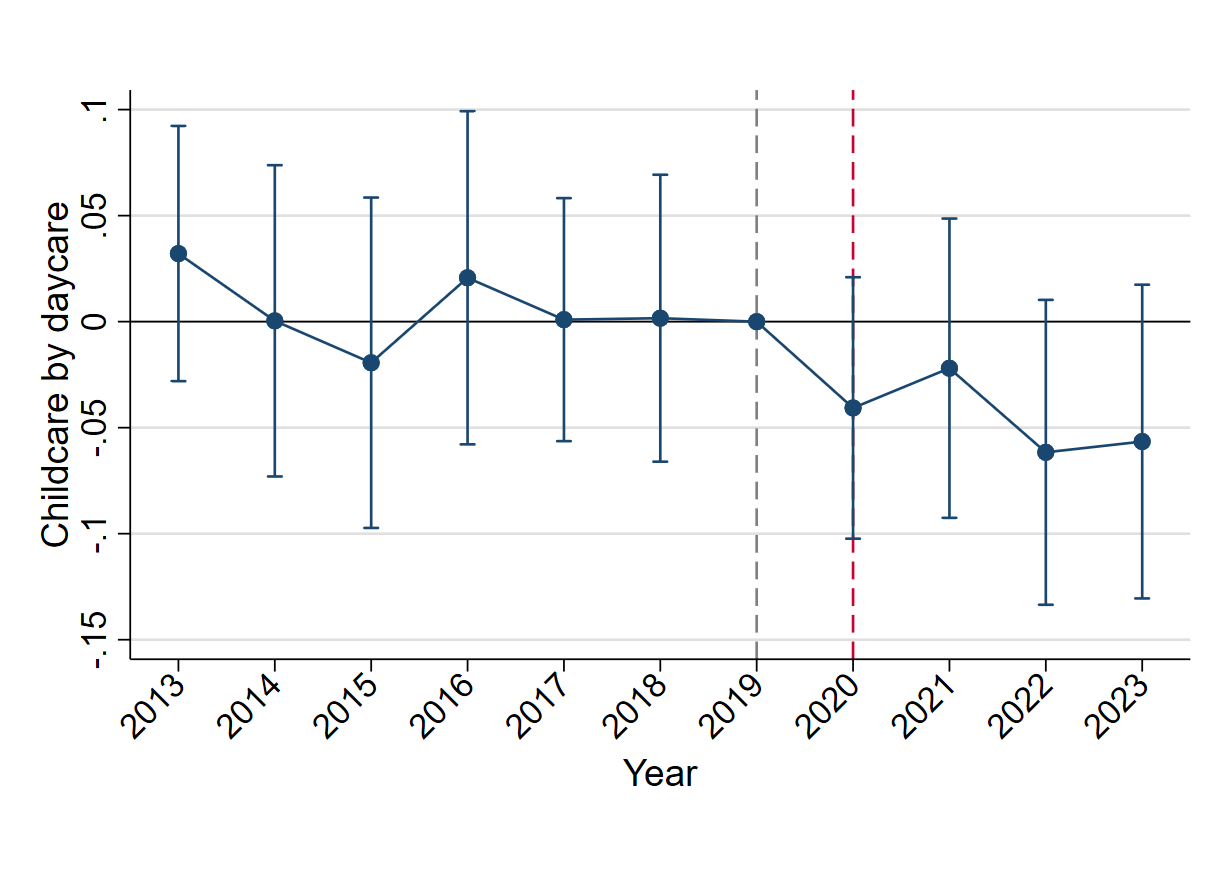} \\
        (b) Log Daycare Expenditure \\ \includegraphics[width=0.8\linewidth]{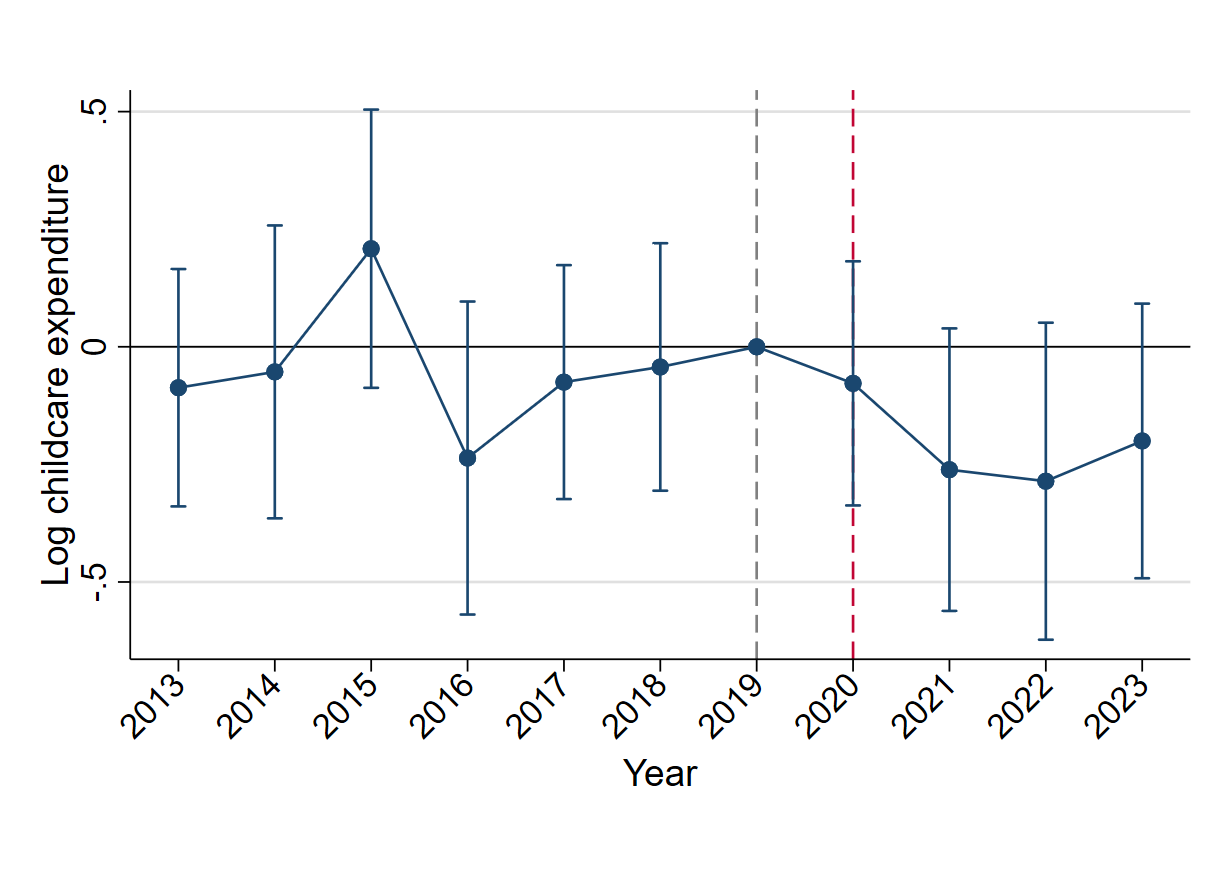} 
    \end{tabular}
    \centering
    \begin{minipage}{0.99\linewidth} \small
   \textit{Notes}: This figure plots yearly coefficients from the dynamic difference-in-differences specification in Equation~\eqref{eq:2}, estimated on the 2014--2024 SIPP sample of dual-parent households with children of child-rearing age. Panel (a) reports the effect on the probability that the household hires paid daycare; panel (b) reports the effect on log self-reported daycare expenditure, both drawn from the SIPP childcare topical modules. The omitted year is 2019. All regressions include state-by-time and husband's-occupation fixed effects, and control for the wife's age, number of children, and education levels of both spouses. Confidence intervals are shown at the 95\% level and standard errors are clustered at the husband's occupation level.
    \end{minipage}
\end{figure}

We also find that treated households are not more likely to rely on grandparents or other relatives for day-to-day childcare. Instead, treated households are 6.4 p.p. less likely to use daycare -- a 25\% drop relative to the baseline mean -- and total childcare expenditure falls by about 20\%. \autoref{fig:daycare} plots estimates of our dynamic DiD specification \eqref{eq:2} for these salient outcomes. Although less precisely estimated, these plots nevertheless corroborate the declines in both the probability of hiring childcare and log childcare spending.

%Overall, these patterns are consistent with fathers becoming more available for childcare in treated households. Accordingly, household substitute away from daycare services. 

    \paragraph{Discussion}  Overall, the results on these section underscore the role of intra-household reallocation in explaining the observed improvements in women’s labor market outcomes. Treatment implied an 11\% reduction in the gender gap in the allocation of the most time-demanding household activity, childcare,  meaning a more equal division at home. Husbands exposed to the WFH shock modestly substitute their wives, alleviating  women's primary childcare time and allowing them to increase their working time, as observed in \autoref{tab:acs_baseline_annual}. Moreover, households substitute resources away from daycare services, consistent with fathers becoming more available for childcare in treated households.

%In Appendix Section \ref{sec:model}, we show that these findings can be rationalized through the lens of a simple model of dual-parent household labor and childcare supply that incorporates the option to work remotely. The model formalizes the idea that work from home expands the household's feasible set of work-care arrangements by allowing some childcare to be produced jointly with market work, thereby relaxing childcare constraints.

%We cast light on the validity of the results in this section by performing a series of robustness tests in \autoref{sec:robusthouse}.

%Here comment on whether commuting changes conditional on WFH and that even within women working and not remotely, they are commuting less... switch job recently?
%Present results of migration (general)
% Present if conditional on migration, if there is a differential migration pattern (more likely to move towards cheaper places)

\subsection{Additional Potential Channels}
\paragraph{Household's Leisure and Personal Time} 

The increase in working time by women could have also been possible due to a crowding out of their personal and leisure time. However, as \autoref{tab:mechatus1} points out, we can rule out that hypothesis, as women do not experience any significant reduction in their time spent in such activities. %Women in treated households experience no significant change in time spent on personal care activities, such as sleeping and grooming; and paradoxically to the hypothesis, exhibit a 3.2 pp increase in time devoted to leisure activities. \footnote{Husbands do not experience any change in their personal or leisure time.}%\footnote{A caveat of this finding is that is based on time-use, which by no means implies that the quality of their leisure of personal time has not changed.}

%Finally, it is plausible that children may be worse off given that our main results suggest that women in our treatment sample are more likely to be working, spend more time at work and, as found in this section, there is an overall reduction in the household's time spent child-caring as a primary activity. Although we cannot fully measure well-being with our current datasets, the SIPP has a handful related variables such as days that parents have dinner or go out with their child and the children's involvement in extracurricular activities. In \autoref{tab:sipp_mechanism_chwell_cross}, we estimate our baseline specification on these outcomes, and we find no significant changes for these variables. This evidence suggests that -- to the extent we can measure well-being -- children are not worse off by changes in household production and labor supply. 

\paragraph{Migration}

A potential alternative mechanism is that increased workplace flexibility among husbands allows households to relocate to areas with better labor market opportunities for women, rather than affecting outcomes through a reallocation of childcare within the household. We do not find evidence consistent with this channel. Appendix \autoref{fig:acs_dynamic_DiD_mig_dummy} shows that treated households are not more likely to migrate. Moreover, among households that do relocate, treatment is not associated with moves to counties offering more favorable labor market conditions for women. Appendix \autoref{tab:acs_migrating_origdest_labor_chars} reports estimates based on changes in labor market characteristics between origin and destination counties and reveals no systematic differences in the destinations chosen by treated households. Together, these findings indicate that differential residential sorting is unlikely to explain our main results.

\section{Conclusion}\label{sec:conclusions}

In this paper, we provide novel evidence on changes in households' market and home production arising from large and persistent changes in husbands’ work-from-home (WFH) arrangements in dual-parent households. Exploiting plausibly exogenous cross-occupational variation in the post-pandemic expansion of remote work in the U.S., we show that increases in husbands’ WFH exposure led to sizable improvements in their wives’ labor market outcomes, including higher earnings and hours worked, as well as a lower likelihood of part-time work. Results are robust to a battery of sensible robustness tests. %but data and power limitations impose a caveat on how far they can be taken 

We empirically isolate the spillover effect from men's increased availability at home and show that a large fraction of these labor market gains can be attributed to it. %as women that are not working from home themselves are most likely to increase their labor supply. %These effects are concentrated among households with young children and among highly educated couples. They are not driven by changes in group composition or assortative matching and are robust across alternative specifications, treatment definitions, and nationally representative datasets.
This is consistent with our findings from time-use data, where we show that these labor supply responses are accompanied by a reallocation of household production -- specifically, a more equitable division in childcare responsibilities.

%Taken together, the results point to the value of analyzing remote work not only as an individual labor market arrangement, but also as a household-level shock, offering new perspective on how WFH access may affect gender gaps in the labor market via intra-household reallocations.

Taken together, our findings demonstrate how large shifts in workplace flexibility can reshape intra-household specialization and generate meaningful labor market spillovers. This points to the value of studying remote work not only as an individual labor market arrangement but as a household-level shock with the potential to narrow gender gaps in the labor market. An important avenue for future research is to examine how the expansion of remote work affects assortative matching in the marriage market and human capital investment over the life cycle.

\vspace{1cm}
\bigskip

\begin{comment}
\begingroup
\onehalfspacing
\noindent
\begin{minipage}[t]{0.32\textwidth}
\centering
\textbf{Isabella Di Filippo}\\
Columbia University\\
Department of Economics\\
id2403@columbia.edu
\end{minipage}
\hfill
\begin{minipage}[t]{0.32\textwidth}
\centering
\textbf{Bruno Escobar}\\
Columbia University\\
Department of Economics\\
be2329@columbia.edu
\end{minipage}
\hfill
\begin{minipage}[t]{0.32\textwidth}
\centering
\textbf{Juan Facal}\\
Columbia University\\
Department of Economics\\
jmf2285@columbia.edu
\end{minipage}
\endgroup
\end{comment}

\clearpage

\bibliographystyle{plainnat.bst}
\singlespacing
\bibliography{wfh_spill_refs.bib}
\clearpage

\section*{Tables}
\begin{table}[H]\centering
\caption{Effect of Occupation-Level WFH Shocks on Husbands' Remote Work Uptake}
\label{tab:acs_first_stage_combined}
\scalebox{0.8}{
\begin{threeparttable}
\begin{tabular}{llcccccc}
\hline
\addlinespace
& & Log                & Usual Weekly  & Weeks  &      Part-time       & WFH   & Commuting \\ 
& & Earnings   & Work Hours    & Worked &      Status          & Uptake& Time \\ 
\cline{3-8} \addlinespace
\addlinespace
$ Exposed_{o(h(i))} \times Post_{t}$&            &      -0.009   &       0.005   &      -0.731***&      -0.003   &       0.124***&      -0.003***\\
                    &            &     (0.009)   &     (0.008)   &     (0.135)   &     (0.002)   &     (0.015)   &     (0.000)   \\
\addlinespace Mean pre-treatment&            &      10.544   &       2.895   &      44.678   &       0.049   &       0.045   &       0.020   \\
Obs                 &            &   1,268,684   &   1,265,821   &   1,265,821   &   1,265,821   &   1,290,256   &   1,290,157   \\
$R^{2}$             &            &        0.35   &        0.35   &        0.10   &        0.06   &        0.15   &        0.06   \\
\addlinespace \hline
\end{tabular}

\end{threeparttable}
}
\par\vspace{0.3cm}
\begin{minipage}{\textwidth} \small
 \textit{Notes}: This table reports estimates from Equation~\eqref{eq:1} using the 2013--2024 ACS sample of dual-parent households with children of child-rearing age (married couples with at least one child aged 0--8). The treatment indicator, $\text{High-}\Delta\text{WFH}_{o(h(i))}$, equals one if the husband's Census 2010 occupation experienced an above-median increase in the share of remote/hybrid job postings between 2019 and 2023--2024 (job postings data from \citet{hansen2023remote}; see Subsection~\ref{subsection:treatment_defn}). The dependent variables in the first four columns are the husband's own annual log earnings, usual weekly hours worked, weeks worked, and part-time status, all measured over the preceding 12 months and conditional on employment. The last two columns use the husband's self-reported work-from-home (WFH) status and commuting time as a share of the day, both measured for the week prior to the interview; the WFH measure is constructed from the ACS commuting-mode question (``How did you usually get to work LAST WEEK? Mark (X) ONE box for the method of transportation used for most of the distance''), which captures the incidence of fully remote work (see \citet{buckman2025measuring}). All regressions control for the wife's age, education, and number of children, as well as the husband's education, and include state-by-time and husband's-occupation fixed effects. Standard errors, clustered at the husband's occupation level, are reported in parentheses. \textsuperscript{*} $p<0.10$; \textsuperscript{**} $p<0.05$; \textsuperscript{***} $p<0.01$.
\end{minipage}
\end{table}

\clearpage

\begin{table}[H]\centering
\caption{Effect of Husbands' WFH Shock on Wives' Labor Market Outcomes}
\label{tab:acs_baseline_annual}
\scalebox{0.8}{
\begin{threeparttable}
\begin{tabular}{llcccc}
\hline
\addlinespace
& &  Log Earnings & Usual Weekly Hours & Weeks Worked & Part-time \\ 
& &  & Worked & & Status \\
\cline{3-6} \addlinespace
\addlinespace
$ Exposed_{o(h(i))} \times Post_{t}$&            &       0.048***&       0.466***&       0.500***&      -0.024***\\
                    &            &     (0.010)   &     (0.096)   &     (0.089)   &     (0.004)   \\
\addlinespace Mean pre-treatment&            &       9.902   &      36.276   &      45.635   &       0.279   \\
Obs                 &            &     949,443   &     948,946   &     949,443   &     948,946   \\
$R^{2}$             &            &        0.15   &        0.03   &        0.03   &        0.03   \\
\addlinespace \hline
\end{tabular}

\end{threeparttable}
}
\par\vspace{0.3cm}
\begin{minipage}{\textwidth}
\small
\textit{Notes}: This table reports estimates from Equation~\eqref{eq:1} using the 2013--2024 ACS sample of dual-parent households with children of child-rearing age (married couples with at least one child aged 0--8). The dependent variables are the wife's annual employment status, log annual earnings, log hourly wage, usual weekly hours worked, number of weeks worked, and part-time status, all measured over the preceding 12 months; intensive-margin outcomes (earnings, wage, hours, weeks, part-time) are defined only for employed wives and coded as missing otherwise. The treatment indicator, $\text{High-}\Delta\text{WFH}_{o(h(i))}$, equals one if the husband's Census 2010 occupation experienced an above-median increase in the share of remote/hybrid job postings between the pre-pandemic (2019) and post-pandemic (2023--2024) periods (see Subsection~\ref{subsection:treatment_defn}). All regressions control for the wife's age, education, and number of children, as well as the husband's education, and include state-by-time and husband's-occupation fixed effects. Standard errors, clustered at the husband's occupation level, are reported in parentheses. \textsuperscript{*} $p<0.10$; \textsuperscript{**} $p<0.05$; \textsuperscript{***} $p<0.01$.
\end{minipage}
\end{table}
\clearpage

\begin{table}[H]\centering
\caption{Decomposition Analysis: Effects of Husbands' Occupation and Wives' College Major WFH Shocks on Wives' Labor Supply}
\label{tab:acs_edfield_estimates}
\scalebox{0.65}{
\begin{threeparttable}
\begin{tabular}{llcccccc}
\hline
\addlinespace
& &  Log Earnings & Usual Weekly & Weeks Worked & Part-time & WFH Uptake & Commuting \\ 
& &  & Hours Worked & & Status & & Time \\
\cline{3-8} \addlinespace
\addlinespace
$ Exposed_{o(h(i))} \times Post_{t} $&            &       0.032***&       0.440***&       0.494***&      -0.023***&       0.019***&      -0.001***\\
                    &            &     (0.011)   &     (0.109)   &     (0.111)   &     (0.005)   &     (0.005)   &     (0.000)   \\
$ Exposed_{e(i)} \times Post_{t} $&            &       0.026   &       0.176   &       0.406** &      -0.018***&       0.094***&      -0.003***\\
                    &            &     (0.017)   &     (0.136)   &     (0.170)   &     (0.005)   &     (0.005)   &     (0.000)   \\
$ Exposed_{o(h(i))} \times Exposed_{e(i)} \times Post_{t} $&            &       0.037*  &       0.071   &      -0.143   &       0.001   &       0.030***&      -0.001** \\
                    &            &     (0.020)   &     (0.167)   &     (0.199)   &     (0.007)   &     (0.007)   &     (0.000)   \\
\addlinespace Mean pre-treatment&            &       9.902   &      36.274   &      45.634   &       0.279   &       0.075   &       0.017   \\
Obs                 &            &     947,311   &     946,809   &     947,311   &     946,809   &     889,378   &     889,310   \\
$R^2$               &            &        0.21   &        0.06   &        0.05   &        0.07   &        0.13   &        0.06   \\
\addlinespace \hline
\end{tabular}
\end{threeparttable}\textit{}
}
\par\vspace{0.3cm}
\begin{minipage}{\textwidth} \small
\textit{Notes}: This table reports estimates from Equation~\eqref{eq:3}, which decomposes the reduced-form effect on wives' labor supply into a \emph{spillover} effect ($\beta_1$, husband's occupation-level WFH shock only), a \emph{direct} effect ($\beta_2$, wife's own WFH exposure via her college major only), and a \emph{matching} effect ($\beta_3$, the interaction of the two). The husband's treatment, $\text{High-}\Delta\text{WFH}_{o(h(i))}$, equals one if his Census 2010 occupation had an above-median increase in the share of remote/hybrid job postings between 2019 and 2023--2024 (\citealt{hansen2023remote}; see Subsection~\ref{subsection:treatment_defn}). The wife's treatment, $\text{High-}\Delta\text{WFH}_{e(i)}$, equals one if her college major is above the median in the observed increase in ACS-reported WFH uptake among graduates of that major between the pre- and post-pandemic periods. The sample is the 2013--2024 ACS sample of dual-parent households with children of child-rearing age. For brevity, we report only the post-COVID (2022 onward) interaction coefficients from Equation~\eqref{eq:3}; during-COVID (2020--2021) interactions are estimated but omitted. All regressions include fixed effects for state-by-time, husband's occupation, and wife's college major, and control for the wife's age, education, and number of children, as well as the husband's education. Discrepancy in the number of observations, in particular between the first four columns and the last two, obey purely to missing data on the latter. Standard errors, clustered at the husband's occupation level, are reported in parentheses. \textsuperscript{*} $p<0.10$; \textsuperscript{**} $p<0.05$; \textsuperscript{***} $p<0.01$.
 \end{minipage}
\end{table}
\clearpage

\begin{table}[H]\centering
\caption{Effect of Husbands' WFH Shock on Time Spent in Labor and Household Supply - Share of the Day}
\label{tab:mechatus1}
\scalebox{0.55}{
\begin{threeparttable}
\begin{tabular}{llcccccclcc}
\hline
\addlinespace
& & \multicolumn{6}{c}{\emph{Time Spent in Labor Production}} & & \multicolumn{2}{c}{\emph{Time Spent for Personal Care and Leisure}} \\ \cline{3-8} \cline{10-11}
& & Working & Main Work & Main Work & Main Work        & Secondary Work        & Commuting &   &  Personal Care & Leisure      \\
& &             &                   & From Home & Outside Home &                                       &                       &       &                                &                      \\ 
\cline{3-8} \cline{10-11} \addlinespace
\multicolumn{11}{l}{\large{\textbf{Wives:}}} \\
\addlinespace
$ Exposed_{o(h(i))} \times Post_{t}$&            &       0.058***&       0.059***&       0.037***&       0.022   &      -0.001   &      -0.000   &            &       0.001   &       0.032***\\
                    &            &     (0.014)   &     (0.014)   &     (0.012)   &     (0.015)   &     (0.001)   &     (0.002)   &            &     (0.010)   &     (0.009)   \\
\addlinespace \addlinespace
\addlinespace Mean pre-treatment&            &       0.162   &       0.160   &       0.020   &       0.140   &       0.001   &       0.012   &            &       0.374   &       0.114   \\
Obs                 &            &      21,216   &      21,216   &      21,216   &      21,216   &      21,216   &      21,216   &            &      21,216   &      21,216   \\
$R^2$               &            &        0.78   &        0.78   &        0.81   &        0.76   &        0.87   &        0.73   &            &        0.74   &        0.78   \\
\addlinespace
\addlinespace
\multicolumn{11}{l}{\large{\textbf{Husbands:}}} \\
\addlinespace
$ Exposed_{o(h(i))} \times Post_{t}$&            &       0.013   &       0.011   &       0.043***&      -0.032*  &       0.003   &      -0.012***&            &       0.001   &      -0.016   \\
                    &            &     (0.014)   &     (0.014)   &     (0.010)   &     (0.017)   &     (0.002)   &     (0.003)   &            &     (0.008)   &     (0.012)   \\
\addlinespace \addlinespace
\addlinespace Mean pre-treatment&            &       0.320   &       0.312   &       0.030   &       0.283   &       0.006   &       0.028   &            &       0.348   &       0.117   \\
Obs                 &            &      20,835   &      20,835   &      20,835   &      20,835   &      20,835   &      20,835   &            &      20,835   &      20,835   \\
$R^2$               &            &        0.79   &        0.80   &        0.82   &        0.80   &        0.72   &        0.76   &            &        0.77   &        0.77   \\
\end{tabular}
\begin{tabular}{llccccccccl} \addlinespace
& & \multicolumn{8}{c}{\emph{Time Spent in Household Production}} &  \\ \cline{3-11} 
& & Housework & Cooking & Shopping & Primary     & Secondary            & Total                & Total                         &  Prop. Secondary CC &         \\
& &           &                 &          & Childcaring & CC while working & Secondary CC     & Childcaring       &  in Total CC &                    \\ 
\cline{3-11} \addlinespace
\multicolumn{11}{l}{\large{\textbf{Wives:}}} \\
$ Exposed_{o(h(i))} \times Post_{t}$&            &      -0.020** &      -0.009*  &      -0.005** &      -0.024** &       0.004   &      -0.001   &      -0.025   &       0.061** &            \\
                    &            &     (0.008)   &     (0.005)   &     (0.003)   &     (0.010)   &     (0.005)   &     (0.017)   &     (0.018)   &     (0.029)   &            \\
\addlinespace \addlinespace
\addlinespace Mean pre-treatment&            &       0.100   &       0.048   &       0.016   &       0.108   &       0.013   &       0.231   &       0.338   &       0.645   &            \\
Obs                 &            &      21,216   &      21,216   &      21,216   &      21,216   &      21,216   &      21,216   &      21,216   &      20,894   &            \\
$R^2$               &            &        0.79   &        0.78   &        0.76   &        0.76   &        0.78   &        0.77   &        0.77   &        0.78   &            \\
\addlinespace
\addlinespace
\multicolumn{11}{l}{\large{\textbf{Husbands:}}} \\
$ Exposed_{o(h(i))} \times Post_{t}$&            &      -0.003   &       0.002   &      -0.001   &      -0.001   &       0.010   &       0.007   &       0.006   &      -0.028   &            \\
                    &            &     (0.007)   &     (0.002)   &     (0.003)   &     (0.007)   &     (0.006)   &     (0.017)   &     (0.017)   &     (0.032)   &            \\
\addlinespace
\addlinespace Mean pre-treatment&            &       0.039   &       0.015   &       0.007   &       0.050   &       0.010   &       0.125   &       0.174   &       0.651   &            \\
Obs                 &            &      20,835   &      20,835   &      20,835   &      20,835   &      20,835   &      20,835   &      20,835   &      19,133   &            \\
$R^2$               &            &        0.79   &        0.83   &        0.80   &        0.79   &        0.76   &        0.77   &        0.77   &        0.78   &            \\
\addlinespace \hline
\end{tabular}

\end{threeparttable}
}
\par\vspace{0.3cm}
\begin{minipage}{\textwidth} \small
  \textit{Notes}: This table reports estimates from Equation~\eqref{eq:1} using the 2013--2024 ATUS sample of dual-parent households with children of child-rearing age, restricted to weekday, non-holiday diary days. Because each ATUS respondent reports time use for only one household member, we use two separate draws: a men-respondent sample (husbands' time use) and a women-respondent sample (wives' time use); see Subsection~\ref{subsection:treatment_defn} and Appendix~\ref{app:data_construction} for details. All dependent variables are expressed as the share of the 24-hour day spent in an activity (individuals who do not perform the activity are coded as zero). Panel A reports labor-production outcomes: total working time, time at the main job, time at the main job spent working from home versus outside the home, time at a secondary job, and commuting time. Panel B reports personal-care (e.g., sleeping, grooming) and leisure (e.g., religious or sport activities) outcomes. Panel C reports household-production outcomes: housework, cooking, shopping, primary childcare (childcare reported as the main activity), childcare while working, secondary childcare (childcare performed alongside another primary activity), total childcare (primary plus secondary), and the share of total childcare that is secondary. All regressions include state-by-time and husband's-occupation fixed effects and control for the wife's age, education, and number of children, as well as the husband's education. Standard errors, clustered at the husband's occupation level, are reported in parentheses. \textsuperscript{*} $p<0.10$; \textsuperscript{**} $p<0.05$; \textsuperscript{***} $p<0.01$.
\end{minipage}
\end{table}

\section*{Figures}

\begin{figure}[H]
    \caption{Evolution of Remote Work Options by Occupation}\label{fig:evolution_WFH}
    \begin{tabular}{c}
    (a) Share of Job Postings with Remote Work Options \\
    \includegraphics[width=0.9\linewidth]{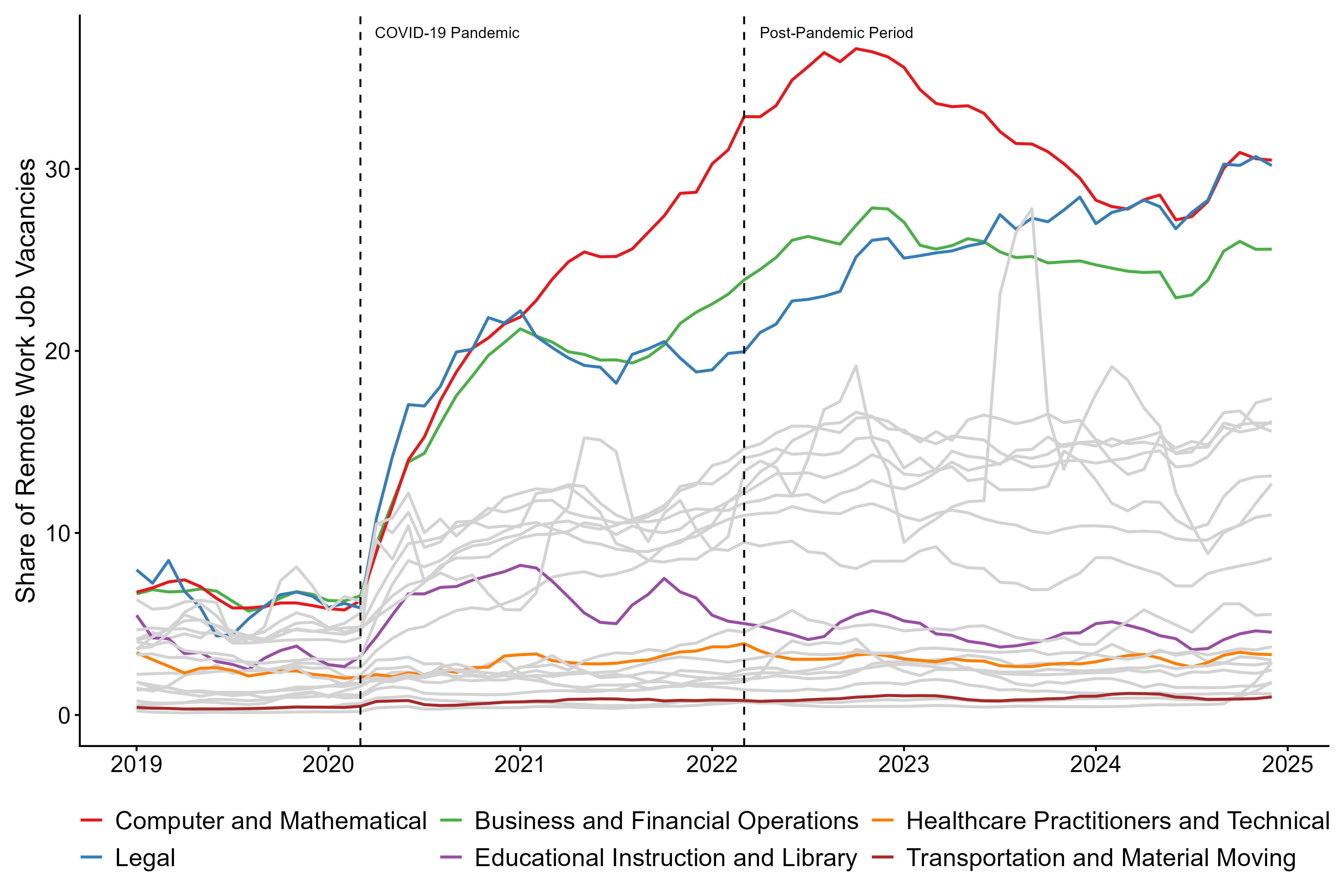} \\
    (b) Percentage of Uptake of Fully Remote Jobs by Men \\
    \includegraphics[width=0.9\linewidth] {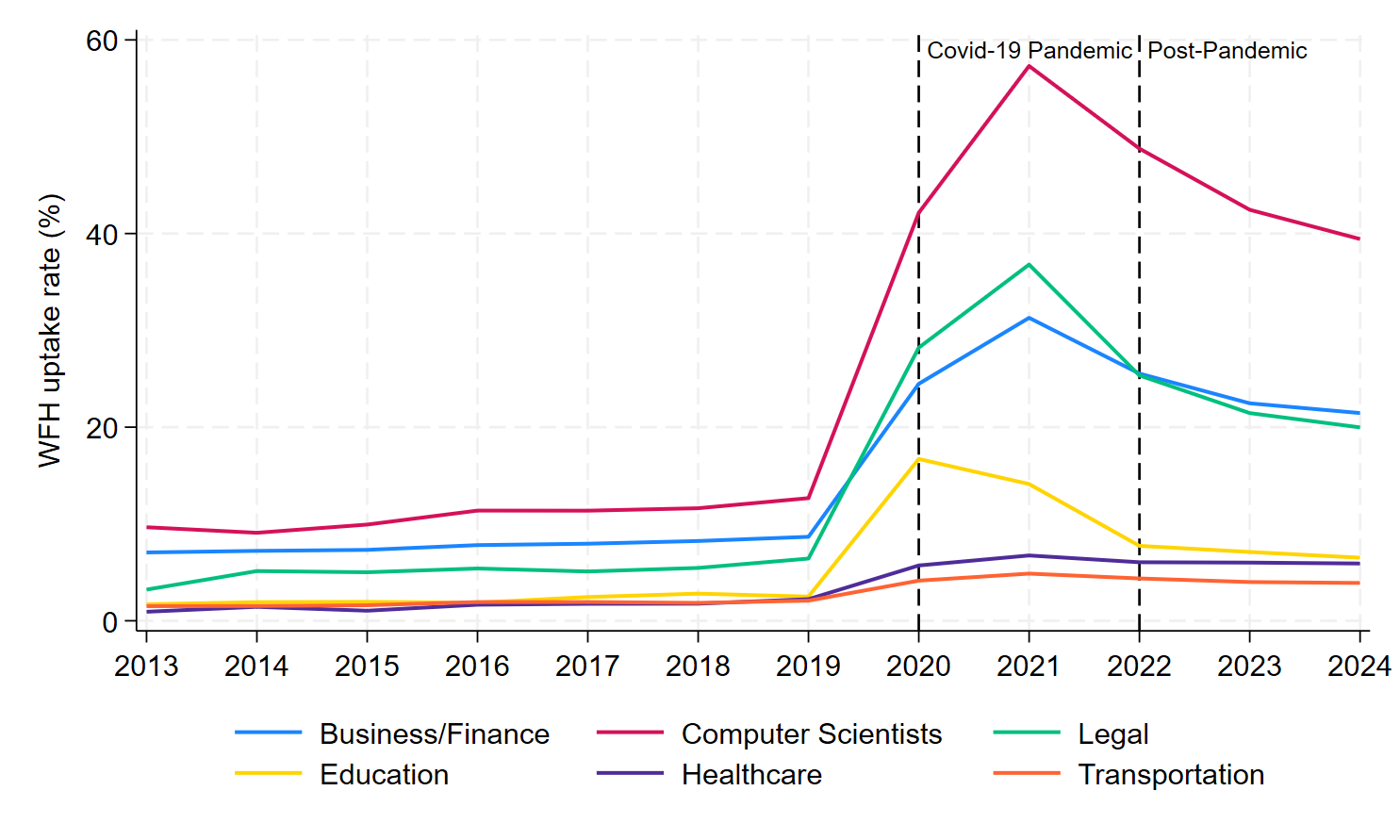}
    \end{tabular}
    \centering
    \begin{minipage}{\textwidth}
    \small
     \textit{Notes}: Author's elaboration. Panel (a) uses monthly WFH job postings data from \citet{hansen2023remote}, with occupations aggregated to the SOC 2018 2-digit classification; the series shows the share of job postings offering remote or hybrid work over time. Panel (b) uses the ACS sample of male parents from dual-parent households with children of child-rearing age (see Section~\ref{sec:data}) and plots the share of husbands in selected broad occupation groups who report having worked from home in the week prior to the interview.
    \end{minipage}
\end{figure}

\clearpage

\clearpage
%%%%%%%%%%%%%%%%%%%%%%%%%%%%%%%%%%%%%%%%%%%%%%%%%%%%%%%%%%%%%%%%%%%%%%%%%%%%%%%%%%%%%%%%%%%%%%
%%% APPENDIX %%%%
%%%%%%%%%%%%%%%%%%%%%%%%%%%%%%%%%%%%%%%%%%%%%%%%%%%%%%%%%%%%%%%%%%%%%%%%%%%%%%%%%%%%%%%%%%%%%%
\appendix
\section*{Online Appendix}\label{appendix_figures}

\section{Tables}
\renewcommand\thetable{\thesection.\arabic{table}} 
\setcounter{table}{0}

\begin{table}[H]
\centering
\caption{Top 15 ACS Occupations by Treatment Status}
\label{tab:top_occ}
\scalebox{0.45}{
\begin{threeparttable}
\begin{tabular}{ll}
\hline
\addlinespace
\textbf{Treated Occupations} &  \textbf{Control Occupations} \\ \midrule \addlinespace
Managers, nec (including Postmasters) (8.9\%)&Driver/Sales Workers and Truck Drivers (6.4\%)\\
Software Developers, Applications and Systems Software (5.0\%)&First-Line Supervisors of Sales Workers (5.5\%)\\
Chief executives and legislators/public administration (2.7\%)&Construction Laborers (3.9\%)\\
Lawyers, and judges, magistrates, and other judicial workers (2.5\%)&Elementary and Middle School Teachers (2.9\%)\\
Sales Representatives, Wholesale and Manufacturing (2.5\%)&Carpenters (2.9\%)\\
Constructions Managers (2.3\%)&Laborers and Freight, Stock, and Material Movers, Hand (2.6\%)\\
General and Operations Managers (2.3\%)&Physicians and Surgeons (2.5\%)\\
Financial Managers (2.2\%)&Retail Salespersons (2.5\%)\\
Computer Scientists and Systems Analysts/Network systems Analysts/Web Developers (2.2\%)&Electricians (2.5\%)\\
Accountants and Auditors (2.1\%)&Grounds Maintenance Workers (2.2\%)\\
Police Officers and Detectives (2.1\%)&First-Line Supervisors of Construction Trades and Extraction Workers (2.2\%)\\
First-Line Supervisors of Production and Operating Workers (1.7\%)&Postsecondary Teachers (2.1\%)\\
Farmers, Ranchers, and Other Agricultural Managers (1.7\%)&Janitors and Building Cleaners (2.1\%)\\
Computer and Information Systems Managers (1.7\%)&Other production workers including semiconductor processors and cooling and freezing equipment operators (2.0\%)\\
Engineers, nec (1.7\%)&Automotive Service Technicians and Mechanics (2.0\%)\\
\addlinespace \hline
\end{tabular}

\end{threeparttable}
}
\par\vspace{0.3cm}
\begin{minipage}{\textwidth}
    \textit{Notes}: This table lists the 15 most common husbands' occupations among treated and control households, respectively, in the 2013--2024 ACS sample of dual-parent households with children of child-rearing age, based on the Census 2010 occupation classification variable \texttt{occ2010}. The treatment indicator is assigned at the occupation level, as described in Subsection~\ref{subsection:treatment_defn}: an occupation is classified as treated if the change in the share of remote/hybrid job postings between 2019 and 2023--2024, measured using data from \citet{hansen2023remote}, is above the median across all occupations. Frequencies are expressed as the percentage of husband-occupation observations within each group (treated/control) accounted for by each occupation. Occupation labels correspond to the value labels associated with \texttt{occ2010}.
 \end{minipage}
\end{table}

\clearpage
\begin{table}[H]\centering
    \caption{Difference between treated and control occupations using Occupational Employment Statistics (OES) 2019}
    \label{tab:occ_oes}
    \scalebox{0.65}{
    \begin{threeparttable}
    \begin{tabular}{>{\raggedright\arraybackslash}p{6.5cm}ccccc}
\toprule
 & Treated & Control & Difference & S.E. & N \\
\midrule
\\[0.5em]
\textit{Panel A: Total Employment and Annual Wage} \\\\
Total Employment & 344,297.33 & 581,275.97 & -236,978.64 & 163,088.65 & 344 \\
Mean annual wage & 83,802.29 & 50,656.48 & 33,145.81*** & 8,442.29 & 343 \\
Annual 10th percentile wage & 42,275.93 & 29,628.75 & 12,647.18*** & 3,497.78 & 343 \\
Annual 25th percentile wage & 55,540.36 & 36,707.87 & 18,832.48*** & 5,277.72 & 343 \\
Annual median wage (50th percentile) & 75,680.69 & 46,598.75 & 29,081.95*** & 7,934.57 & 340 \\
Annual 75th percentile wage & 94,318.43 & 60,786.59 & 33,531.84*** & 8,360.21 & 338 \\
Annual 90th percentile wage & 108,760.01 & 77,503.22 & 31,256.80*** & 8,588.67 & 329 \\
\\[0.5em]
\textit{Panel B: Hourly Wage} \\\\
Mean hourly wage & 40.49 & 23.70 & 16.79*** & 4.19 & 336 \\
Hourly 10th percentile wage & 20.67 & 13.93 & 6.74*** & 1.72 & 336 \\
Hourly 25th percentile wage & 27.21 & 17.23 & 9.99*** & 2.59 & 336 \\
Hourly median wage (50th percentile) & 36.94 & 21.82 & 15.12*** & 3.92 & 333 \\
Hourly 75th percentile wage & 45.72 & 28.38 & 17.34*** & 4.17 & 332 \\
Hourly 90th percentile wage & 52.54 & 36.26 & 16.28*** & 4.29 & 324 \\
\bottomrule
\end{tabular}

    \end{threeparttable}
    }
    \par\vspace{0.3cm}
    \begin{minipage}{\textwidth}
    \small
     \textit{Notes}: This table is a balance table comparing worker and job characteristics between treated and control occupations, as classified in Subsection~\ref{subsection:treatment_defn}, using 2019 Occupational Employment Statistics (OES) data. Statistics are weighted by ACS husband-employment counts at the Census 2010 occupation level, using husbands' occupations from survey year 2019. OES variables are reported at the SOC 2018 (six-digit) level in the source data and were aggregated to Census 2010 occupations using employment-share weights; the OES reference period is May 2019. For each variable, we report the treated-group mean, the control-group mean, their difference (treated $-$ control), the standard error of the difference, and the number of Census 2010 occupations (N) underlying the estimate; N varies across rows due to missing values for some occupation-level wage percentiles. Panel A reports total employment and annual wage levels; Panel B reports hourly wage levels. Significance of the treated-control difference is based on a two-sided $t$-test: \textsuperscript{*} $p<0.10$; \textsuperscript{**} $p<0.05$; \textsuperscript{***} $p<0.01$.
 \end{minipage}
\end{table}

\clearpage

\begin{table}[H]\centering
    \caption{Difference between treated and control occupations using ACS 2019}
    \label{tab:occ_acs}
    \scalebox{0.65}{
\begin{threeparttable}
\begin{tabular}{>{\raggedright\arraybackslash}p{6.5cm}ccccc}
\toprule
 & Treated & Control & Difference & S.E. & N \\
\midrule
\\[0.5em]
\textit{Panel A: Occupational Characteristics (ACS)} \\\\
Share of Women & 0.46 & 0.44 & 0.01*** & 0.00 & 212,671 \\
Age & 36.95 & 35.70 & 1.25*** & 0.04 & 212,671 \\
Share of Below Any College & 0.11 & 0.37 & -0.26*** & 0.00 & 212,671 \\
Share of Below Any High-School & 0.02 & 0.12 & -0.10*** & 0.00 & 212,671 \\
Share of College Graduates & 0.66 & 0.31 & 0.35*** & 0.00 & 212,671 \\
Potential Years of Education & 15.09 & 13.77 & 1.32*** & 0.01 & 212,671 \\
Potential Experience & 15.87 & 15.93 & -0.07* & 0.04 & 212,671 \\
Earnings & 50,839.08 & 31,103.25 & 19,735.83*** & 255.51 & 212,671 \\
Hourly Wage & 28.57 & 19.48 & 9.10*** & 0.26 & 187,207 \\
Usual Weekly Work Hours & 41.37 & 40.67 & 0.70*** & 0.07 & 187,207 \\
Weeks Worked & 49.30 & 48.36 & 0.93*** & 0.06 & 187,475 \\
Share Working Part-Time & 0.13 & 0.15 & -0.02*** & 0.00 & 187,207 \\
\\[0.5em]
\bottomrule
\end{tabular}

\end{threeparttable}
}
\par\vspace{0.3cm}
\begin{minipage}{\textwidth}
    \small
\textit{Notes}: This table is a balance table comparing husbands in treated and control occupations, as classified in Subsection~\ref{subsection:treatment_defn}, using ACS survey year 2019 (the pre-pandemic reference year). All statistics are weighted by ACS sampling weights. The reference period is survey year 2019 for all variables except earnings, hourly wage, usual weekly work hours, weeks worked, and part-time status, which refer to earnings in the previous calendar year (2018). For each variable, we report the treated-group mean, the control-group mean, their difference (treated $-$ control), the standard error of the difference, and the number of observations (N) underlying the estimate. Significance of the treated-control difference is based on a two-sided $t$-test: \textsuperscript{*} $p<0.10$; \textsuperscript{**} $p<0.05$; \textsuperscript{***} $p<0.01$.
 \end{minipage}
\end{table}

\begin{table}[H]\centering
\caption{Difference Between Treated and Control Occupations using O*NET 2025}
\label{tab:desc_stat_onet}
\scalebox{0.8}{
\begin{threeparttable}
\begin{tabular}{>{\raggedright\arraybackslash}p{6.5cm}ccccc}
\toprule
 & Treated & Control & Difference & S.E. & N \\
\midrule
\\[0.5em]
\textit{Panel A: Work Context} \\\\
Time Pressure & 0.02 & 0.26 & -0.25 & 0.19 & 342 \\
Frequency of Conflict Situations & 0.60 & -0.03 & 0.64*** & 0.21 & 342 \\
Contact with Others & 0.45 & -0.02 & 0.47** & 0.22 & 342 \\
Coordinate or Lead Others & 0.66 & -0.07 & 0.73** & 0.30 & 342 \\
Degree of Automation & 0.19 & -0.37 & 0.56*** & 0.15 & 342 \\
Structured Work & -0.55 & 0.21 & -0.76*** & 0.18 & 342 \\
Duration of Typical Work Week & 0.50 & 0.23 & 0.27 & 0.22 & 342 \\
Face-to-Face Discussions & 0.32 & 0.06 & 0.26 & 0.20 & 342 \\
Freedom to Make Decisions & 0.44 & -0.24 & 0.68*** & 0.19 & 342 \\
Frequency of Decision Making & 0.28 & 0.24 & 0.04 & 0.19 & 342 \\
Impact of Decisions on Coworkers or Company Results & 0.38 & 0.13 & 0.26 & 0.20 & 342 \\
Level of Competition & 0.41 & -0.09 & 0.49*** & 0.18 & 342 \\
Physical Proximity & -0.25 & 0.19 & -0.44* & 0.23 & 342 \\
Regular Work Schedule & -0.32 & 0.14 & -0.45** & 0.20 & 342 \\
Work with a Workgroup or Team & 0.48 & -0.32 & 0.81*** & 0.27 & 342 \\
\\[0.5em]
\textit{Panel B: Work Activities} \\\\
Communicating with Supervisors, Peers, or Subordinates & 0.74 & -0.30 & 1.04*** & 0.20 & 342 \\
Coordinating the Work and Activities of Others & 0.71 & -0.23 & 0.94*** & 0.23 & 336 \\
Developing and Building Teams & 0.76 & -0.17 & 0.93*** & 0.22 & 338 \\
Establishing and Maintaining Interpersonal Relationships & 0.76 & -0.25 & 1.00*** & 0.18 & 342 \\
\\[0.5em]
\textit{Panel C: O*Net Indexes (Goldin, 2014)} \\\\
Avg. of 5 normalized characteristics & 0.22 & -0.01 & 0.23*** & 0.08 & 342 \\
Avg. of 7 normalized characteristics & 0.24 & 0.04 & 0.21** & 0.10 & 342 \\
\bottomrule
\end{tabular}

\end{threeparttable}
}
\par\vspace{0.3cm}
\begin{minipage}{\textwidth} \small
 \textit{Notes}: This table is a balance table comparing job-skill characteristics between treated and control occupations, as classified in Subsection~\ref{subsection:treatment_defn}, using O*NET 2025 data at the Census 2010 occupation level. All statistics are weighted by ACS 2019 male-employment counts at the Census 2010 occupation level. All O*NET variables are standardized to $z$-scores (demeaned and divided by the cross-occupation standard deviation). Following \citet{goldin_grand_2014}, the five-characteristic index is the average of the normalized dimensions \emph{Time pressure}, \emph{Contact with others}, \emph{Establishing and maintaining interpersonal relationships}, \emph{Structured work}, and \emph{Freedom to make decisions}; the seven-characteristic index additionally includes \emph{Face-to-face discussions} and \emph{Frequency of decision making}. Higher values of both indices indicate lower worker substitutability and greater job inflexibility. \textit{Measurement note}: the current ``Structured work'' item differs from that used in \citet{goldin_grand_2014}'s original period due to O*NET wording changes; the present question asks ``How much freedom does the worker have in determining the tasks, priorities, or goals of the job?'' We use this item and take its complement so that higher values correspond to more structure, aligning with the original construct. Significance of the treated-control difference (if reported) follows a two-sided $t$-test: \textsuperscript{*} $p<0.10$; \textsuperscript{**} $p<0.05$; \textsuperscript{***} $p<0.01$.
\end{minipage}
\end{table}

\clearpage
\begin{table}[H]\centering
\caption{Descriptive Statistics of Main Labor Market Outcomes, by Treatment Group and Pre/Post-COVID Period}
\label{tab:desc_labor}
\scalebox{0.99}{
\begin{threeparttable}
\renewcommand{\baselinestretch}{1} \normalsize \scriptsize
\begin{tabular}{lcccccc}
\hline \\
& \multicolumn{3}{c}{Pre-COVID} & \multicolumn{3}{c}{Post-COVID} \\\\
\cmidrule(lr){2-4} \cmidrule(lr){5-7} \\
& Mean & SD & N & Mean & SD & N \\\\
\addlinespace \\
\textit{Panel A: Control Sample} & & & & & & \\\\
\hline \addlinespace \\
Spouse Employed (Last Week) & 0.94 & 0.24 & 401,867 & 0.93 & 0.25 & 249,246 \\
Spouse LFP (Last Week) & 0.97 & 0.18 & 401,867 & 0.96 & 0.20 & 249,246 \\
Spouse Employed (Last Year) & 0.91 & 0.28 & 401,867 & 0.90 & 0.30 & 249,246 \\
Spouse Log Earnings & 10.21 & 0.85 & 401,867 & 10.26 & 0.88 & 249,246 \\
Spouse Log Wage (Last Year) & 2.61 & 0.70 & 401,867 & 2.68 & 0.75 & 249,246 \\
Spouse Usual Weekly Hours Worked (Last Year) & 43.75 & 10.46 & 401,867 & 43.23 & 10.65 & 249,246 \\
Spouse Weeks Worked (Last Year) & 48.71 & 7.86 & 401,867 & 49.00 & 9.23 & 249,246 \\
Spouse Part-time Status (Last Year) & 0.07 & 0.26 & 401,867 & 0.07 & 0.26 & 249,246 \\
Spouse Travel Time to Work & 28.36 & 24.48 & 401,867 & 27.79 & 25.10 & 249,246 \\
Spouse Usually Works from Home & 0.02 & 0.15 & 401,867 & 0.06 & 0.24 & 249,246 \\
Wife Employed (Last Week) & 0.63 & 0.48 & 401,867 & 0.66 & 0.47 & 249,246 \\
Wife LFP (Last Week) & 0.66 & 0.47 & 401,867 & 0.69 & 0.46 & 249,246 \\
Wife Employed (Last Year) & 0.66 & 0.47 & 401,867 & 0.68 & 0.47 & 249,246 \\
Wife Log Earnings & 9.72 & 1.14 & 401,867 & 9.81 & 1.16 & 249,246 \\
Wife Log Wage (Last Year) & 2.47 & 0.75 & 401,867 & 2.57 & 0.80 & 249,246 \\
Wife Usual Weekly Hours Worked (Last Year) & 36.31 & 10.93 & 401,867 & 36.41 & 11.16 & 249,246 \\
Wife Weeks Worked (Last Year) & 45.47 & 12.21 & 401,867 & 46.18 & 12.89 & 249,246 \\
Wife Part-time Status (Last Year) & 0.28 & 0.45 & 401,867 & 0.26 & 0.44 & 249,246 \\
Wife Travel Time to Work & 23.50 & 20.06 & 401,867 & 20.92 & 20.50 & 249,246 \\
 Wife Usually Works from Home & 0.05 & 0.23 & 401,867 & 0.16 & 0.37 & 249,246 \\
\addlinespace \addlinespace \\
\textit{Panel B: Treated Sample} & & & & & & \\\\
\hline \addlinespace \\
Spouse Employed (Last Week) & 0.97 & 0.18 & 413,548 & 0.96 & 0.19 & 306,544 \\
Spouse LFP (Last Week) & 0.98 & 0.14 & 413,548 & 0.98 & 0.15 & 306,544 \\
Spouse Employed (Last Year) & 0.94 & 0.23 & 413,548 & 0.94 & 0.24 & 306,544 \\
Spouse Log Earnings & 10.82 & 0.79 & 413,548 & 10.83 & 0.82 & 306,544 \\
Spouse Log Wage (Last Year) & 3.13 & 0.71 & 413,548 & 3.17 & 0.75 & 306,544 \\
Spouse Usual Weekly Hours Worked (Last Year) & 45.28 & 9.33 & 413,548 & 44.06 & 9.29 & 306,544 \\
Spouse Weeks Worked (Last Year) & 50.06 & 5.38 & 413,548 & 50.57 & 6.59 & 306,544 \\
Spouse Part-time Status (Last Year) & 0.03 & 0.17 & 413,548 & 0.03 & 0.18 & 306,544 \\
Spouse Travel Time to Work & 28.91 & 24.65 & 413,548 & 22.44 & 24.47 & 306,544 \\
Spouse Usually Works from Home & 0.07 & 0.25 & 413,548 & 0.25 & 0.44 & 306,544 \\
Wife Employed (Last Week) & 0.67 & 0.47 & 413,548 & 0.70 & 0.46 & 306,544 \\
Wife LFP (Last Week) & 0.69 & 0.46 & 413,548 & 0.73 & 0.45 & 306,544 \\
Wife Employed (Last Year) & 0.68 & 0.46 & 413,548 & 0.72 & 0.45 & 306,544 \\
Wife Log Earnings & 10.07 & 1.19 & 413,548 & 10.17 & 1.17 & 306,544 \\
Wife Log Wage (Last Year) & 2.80 & 0.76 & 413,548 & 2.87 & 0.80 & 306,544 \\
Wife Usual Weekly Hours Worked (Last Year) & 36.57 & 11.88 & 413,548 & 37.05 & 11.40 & 306,544 \\
Wife Weeks Worked (Last Year) & 45.98 & 11.58 & 413,548 & 46.95 & 11.98 & 306,544 \\
Wife Part-time Status (Last Year) & 0.27 & 0.44 & 413,548 & 0.24 & 0.42 & 306,544 \\
Wife Travel Time to Work & 24.29 & 20.97 & 413,548 & 18.74 & 20.72 & 306,544 \\
 Wife Usually Works from Home & 0.09 & 0.29 & 413,548 & 0.28 & 0.45 & 306,544 \\
\addlinespace \\
\hline
\end{tabular}

\end{threeparttable}
}
\par\vspace{0.3cm}
\begin{minipage}{\textwidth} \small
  \textit{Notes}: This table reports summary statistics for key labor market outcomes, separately for treated and control households (as classified in Subsection~\ref{subsection:treatment_defn}) and for the pre-COVID (before 2020) and post-COVID (2020 onward) periods. The sample is the pooled 2013--2024 ACS sample of dual-parent households with children of child-rearing age. All statistics are computed using ACS sampling weights.
\end{minipage}
\end{table}

\begin{table}[H]\centering
\caption{Descriptive Statistics of Main Control Variables, by Treatment Group and Pre/Post-COVID Period}
\label{tab:desc_hh}
\scalebox{0.99}{
\begin{threeparttable}
\renewcommand{\baselinestretch}{1} \normalsize \scriptsize
\begin{tabular}{lcccccc}
\hline  \\
& \multicolumn{3}{c}{Pre-COVID} & \multicolumn{3}{c}{Post-COVID} \\\\
\cmidrule(lr){2-4} \cmidrule(lr){5-7} \\
& Mean & SD & N & Mean & SD & N \\\\
\addlinespace \\
\textit{Panel A: Control Sample} & & & & & & \\\\
\hline \addlinespace \\
Spouse - Below HS Level & 0.16 & 0.37 & 401,867 & 0.13 & 0.34 & 249,246 \\
Spouse - HS Level & 0.31 & 0.46 & 401,867 & 0.30 & 0.46 & 249,246 \\
Spouse - Above HS Level & 0.53 & 0.50 & 401,867 & 0.56 & 0.50 & 249,246 \\
Spouse Age & 36.65 & 7.11 & 401,867 & 37.16 & 7.03 & 249,246 \\
Wife - Below HS Level & 0.13 & 0.33 & 401,867 & 0.10 & 0.29 & 249,246 \\
Wife - HS Level & 0.22 & 0.41 & 401,867 & 0.21 & 0.41 & 249,246 \\
Wife - Above HS Level & 0.66 & 0.48 & 401,867 & 0.69 & 0.46 & 249,246 \\
Wife Age & 34.15 & 6.13 & 401,867 & 34.72 & 5.99 & 249,246 \\
Number of own children in the household & 2.17 & 0.92 & 401,867 & 2.16 & 0.92 & 249,246 \\
Age of Youngest Child & 3.46 & 2.58 & 401,867 & 3.47 & 2.59 & 249,246 \\
Age of Oldest Child & 8.07 & 5.29 & 401,867 & 8.07 & 5.39 & 249,246 \\
Youngest Child 0-4 & 0.64 & 0.48 & 401,867 & 0.64 & 0.48 & 249,246 \\
Youngest Child 5-8 & 0.36 & 0.48 & 401,867 & 0.36 & 0.48 & 249,246 \\
\addlinespace \addlinespace \\
\textit{Panel B: Treated Sample} & & & & & & \\\\
\hline \addlinespace \\
Spouse - Below HS Level & 0.02 & 0.14 & 413,548 & 0.02 & 0.14 & 306,544 \\
Spouse - HS Level & 0.10 & 0.30 & 413,548 & 0.10 & 0.30 & 306,544 \\
Spouse - Above HS Level & 0.88 & 0.32 & 413,548 & 0.88 & 0.32 & 306,544 \\
Spouse Age & 37.87 & 6.44 & 413,548 & 38.19 & 6.24 & 306,544 \\
Wife - Below HS Level & 0.02 & 0.15 & 413,548 & 0.02 & 0.15 & 306,544 \\
Wife - HS Level & 0.09 & 0.29 & 413,548 & 0.09 & 0.29 & 306,544 \\
Wife - Above HS Level & 0.88 & 0.32 & 413,548 & 0.89 & 0.32 & 306,544 \\
Wife Age & 35.70 & 5.64 & 413,548 & 36.12 & 5.42 & 306,544 \\
Number of own children in the household & 2.03 & 0.86 & 413,548 & 2.03 & 0.87 & 306,544 \\
Age of Youngest Child & 3.39 & 2.59 & 413,548 & 3.41 & 2.60 & 306,544 \\
Age of Oldest Child & 6.99 & 4.79 & 413,548 & 7.00 & 4.81 & 306,544 \\
Youngest Child 0-4 & 0.65 & 0.48 & 413,548 & 0.65 & 0.48 & 306,544 \\
Youngest Child 5-8 & 0.35 & 0.48 & 413,548 & 0.35 & 0.48 & 306,544 \\
\addlinespace \\
\hline 
\end{tabular}

\end{threeparttable}
}
\par\vspace{0.3cm}
\begin{minipage}{\textwidth} \small
 \textit{Notes}: This table reports summary statistics for the household- and individual-level control variables used throughout the paper, separately for treated and control households (as classified in Subsection~\ref{subsection:treatment_defn}) and for the pre-COVID (before 2020) and post-COVID (2020 onward) periods. The sample is the pooled 2013--2024 ACS sample of dual-parent households with children of child-rearing age. All statistics are computed using ACS sampling weights.
\end{minipage}
\end{table}

\clearpage

\begin{table}[H]\centering
\caption{Descriptive Statistics of Main Time Use Variables, by Treatment Group and Pre/Post-COVID Period}
\label{tab:desc_atus_out}
\scalebox{0.65}{
\begin{threeparttable}
\renewcommand{\baselinestretch}{1} \normalsize \scriptsize
\begin{tabular}{lcccccc}
\hline \\
& \multicolumn{3}{c}{Pre-COVID} & \multicolumn{3}{c}{Post-COVID} \\\\
\cmidrule(lr){2-4} \cmidrule(lr){5-7} \\
& Mean & SD & N & Mean & SD & N \\\\
\addlinespace \\
\textit{Panel A: Control Sample} & & & & & & \\\\
\hline \addlinespace \\
Husband's Time Spent Working & 0.31 & 0.17 & 12,872 & 0.28 & 0.17 & 5,743 \\
Husband's Time Spent in Main Work & 0.30 & 0.17 & 12,872 & 0.28 & 0.17 & 5,743 \\
Husband's Time Spent in Main Work From Home & 0.01 & 0.05 & 12,872 & 0.02 & 0.08 & 5,743 \\
Husband's Time Spent in Main Work Outside Home & 0.29 & 0.17 & 12,872 & 0.26 & 0.17 & 5,743 \\
Husband's Time Spent in Secondary Work & 0.01 & 0.04 & 12,872 & 0.00 & 0.02 & 5,743 \\
Husband's Time Spent Commuting to Work & 0.03 & 0.03 & 12,872 & 0.03 & 0.03 & 5,743 \\
Husband's Time Spent in Personal Care & 0.35 & 0.08 & 12,872 & 0.37 & 0.08 & 5,743 \\
Husband's Time Spent in Leisure & 0.13 & 0.10 & 12,872 & 0.13 & 0.11 & 5,743 \\
Husband's Time Spent in Housework & 0.04 & 0.07 & 12,872 & 0.05 & 0.08 & 5,743 \\
Husband's Time Spent Cooking & 0.02 & 0.02 & 12,872 & 0.02 & 0.05 & 5,743 \\
Husband's Time Spent Shopping & 0.01 & 0.02 & 12,872 & 0.01 & 0.02 & 5,743 \\
Husband's Time Spent in Primary Child Care & 0.04 & 0.06 & 12,872 & 0.05 & 0.06 & 5,743 \\
Husband's Time Spent Child Caring while Working & 0.01 & 0.04 & 12,872 & 0.01 & 0.05 & 5,743 \\
Husband's Time Spent in Secondary Child Care & 0.13 & 0.14 & 12,872 & 0.16 & 0.15 & 5,743 \\
Husband's Total Time Spent in Child Care & 0.18 & 0.16 & 12,872 & 0.20 & 0.17 & 5,743 \\
Husband's Share of Time in Secondary Child Care over Total Chilcare & 0.69 & 0.32 & 12,872 & 0.71 & 0.31 & 5,743 \\
Wife's Time Spent Working & 0.15 & 0.17 & 12,872 & 0.15 & 0.17 & 5,743 \\
Wife's Time Spent in Main Work & 0.15 & 0.17 & 12,872 & 0.15 & 0.17 & 5,743 \\
Wife's Time Spent in Main Work From Home & 0.01 & 0.06 & 12,872 & 0.04 & 0.10 & 5,743 \\
Wife's Time Spent in Main Work Outside Home & 0.14 & 0.16 & 12,872 & 0.11 & 0.16 & 5,743 \\
Wife's Time Spent in Secondary Work & 0.00 & 0.01 & 12,872 & 0.00 & 0.02 & 5,743 \\
Wife's Time Spent Commuting to Work & 0.01 & 0.02 & 12,872 & 0.01 & 0.02 & 5,743 \\
Wife's Time Spent in Personal Care & 0.38 & 0.08 & 12,872 & 0.38 & 0.08 & 5,743 \\
Wife's Time Spent in Leisure & 0.11 & 0.10 & 12,872 & 0.11 & 0.09 & 5,743 \\
Wife's Time Spent in Housework & 0.11 & 0.10 & 12,872 & 0.12 & 0.10 & 5,743 \\
Wife's Time Spent Cooking & 0.05 & 0.05 & 12,872 & 0.06 & 0.06 & 5,743 \\
Wife's Time Spent Shopping & 0.01 & 0.03 & 12,872 & 0.01 & 0.03 & 5,743 \\
Wife's Time Spent in Primary Child Care & 0.10 & 0.09 & 12,872 & 0.11 & 0.09 & 5,743 \\
Wife's Time Spent Child Caring while Working & 0.01 & 0.05 & 12,872 & 0.03 & 0.08 & 5,743 \\
Wife's Time Spent in Secondary Child Care & 0.24 & 0.16 & 12,872 & 0.26 & 0.18 & 5,743 \\
Wife's Total Time Spent in Child Care & 0.34 & 0.19 & 12,872 & 0.37 & 0.19 & 5,743 \\
Wife's Share of Time in Secondary Child Care over Total Chilcare & 0.66 & 0.25 & 12,872 & 0.66 & 0.26 & 5,743 \\
\addlinespace \addlinespace \\
\textit{Panel B: Treated Sample} & & & & & & \\\\
\hline \addlinespace \\
Husband's Time Spent Working & 0.33 & 0.14 & 15,574 & 0.31 & 0.14 & 8,354 \\
Husband's Time Spent in Main Work & 0.32 & 0.15 & 15,574 & 0.31 & 0.14 & 8,354 \\
Husband's Time Spent in Main Work From Home & 0.04 & 0.10 & 15,574 & 0.12 & 0.16 & 8,354 \\
Husband's Time Spent in Main Work Outside Home & 0.28 & 0.16 & 15,574 & 0.19 & 0.18 & 8,354 \\
Husband's Time Spent in Secondary Work & 0.01 & 0.03 & 15,574 & 0.00 & 0.03 & 8,354 \\
Husband's Time Spent Commuting to Work & 0.03 & 0.03 & 15,574 & 0.02 & 0.02 & 8,354 \\
Husband's Time Spent in Personal Care & 0.34 & 0.06 & 15,574 & 0.36 & 0.07 & 8,354 \\
Husband's Time Spent in Leisure & 0.11 & 0.09 & 15,574 & 0.11 & 0.09 & 8,354 \\
Husband's Time Spent in Housework & 0.04 & 0.06 & 15,574 & 0.05 & 0.06 & 8,354 \\
Husband's Time Spent Cooking & 0.01 & 0.02 & 15,574 & 0.02 & 0.02 & 8,354 \\
Husband's Time Spent Shopping & 0.01 & 0.02 & 15,574 & 0.01 & 0.02 & 8,354 \\
Husband's Time Spent in Primary Child Care & 0.05 & 0.05 & 15,574 & 0.06 & 0.06 & 8,354 \\
Husband's Time Spent Child Caring while Working & 0.01 & 0.05 & 15,574 & 0.04 & 0.10 & 8,354 \\
Husband's Time Spent in Secondary Child Care & 0.12 & 0.13 & 15,574 & 0.17 & 0.17 & 8,354 \\
Husband's Total Time Spent in Child Care & 0.17 & 0.14 & 15,574 & 0.22 & 0.18 & 8,354 \\
Husband's Share of Time in Secondary Child Care over Total Chilcare & 0.63 & 0.30 & 15,574 & 0.68 & 0.30 & 8,354 \\
Wife's Time Spent Working & 0.16 & 0.17 & 15,574 & 0.20 & 0.17 & 8,354 \\
Wife's Time Spent in Main Work & 0.16 & 0.17 & 15,574 & 0.19 & 0.17 & 8,354 \\
Wife's Time Spent in Main Work From Home & 0.02 & 0.07 & 15,574 & 0.08 & 0.14 & 8,354 \\
Wife's Time Spent in Main Work Outside Home & 0.14 & 0.16 & 15,574 & 0.12 & 0.16 & 8,354 \\
Wife's Time Spent in Secondary Work & 0.00 & 0.01 & 15,574 & 0.00 & 0.02 & 8,354 \\
Wife's Time Spent Commuting to Work & 0.01 & 0.02 & 15,574 & 0.01 & 0.02 & 8,354 \\
Wife's Time Spent in Personal Care & 0.37 & 0.07 & 15,574 & 0.38 & 0.07 & 8,354 \\
Wife's Time Spent in Leisure & 0.11 & 0.09 & 15,574 & 0.10 & 0.09 & 8,354 \\
Wife's Time Spent in Housework & 0.10 & 0.08 & 15,574 & 0.09 & 0.08 & 8,354 \\
Wife's Time Spent Cooking & 0.05 & 0.04 & 15,574 & 0.05 & 0.04 & 8,354 \\
Wife's Time Spent Shopping & 0.02 & 0.03 & 15,574 & 0.01 & 0.03 & 8,354 \\
Wife's Time Spent in Primary Child Care & 0.11 & 0.09 & 15,574 & 0.11 & 0.10 & 8,354 \\
Wife's Time Spent Child Caring while Working & 0.01 & 0.05 & 15,574 & 0.03 & 0.09 & 8,354 \\
Wife's Time Spent in Secondary Child Care & 0.23 & 0.16 & 15,574 & 0.23 & 0.17 & 8,354 \\
Wife's Total Time Spent in Child Care & 0.34 & 0.18 & 15,574 & 0.34 & 0.20 & 8,354 \\
Wife's Share of Time in Secondary Child Care over Total Chilcare & 0.64 & 0.23 & 15,574 & 0.64 & 0.26 & 8,354 \\
\addlinespace \\
\hline
\end{tabular}

\end{threeparttable}
}
\par\vspace{0.3cm}
\begin{minipage}{\textwidth} \small
 \textit{Notes}: This table reports summary statistics for individual-level time-use variables, separately for treated and control households (as classified in Subsection~\ref{subsection:treatment_defn}) and for the pre-COVID (before 2020) and post-COVID (2020 onward) periods. Unless noted as a ratio, each variable is measured as the share of the 24-hour day spent in the activity; see \autoref{tab:mechatus1} for full outcome definitions. The sample is the pooled 2013--2024 ATUS sample of dual-parent households with children of child-rearing age, restricted to weekday, non-holiday diary days. All statistics are computed using ATUS sampling weights.
\end{minipage}
\end{table}

\clearpage

%\begin{table}[H]\centering
%\caption{Effect of Husbands' WFH Shock on Their Own Labor Market Outcomes}
%\label{tab:acs_men_ownshock}
%\scalebox{0.65}{
%\begin{threeparttable}
%\input{tables/acs_men_ownshock}
%\end{threeparttable}
%}
%\par\vspace{0.3cm}
%\begin{minipage}{\textwidth} \small
% \textit{Notes}: We estimate Equation \eqref{eq:1} using male spouse's intensive margin of labor supply as outcomes. The sample is the ACS 2013–2024. All specifications control for the female spouse’s age, both spouses' education, and number of children. State-by-time and husband-occupation fixed effects are included. Standard errors are clustered at the husband occupation level.
% \end{minipage}
%\end{table}
%\clearpage

\begin{table}[H]\centering
\caption{Heterogeneous Effects on Wives' Labor Market Outcomes by Quartile of Husbands' WFH Shock}
\label{tab:acs_intensity}
\scalebox{0.8}{
\begin{threeparttable}
\begin{tabular}{llcccc}
\hline
\addlinespace
& &  Log Earnings & Usual Weekly Hours & Weeks Worked & Part-time \\ 
& &  & Worked & & Status \\
\cline{3-6} \addlinespace
\addlinespace
$ \text{Exposure: 2nd quartile}_{o(h(i))} \times Post_{t}$&            &       0.021   &       0.133   &       0.109   &      -0.004   \\
                    &            &     (0.014)   &     (0.130)   &     (0.142)   &     (0.006)   \\
$ \text{Exposure: 3rd quartile}_{o(h(i))} \times Post_{t}$&            &       0.043***&       0.494***&       0.380***&      -0.024***\\
                    &            &     (0.015)   &     (0.134)   &     (0.113)   &     (0.006)   \\
$ \text{Exposure: 4th quartile}_{o(h(i))} \times Post_{t}$&            &       0.067***&       0.522***&       0.695***&      -0.027***\\
                    &            &     (0.012)   &     (0.118)   &     (0.109)   &     (0.005)   \\
\addlinespace Mean pre-treatment&            &       9.915   &      36.294   &      45.671   &       0.277   \\
Obs                 &            &     949,443   &     948,946   &     949,443   &     948,946   \\
$R^2$               &            &        0.15   &        0.03   &        0.03   &        0.03   \\
\addlinespace \hline
\end{tabular}

\end{threeparttable}
}
\par\vspace{0.3cm}
\begin{minipage}{\textwidth} \small
 \textit{Notes}: This table reports estimates from a version of Equation~\eqref{eq:1} in which the binary treatment $\text{High-}\Delta\text{WFH}_{o(h(i))}$ is replaced by indicators for the second, third, and fourth quartiles of the husband's occupation-level change in WFH probability (as defined in Subsection~\ref{subsection:treatment_defn}), each interacted with the post- and during-COVID dummies; the omitted category is the bottom quartile of WFH-exposure change. The sample is the 2013--2024 ACS sample of dual-parent households with children of child-rearing age. For brevity, we report only the post-COVID (2022 onward) interaction coefficients. All regressions control for the wife's age, education, and number of children, as well as the husband's education, and include state-by-time and husband's-occupation fixed effects. Standard errors, clustered at the husband's occupation level, are reported in parentheses. \textsuperscript{*} $p<0.10$; \textsuperscript{**} $p<0.05$; \textsuperscript{***} $p<0.01$.
\end{minipage}
\end{table}
\clearpage

\clearpage

%\begin{table}[H]\centering
%\caption{Effect of Husbands' WFH Uptake on Wives' Labor Supply with a 2SLS Approach}
%\label{tab:acs_baseline_2sls}
%\scalebox{0.9}{
%\begin{threeparttable}
%\input{tables/acs_X_spwfh_only_Z_spwfh_rate_2sls}
%\end{threeparttable}
%}
%\par\vspace{0.3cm}
%\begin{minipage}{\textwidth} \small
% \textit{Notes}: This table reports second-stage estimates from a 2SLS variant of Equation~\eqref{eq:1}, using the 2013--2024 ACS sample of dual-parent households with children of child-rearing age. The endogenous regressor is the husband's observed post/during-COVID interaction with his own WFH uptake, $WFH_{h(i),t}$ (replacing the exogenous occupation-level treatment $\text{High-}\Delta\text{WFH}_{o(h(i))}$ used in Equation~\eqref{eq:1}), instrumented with the average WFH rate in the husband's occupation, $WFH_{o(h(i)),t}$, interacted with the post/during-COVID dummies. The corresponding reduced-form first-stage relationship is shown in \autoref{fig:acs_dynamic_DiD_spwfh_only}; first-stage coefficients are not reported here for brevity. All regressions control for the wife's age, education, and number of children, as well as the husband's education, and include state-by-time and husband's-occupation fixed effects. Standard errors, clustered at the husband's occupation level, are reported in parentheses. \textsuperscript{*} $p<0.10$; \textsuperscript{**} $p<0.05$; \textsuperscript{***} $p<0.01$.
%\end{minipage}
%\end{table}
%\clearpage

\begin{table}[H]\centering
\caption{Effect of Husbands' WFH Uptake on Wives' Labor Supply with a 2SLS Approach}
\label{tab:acs_baseline_2sls_loo}
\scalebox{0.9}{
\begin{threeparttable}
\begin{tabular}{llcccc}
\hline
\addlinespace
& &   Log Earnings  & Usual Weekly Hours & Weeks Worked & Part-time \\
& &      & Worked & & Status \\
\cline{3-6} \addlinespace
\addlinespace
$ WFH_{h(i),t} \times Post_{t}$&            &       0.187***&       1.213***&       1.030** &      -0.065***\\
                    &            &     (0.042)   &     (0.399)   &     (0.472)   &     (0.020)   \\
\addlinespace Mean pre-treatment&            &       9.902   &      36.147   &      45.643   &       0.282   \\
Obs                 &            &     889,634   &     889,200   &     889,634   &     889,200   \\
$R^{2}$             &            &        0.07   &        0.02   &        0.02   &        0.02   \\
\addlinespace \hline
\end{tabular}

\end{threeparttable}
}
\par\vspace{0.3cm}
\begin{minipage}{\textwidth} \small
 \textit{Notes}: This table reports second-stage estimates from a 2SLS variant of Equation~\eqref{eq:1}, using the 2013--2024 ACS sample of dual-parent households with children of child-rearing age. The endogenous regressor is the husband's observed post/during-COVID interaction with his own WFH uptake, $WFH_{h(i),t}$ (replacing the exogenous occupation-level treatment $\text{High-}\Delta\text{WFH}_{o(h(i))}$ used in Equation~\eqref{eq:1}), instrumented with a leave-one-out average WFH rate in the husband's occupation where we leave the individual's own WFH rate, $\overline{WFH}_{o(h(i)),-i,t}$, interacted with the post/during-COVID dummies. The corresponding reduced-form first-stage relationship is shown in \autoref{fig:acs_dynamic_DiD_spwfh_only}; first-stage coefficients are not reported here for brevity. All regressions control for the wife's age, education, and number of children, as well as the husband's education, and include state-by-time and husband's-occupation fixed effects. Standard errors, clustered at the husband's occupation level, are reported in parentheses. \textsuperscript{*} $p<0.10$; \textsuperscript{**} $p<0.05$; \textsuperscript{***} $p<0.01$.
\end{minipage}
\end{table}
\clearpage

\clearpage

%\begin{table}[H]\centering
%\caption{Heterogeneous Effects of Husbands' WFH Shock on Wives' Labor Market %Outcomes by Wives' Relative Education Level}
%\label{tab:acs_het_reledlevelquad}
%\scalebox{0.65}{
%\begin{threeparttable}
%\input{tables/acs_het_reledu_quadratic}
%\begin{tablenotes}\scriptsize
%\textit{Notes}: We replicate the analysis \autoref{tab:acs_baseline_annual} %using the 2017-2023 ACS sample. For simplicity, we only present the %coefficient for the post-Covid--by-treatment interaction. In this Table, %present the differential treatment effect when we account for a quadratic %function of a 1 standard deviation increase in the ratio of years of education %between women and their husbands. 
%\end{tablenotes} 
%\end{threeparttable}
%}
%\end{table}

%\begin{table}[H]\centering
%\caption{Heterogeneous Effects of Husbands' WFH Shock on Wives' Labor Market %Outcomes by Wives' Relative Education Level}
%\label{tab:acs_het_reledlevel}
%\scalebox{0.65}{
%\begin{threeparttable}
%\input{tables/acs_het_reledu_categorical}
%\begin{tablenotes}\scriptsize
%\textit{Notes}: We replicate the analysis \autoref{tab:acs_baseline_annual} %using different subsamples of the 2017-2023 ACS sample. For simplicity, we %only present the coefficient for the post-Covid--by-treatment interaction. In %this Table, we split the sample by whether wives have less, equal, or more %years of education than their husbands. 
%\end{tablenotes} 
%\end{threeparttable}
%}
%\end{table}

\clearpage

\begin{table}[H]\centering
\caption{Effect of Husbands' WFH Shock on Women's Labor Market Outcomes in Samples Without Children and with Older Children}
\label{tab:acs_placebo_allchild}
\scalebox{0.8}{
\begin{threeparttable}
\begin{tabular}{llcccc}
\hline
\addlinespace
& &  Log Earnings & Usual Weekly Hours & Weeks Worked & Part-time \\ 
& &  & Worked & & Status \\
\cline{3-6} \addlinespace
\multicolumn{6}{l}{A. No Children} \\
\addlinespace
$ Exposed_{o(h(i))} \times Post_{t}$&            &       0.017*  &      -0.112   &       0.103   &      -0.003   \\
                    &            &     (0.010)   &     (0.105)   &     (0.095)   &     (0.003)   \\
\addlinespace Mean pre-treatment&            &      10.014   &      39.525   &      47.502   &       0.170   \\
$\beta_1+\beta_2=0$ (p-val)&            &       0.559   &       0.134   &       0.005   &       0.886   \\
Obs                 &            &     647,467   &     647,007   &     647,467   &     647,007   \\
$R^{2}$             &            &        0.16   &        0.04   &        0.03   &        0.03   \\
\addlinespace
\multicolumn{6}{l}{B. Youngest Child is Teenager (12-17 Years Old)} \\
\addlinespace
$ Exposed_{o(h(i))} \times Post_{t}$&            &       0.046***&       0.515***&       0.178   &      -0.020***\\
                    &            &     (0.011)   &     (0.114)   &     (0.118)   &     (0.005)   \\
\addlinespace Mean pre-treatment&            &       9.940   &      37.390   &      47.238   &       0.248   \\
$\beta_1+\beta_2=0$ (p-val)&            &       0.020   &       0.210   &       0.019   &       0.001   \\
Obs                 &            &     392,468   &     392,240   &     392,468   &     392,240   \\
$R^{2}$             &            &        0.10   &        0.03   &        0.02   &        0.03   \\
\addlinespace
\multicolumn{6}{l}{C. Youngest Child is Adult (18+ Years Old)} \\
\addlinespace
$ Exposed_{o(h(i))} \times Post_{t}$&            &       0.011   &      -0.134   &      -0.006   &      -0.004   \\
                    &            &     (0.009)   &     (0.089)   &     (0.085)   &     (0.003)   \\
\addlinespace Mean pre-treatment&            &       9.963   &      39.133   &      47.371   &       0.182   \\
$\beta_1+\beta_2=0$ (p-val)&            &       0.463   &       0.412   &       0.019   &       0.761   \\
Obs                 &            &     839,451   &     838,870   &     839,451   &     838,870   \\
$R^{2}$             &            &        0.16   &        0.04   &        0.03   &        0.03   \\
\addlinespace \hline
\end{tabular}

\end{threeparttable}
}
\par\vspace{0.3cm}
\begin{minipage}{\textwidth} \small
 \textit{Notes}: This table reports estimates from Equation~\eqref{eq:1}, using the 2013--2024 ACS sample of married dual-parent households (dropping the child-age restriction of the main sample) for three placebo subsamples in which the husband's WFH shock should not operate through young-child childcare: Panel A restricts to households with no children; Panel B restricts to households whose youngest child is 12--17 years old; Panel C restricts to households whose youngest child is over 17. Treatment is defined as in the main sample (Subsection~\ref{subsection:treatment_defn}). For brevity, we report only the post-COVID (2022 onward) interaction coefficient. All regressions include standard demographic controls (wife's age, number of children, and both spouses' education) and state-by-time and husband's-occupation fixed effects. Standard errors, clustered at the husband's occupation level, are reported in parentheses. \textsuperscript{*} $p<0.10$; \textsuperscript{**} $p<0.05$; \textsuperscript{***} $p<0.01$.
\end{minipage}
\end{table}

\clearpage
\begin{table}[H]\centering
\caption{Effect of Wives' WFH Shocks on Their Husbands' Labor Market Outcomes}
\label{tab:acs_men_annual}
\scalebox{0.6}{
\begin{threeparttable}
\begin{tabular}{llcccccc}
\hline
\addlinespace
& &  Annual Employment & Log Earnings & Log Wage & Weekly Work Hours & Weeks Worked & Part-time Status \\ 
\cline{3-8} \addlinespace
\addlinespace
$ Exposed_{o(h(i))} \times Post_{t}$&            &       0.002   &      -0.009   &      -0.003   &      -0.265***&      -0.060   &      -0.000   \\
                    &            &     (0.002)   &     (0.007)   &     (0.006)   &     (0.094)   &     (0.052)   &     (0.002)   \\
\addlinespace Mean pre-treatment&            &       0.914   &      10.541   &       2.898   &      44.518   &      49.396   &       0.049   \\
Obs                 &            &     980,278   &     895,563   &     893,670   &     893,670   &     895,563   &     893,670   \\
$R^2$               &            &        0.02   &        0.20   &        0.22   &        0.02   &        0.02   &        0.01   \\
\addlinespace \hline
\end{tabular}

\end{threeparttable}
}
\par\vspace{0.3cm}
\begin{minipage}{\textwidth} \small
\textit{Notes}: This table reports estimates from a version of Equation~\eqref{eq:1} in which the roles of the spouses are reversed: the treatment indicator, $\text{High-}\Delta\text{WFH}_{o(i)}$, is defined at the level of the \emph{wife's} own occupation (above-median change in WFH probability between 2019 and 2023--2024, as in Subsection~\ref{subsection:treatment_defn}), and the outcomes are the husband's labor market outcomes. The sample is the 2013--2024 ACS sample of dual-parent households with children of child-rearing age. For brevity, we report only the post-COVID (2022 onward) interaction coefficient. All regressions include standard demographic controls (wife's age, number of children, and both spouses' education) and state-by-time and wife's-occupation fixed effects. Standard errors, clustered at the wife's occupation level, are reported in parentheses. \textsuperscript{*} $p<0.10$; \textsuperscript{**} $p<0.05$; \textsuperscript{***} $p<0.01$.
\end{minipage}
\end{table}

\clearpage
\begin{table}[H]\centering
\caption{Effects of Husbands' WFH Shock on Labor Gender Gaps}
\label{tab:acs_gender_gap}
\scalebox{0.85}{
\begin{threeparttable}
\begin{tabular}{lcccc}
\hline \addlinespace
 & Log Earnings & Usual Weekly Hours Worked & Weeks Worked & Part-time Status \\
\cline{2-5} \addlinespace
Counterfactual gender gap & 0.497 & 0.822 & 0.925 & 6.951 \\
Post-treatment gender gap & 0.527 & 0.846 & 0.936 & 6.852 \\
\addlinespace Change in gap (\%) &   5.9 &   3.0 &   1.2 &  -1.4 \\
\addlinespace \hline
\end{tabular}

\end{threeparttable}
}
\par\vspace{0.3cm}
\begin{minipage}{\textwidth} \small
\textit{Notes}: This table presents an accounting exercise that quantifies the variation of labor gender gaps due to treatment, following \cite{kleven2019children}. Treatment, $\text{High-}\Delta\text{WFH}_{o(h(i))}$, is defined based on the change in the probability of WFH before and after the COVID-19 pandemic using job postings data at the husband's occupation level (Census 2010; see Subsection~\ref{subsection:treatment_defn}). The sample is the 2013--2024 ACS sample of dual-parent households with children of child-rearing age. Appendix \ref{app:gender_gap_accounting} describes the step-by-step construction.
 \end{minipage}
\end{table}
\clearpage

\clearpage
\begin{table}[H]\centering
\caption{Effects of Husbands' WFH Shock on Wives' Intensive Margin of Labor Supply after Adding Wives' Occupation Fixed Effects}
\label{tab:acs_intensive_with_WomenOccFE}
\scalebox{0.85}{
\begin{threeparttable}
\begin{tabular}{llcccc}
\hline
\addlinespace
& &  Log Earnings & Usual Weekly Hours & Weeks Worked & Part-time \\ 
& &  & Worked & & Status \\
\cline{3-6} \addlinespace
\addlinespace
$ Exposed_{o(h(i))} \times Post_{t}$&            &       0.050***&       0.507***&       0.501***&      -0.028***\\
                    &            &     (0.010)   &     (0.092)   &     (0.085)   &     (0.004)   \\
\addlinespace Mean pre-treatment&            &       9.902   &      36.276   &      45.635   &       0.279   \\
Obs                 &            &     949,441   &     948,944   &     949,441   &     948,944   \\
$R^{2}$             &            &        0.34   &        0.16   &        0.09   &        0.16   \\
\addlinespace \hline
\end{tabular}

\end{threeparttable}
}
\par\vspace{0.3cm}
\begin{minipage}{\textwidth} \small
\textit{Notes}: This table presents estimates from Equation~\eqref{eq:1} on the sample of employed wives, additionally including wife's-occupation fixed effects alongside the baseline state-by-time and husband's-occupation fixed effects. Treatment, $\text{High-}\Delta\text{WFH}_{o(h(i))}$, is defined based on the change in the probability of WFH before and after the COVID-19 pandemic using job postings data at the husband's occupation level (Census 2010; see Subsection~\ref{subsection:treatment_defn}). The sample is the 2013--2024 ACS sample of dual-parent households with children of child-rearing age. For brevity, we report only the post-COVID (2022 onward) interaction coefficient. All regressions control for the wife's age, education, and number of children, as well as the husband's education. Standard errors, clustered at the husband's occupation level, are reported in parentheses. \textsuperscript{*} $p<0.10$; \textsuperscript{**} $p<0.05$; \textsuperscript{***} $p<0.01$.
 \end{minipage}
\end{table}
\clearpage

%\begin{table}[H]\centering
%\caption{Effect of Wives' WFH Occupation Shock on Their Own Labor Market Outcomes}
%\label{tab:acs_ownshock_annual}
%\scalebox{0.9}{
%\begin{threeparttable}
%\input{tables/acs_ownshock_annual}\end{threeparttable}
%}
%\par\vspace{0.3cm}
%\begin{minipage}{\textwidth} \small
%\textit{Notes}: This table adapts our baseline cross-sectional specification to estimate the effect of women's WFH shocks on their own labor market outcomes, using the 2013–2024 ACS sample of dual-parent households with children of child-rearing age. For simplicity we only present the coefficient of the post-Covid interaction. All regressions include fixed effects for state-by-time and women's occupation fixed effects. Controls include the wife’s age, education, and number of children, as well as the husband’s education. Standard errors are clustered at the wife's occupation level.
%\end{minipage}
%\end{table}
%\clearpage

\begin{table}[H]\centering
\caption{Decomposition Analysis: Effects of Male and Female Spouses' Occupation and  WFH Shocks on Women Labor Supply}
\label{tab:acs_sp_own_occ_estimates}
\scalebox{0.7}{
\begin{threeparttable}
\begin{tabular}{llcccccc}
\hline
\addlinespace
& &  Log               & Usual Weekly & Weeks  & Part-time & WFH       & Commuting     \\ 
& &  Earnings  & Hours Worked & Worked & Status        & Uptake& Time          \\
\cline{3-8} \addlinespace
\addlinespace
$ Exposed_{o(h(i))} \times Post_{t} $&            &       0.032***&       0.630***&       0.519***&      -0.029***&       0.025***&      -0.001***\\
                    &            &     (0.012)   &     (0.109)   &     (0.136)   &     (0.005)   &     (0.004)   &     (0.000)   \\
$ Exposed_{o(i)} \times Post_{t} $&            &       0.009   &       0.163*  &       0.326** &      -0.009** &       0.123***&      -0.003***\\
                    &            &     (0.011)   &     (0.094)   &     (0.153)   &     (0.004)   &     (0.004)   &     (0.000)   \\
$ Exposed_{o(h(i))} \times Exposed_{o(i)} \times Post_{t} $&            &       0.018   &      -0.282** &      -0.186   &       0.005   &       0.035***&      -0.001***\\
                    &            &     (0.015)   &     (0.136)   &     (0.190)   &     (0.006)   &     (0.008)   &     (0.000)   \\
\addlinespace Mean pre-treatment&            &       9.905   &      36.275   &      45.652   &       0.279   &       0.075   &       0.017   \\
Obs                 &            &     919,450   &     918,961   &     919,450   &     918,961   &     861,911   &     861,845   \\
$R^2$               &            &        0.40   &        0.23   &        0.17   &        0.23   &        0.25   &        0.14   \\
\addlinespace \hline
\end{tabular}
\end{threeparttable}
}
\par\vspace{0.3cm}
\begin{minipage}{\textwidth} \small
    \textit{Notes}: This table reports estimates from Equation~\eqref{eq:3}, replacing the wife's college-major-based treatment used in \autoref{tab:acs_edfield_estimates} with a treatment defined at the wife's own occupation, $\text{High-}\Delta\text{WFH}_{o(i)}$. Both treatment indicators equal one if the corresponding occupation ($o(h(i))$ for the husband, $o(i)$ for the wife) had an above-median increase in the share of remote/hybrid job postings between 2019 and 2023--2024 (\citealt{hansen2023remote}; see Subsection~\ref{subsection:treatment_defn}). This specification is estimated only on employed wives and unemployed wives reporting their most recent occupation (50.6\% of unemployed wives). The sample is the 2013--2024 ACS sample of dual-parent households with children of child-rearing age. For brevity, we report only the post-COVID (2022 onward) interaction coefficients. All regressions include fixed effects for state-by-time, husband's occupation, and wife's occupation, and control for the wife's age, education, and number of children, as well as the husband's education. Standard errors, clustered at the husband's occupation level, are reported in parentheses. \textsuperscript{*} $p<0.10$; \textsuperscript{**} $p<0.05$; \textsuperscript{***} $p<0.01$.ors are clustered at the husband's occupation level. 
 \end{minipage}
\end{table}

%\begin{table}[H]\centering
%\caption{Effect of Husbands' WFH Shock on Wives' Probability of Working Remotely}
%\label{tab:cps_telwrk_occ_combined}
%\scalebox{0.55}{
%\begin{threeparttable}
%\input{tables/cps_telwrk_occ_combined}
%\begin{tablenotes}{\em{Notes:}} We regress women's self-reported teleworking against their spouses' treatment status using our cross-section and panel specifications. Data comes from our household sample of the 2023-2024 monthly CPS. Respondents are asked if they work from home in their current occupation (post-Covid observation) and are asked whether they worked from home on February 2020 (pre-Covid observation). We use this two data points to build a pseudo-panel. It is implicitly assumed that they are working on same occupation post-Covid as they did pre-Covid. In the first three columns, we include fixed effects for month-year-state and spouse occupation, hence we label them as the cross-section regression. In the last three columns, we replace occupation with household fixed effects to exploit the panel dimension of the recalled data, hence we label them as the panel regressions. For each specification, we look at results in our regular sample of households, households where the male spouse is on an employee arrangement, and where they are self-employed.    \end{tablenotes}
%\end{threeparttable}
%}
%\end{table}

\clearpage

\begin{table}[H]\centering
\caption{Effect of Husbands' WFH Shock on Wives' Labor Market Outcomes: 
         Robustness to Husband's Education Trends}
\label{tab:acs_appendix_trends_spedlevel}
\scalebox{0.85}{
\begin{threeparttable}
\begin{tabular}{lcccc}
\hline
\addlinespace
& Log Earnings & Usual Weekly Hours & Weeks Worked & Part-time \\
& & Worked & & Status \\
\cline{2-5}
\addlinespace
\multicolumn{5}{l}{A. Baseline} \\
\addlinespace
$Exposed_{o(h(i))} \times Post_{t}$&       0.048***&       0.466***&       0.500***&      -0.024***\\
                    &     (0.010)   &     (0.096)   &     (0.089)   &     (0.004)   \\
\addlinespace Mean pre-treatment&       9.915   &      36.294   &      45.671   &       0.277   \\
Obs                 &     949,443   &     948,946   &     949,443   &     948,946   \\
$R^{2}$             &        0.15   &        0.03   &        0.03   &        0.03   \\
\addlinespace
\multicolumn{5}{l}{B. Non-parametric Spouse Education Trends} \\
\addlinespace
$Exposed_{o(h(i))} \times Post_{t}$&       0.044***&       0.380***&       0.456***&      -0.018***\\
                    &     (0.011)   &     (0.101)   &     (0.092)   &     (0.004)   \\
\addlinespace Mean pre-treatment&       9.915   &      36.294   &      45.671   &       0.277   \\
Obs                 &     949,443   &     948,946   &     949,443   &     948,946   \\
$R^{2}$             &        0.15   &        0.03   &        0.03   &        0.03   \\
\addlinespace
\multicolumn{5}{l}{C. Linear Spouse Education Trends} \\
\addlinespace
$Exposed_{o(h(i))} \times Post_{t}$&       0.044***&       0.395***&       0.449***&      -0.020***\\
                    &     (0.010)   &     (0.101)   &     (0.088)   &     (0.004)   \\
\addlinespace Mean pre-treatment&       9.915   &      36.294   &      45.671   &       0.277   \\
Obs                 &     949,443   &     948,946   &     949,443   &     948,946   \\
$R^{2}$             &        0.15   &        0.03   &        0.03   &        0.03   \\
\addlinespace
\multicolumn{5}{l}{D. \citet{jakobsen2020wealth} Pre-trend Adjusted (Spouse Education)} \\
\addlinespace
$Exposed_{o(h(i))} \times Post_{t}$&       0.047***&       0.459***&       0.468***&      -0.023***\\
                    &     (0.010)   &     (0.095)   &     (0.088)   &     (0.004)   \\
\addlinespace Mean pre-treatment&       9.920   &      36.316   &      45.784   &       0.272   \\
Obs                 &     949,443   &     948,946   &     949,443   &     948,946   \\
$R^{2}$             &        0.15   &        0.03   &        0.03   &        0.03   \\
\addlinespace \hline
\end{tabular}

\end{threeparttable}
}
\par\vspace{0.3cm}
\begin{minipage}{\textwidth} \small
 \textit{Notes}: This table assesses the robustness of the baseline estimates in \autoref{tab:acs_baseline_annual} to the inclusion of trends in the husband's education level. Education categories are below high school, high school, and above high school. Panel A replicates the baseline estimates from \autoref{tab:acs_baseline_annual}. Panel B includes non-parametric trends in the form of education-by-year fixed effects. Panel C adds linear time trends interacted with education category. Panel D controls for pre-trends in a linear fashion following \citet{jakobsen2020wealth}, as described in Section~\ref{section:robustness}. All specifications use the 2013--2024 ACS sample of dual-parent households with children of child-rearing age and control for the wife's age, education, and number of children, as well as the husband's education. Specifications include state-by-year and husband's-occupation fixed effects. Standard errors, clustered at the husband's occupation level, are reported in parentheses. \textsuperscript{*} $p<0.10$; \textsuperscript{**} $p<0.05$; \textsuperscript{***} $p<0.01$.
\end{minipage}
\end{table}
\clearpage

\begin{table}[H]\centering
\caption{Effect of Husbands' WFH Shock on Wives' Labor Market Outcomes: 
         Robustness to Husband's Broad Occupation Group Trends}
\label{tab:acs_appendix_trends_broadspocc}
\scalebox{0.85}{
\begin{threeparttable}
\begin{tabular}{lcccc}
\hline
\addlinespace
& Log Earnings & Usual Weekly Hours & Weeks Worked & Part-time \\
& & Worked & & Status \\
\cline{2-5}
\addlinespace
\multicolumn{5}{l}{A. Baseline} \\
\addlinespace
$Exposed_{o(h(i))} \times Post_{t}$&       0.048***&       0.466***&       0.500***&      -0.024***\\
                    &     (0.010)   &     (0.096)   &     (0.089)   &     (0.004)   \\
\addlinespace Mean pre-treatment&       9.915   &      36.294   &      45.671   &       0.277   \\
Obs                 &     949,443   &     948,946   &     949,443   &     948,946   \\
$R^{2}$             &        0.15   &        0.03   &        0.03   &        0.03   \\
\addlinespace
\multicolumn{5}{l}{B. Non-parametric Broad Occupation Group Trends} \\
\addlinespace
$Exposed_{o(h(i))} \times Post_{t}$&       0.035***&       0.285** &       0.427***&      -0.012***\\
                    &     (0.013)   &     (0.114)   &     (0.108)   &     (0.005)   \\
\addlinespace Mean pre-treatment&       9.915   &      36.294   &      45.671   &       0.277   \\
Obs                 &     949,443   &     948,946   &     949,443   &     948,946   \\
$R^{2}$             &        0.15   &        0.03   &        0.03   &        0.03   \\
\addlinespace
\multicolumn{5}{l}{C. Linear Broad Occupation Group Trends} \\
\addlinespace
$Exposed_{o(h(i))} \times Post_{t}$&       0.036***&       0.320***&       0.416***&      -0.015***\\
                    &     (0.012)   &     (0.100)   &     (0.098)   &     (0.004)   \\
\addlinespace Mean pre-treatment&       9.915   &      36.294   &      45.671   &       0.277   \\
Obs                 &     949,443   &     948,946   &     949,443   &     948,946   \\
$R^{2}$             &        0.15   &        0.03   &        0.03   &        0.03   \\
\addlinespace
\multicolumn{5}{l}{D. \citet{jakobsen2020wealth} Pre-trend Adjusted (Broad Occupation Group)} \\
\addlinespace
$Exposed_{o(h(i))} \times Post_{t}$&       0.038***&       0.424***&       0.440***&      -0.021***\\
                    &     (0.010)   &     (0.094)   &     (0.087)   &     (0.004)   \\
\addlinespace Mean pre-treatment&       9.905   &      36.256   &      45.616   &       0.280   \\
Obs                 &     949,443   &     948,946   &     949,443   &     948,946   \\
$R^{2}$             &        0.15   &        0.03   &        0.03   &        0.04   \\
\addlinespace \hline
\end{tabular}

\end{threeparttable}
}
\par\vspace{0.3cm}
\begin{minipage}{\textwidth} \small
 \textit{Notes}: This table assesses the robustness of the baseline estimates in \autoref{tab:acs_baseline_annual} to the inclusion of trends in the husband's broad occupation group. Broad occupation groups are high-skill/professional, services and sales, and blue-collar/manual. Panel A replicates the baseline estimates from \autoref{tab:acs_baseline_annual}. Panel B includes non-parametric trends in the form of broad-occupation-by-year fixed effects. Panel C adds linear time trends interacted with broad occupation group. Panel D controls for pre-trends in a linear fashion following \citet{jakobsen2020wealth}, as described in Section~\ref{section:robustness}. All specifications use the 2013--2024 ACS sample of dual-parent households with children of child-rearing age and control for the wife's age, education, and number of children, as well as the husband's education. Specifications include state-by-year and husband's-occupation fixed effects. Standard errors, clustered at the husband's occupation level, are reported in parentheses. \textsuperscript{*} $p<0.10$; \textsuperscript{**} $p<0.05$; \textsuperscript{***} $p<0.01$.
\end{minipage}
\end{table}
\clearpage

\begin{table}[H]\centering
\caption{Robustness of ACS Results to Alternative Fixed Effects Specifications}
\label{tab:acs_estimates_change_FE}
\scalebox{0.6}{
\begin{threeparttable}
\begin{tabular}{llcccc}
\hline
\addlinespace
& & Log Earnings & Usual work hours & Weeks Worked & Part-time status  \\ \cline{3-6} \\
\addlinespace
\multicolumn{6}{l}{\emph{A. FE: Year + Spouse Occupation (Group of 16) + State}}\\
\addlinespace
$ Exposed_{o(h(i))} \times Post_{t}$&            &       0.075***&       0.332***&       0.359***&      -0.018***\\
                    &            &     (0.010)   &     (0.105)   &     (0.093)   &     (0.005)   \\
\addlinespace Mean pre-treatment&            &       9.902   &      36.276   &      45.635   &       0.279   \\
Obs                 &            &     788,306   &     787,899   &     788,306   &     787,899   \\
$R^2$               &            &        0.14   &        0.03   &        0.03   &        0.03   \\
\multicolumn{6}{l}{\emph{B. FE: Year-state + Spouse Occupation (Group of 16)}}\\
\addlinespace
$ Exposed_{o(h(i))} \times Post_{t}$&            &       0.073***&       0.314***&       0.352***&      -0.017***\\
                    &            &     (0.010)   &     (0.106)   &     (0.093)   &     (0.005)   \\
\addlinespace Mean pre-treatment&            &       9.902   &      36.276   &      45.635   &       0.279   \\
Obs                 &            &     788,306   &     787,899   &     788,306   &     787,899   \\
$R^2$               &            &        0.15   &        0.03   &        0.03   &        0.03   \\
\multicolumn{6}{l}{\emph{C. FE: Year-state + Spouse Occupation (Group of 16)-by-state}}\\
\addlinespace
$ Exposed_{o(h(i))} \times Post_{t}$&            &       0.075***&       0.328***&       0.371***&      -0.017***\\
                    &            &     (0.010)   &     (0.108)   &     (0.090)   &     (0.005)   \\
\addlinespace Mean pre-treatment&            &       9.902   &      36.276   &      45.635   &       0.279   \\
Obs                 &            &     788,305   &     787,898   &     788,305   &     787,898   \\
$R^2$               &            &        0.15   &        0.03   &        0.03   &        0.03   \\
\multicolumn{6}{l}{\emph{D. FE: Year + Spouse Occupation (4d) + State}}\\
\addlinespace
$ Exposed_{o(h(i))} \times Post_{t}$&            &       0.051***&       0.481***&       0.508***&      -0.025***\\
                    &            &     (0.010)   &     (0.094)   &     (0.089)   &     (0.004)   \\
\addlinespace Mean pre-treatment&            &       9.902   &      36.276   &      45.635   &       0.279   \\
Obs                 &            &     788,306   &     787,899   &     788,306   &     787,899   \\
$R^2$               &            &        0.15   &        0.03   &        0.03   &        0.03   \\
\multicolumn{6}{l}{\emph{E. FE: Year-state + Spouse Occupation (4d) (Baseline)}}\\
\addlinespace
$ Exposed_{o(h(i))} \times Post_{t}$&            &       0.048***&       0.463***&       0.500***&      -0.024***\\
                    &            &     (0.010)   &     (0.096)   &     (0.089)   &     (0.004)   \\
\addlinespace Mean pre-treatment&            &       9.902   &      36.276   &      45.635   &       0.279   \\
Obs                 &            &     788,306   &     787,899   &     788,306   &     787,899   \\
$R^2$               &            &        0.15   &        0.03   &        0.03   &        0.03   \\
\multicolumn{6}{l}{\emph{F. FE: Year-state + Spouse Occupation (4d)-by-state}}\\
\addlinespace
$ Exposed_{o(h(i))} \times Post_{t}$&            &       0.050***&       0.497***&       0.503***&      -0.025***\\
                    &            &     (0.010)   &     (0.099)   &     (0.089)   &     (0.004)   \\
\addlinespace Mean pre-treatment&            &       9.902   &      36.277   &      45.636   &       0.279   \\
Obs                 &            &     785,635   &     785,228   &     785,635   &     785,228   \\
$R^2$               &            &        0.18   &        0.07   &        0.07   &        0.07   \\
\multicolumn{6}{l}{\emph{G. FE: Year-MSA + Spouse Occupation (4d)}}\\
\addlinespace
$ Exposed_{o(h(i))} \times Post_{t}$&            &       0.050***&       0.464***&       0.487***&      -0.025***\\
                    &            &     (0.012)   &     (0.106)   &     (0.110)   &     (0.004)   \\
\addlinespace Mean pre-treatment&            &       9.984   &      36.372   &      45.696   &       0.277   \\
Obs                 &            &     596,537   &     596,241   &     596,537   &     596,241   \\
$R^2$               &            &        0.16   &        0.04   &        0.04   &        0.04   \\
\addlinespace \hline
\end{tabular}

\end{threeparttable}
}
\par\vspace{0.3cm}
\begin{minipage}{\textwidth}
    \small
 \textit{Notes}: This table presents estimates from alternative versions of Equation~\eqref{eq:1} using the 2013--2024 ACS sample of dual-parent households with children of child-rearing age, varying the fixed-effect (FE) structure. For brevity, we report only the post-COVID (2022 onward) interaction coefficient. Panel E reports the baseline specification (state-by-time and husband's-occupation fixed effects). Panel A includes separate fixed effects for month-year, state, and broad occupation group (16 categories, e.g., Business/Finance, Education Professionals, Healthcare, Sales, Construction). Panel B adds state-by-time fixed effects to Panel A, and Panel C uses state-by-time and broad-occupation-group fixed effects. Panels D--F replicate Panels A--C, respectively, replacing the broad occupation groups with detailed 4-digit Census occupation fixed effects. Panel G uses MSA fixed effects as an alternative to state fixed effects. All specifications control for the wife's age, education, and number of children, as well as the husband's education. Standard errors, clustered at the husband's occupation level, are reported in parentheses. \textsuperscript{*} $p<0.10$; \textsuperscript{**} $p<0.05$; \textsuperscript{***} $p<0.01$.
\end{minipage}
\end{table}

\clearpage

\begin{table}[H]\centering
\caption{Effect of Husbands' WFH Shock on Wives Labor Market Outcomes, Controlling for Spouse Employment Status}
\label{tab:acs_baseline_annual_spemp_annual_control}
\scalebox{0.9}{
\begin{threeparttable}
\begin{tabular}{llcccc}
\hline
\addlinespace
& &  Log Earnings & Usual Weekly Hours & Weeks Worked & Part-time \\ 
& &  & Worked & & Status \\
\cline{3-6} \addlinespace
\addlinespace
$ Exposed_{o(h(i))} \times Post_{t}$&            &       0.049***&       0.475***&       0.504***&      -0.024***\\
                    &            &     (0.010)   &     (0.096)   &     (0.089)   &     (0.004)   \\
\addlinespace Mean pre-treatment&            &       9.902   &      36.276   &      45.635   &       0.279   \\
Obs                 &            &     949,443   &     948,946   &     949,443   &     948,946   \\
$R^{2}$             &            &        0.15   &        0.03   &        0.03   &        0.04   \\
\addlinespace \hline
\end{tabular}

\end{threeparttable}
}
\par\vspace{0.3cm}
\begin{minipage}{\textwidth} \small
  \textit{Notes}: This table reports estimates from Equation~\eqref{eq:1} using the 2013--2024 ACS sample of dual-parent households with children of child-rearing age, additionally controlling for the husband's annual employment status. The dependent variables and treatment definition are as in \autoref{tab:acs_baseline_annual}: annual employment, log annual earnings, log hourly wage, usual weekly hours worked, weeks worked, and part-time status for the wife, all measured over the preceding 12 months; intensive-margin outcomes are defined only for employed wives. All regressions control for the wife's age, education, and number of children, as well as the husband's education and employment status. Specifications include state-by-time and husband's-occupation fixed effects. Standard errors, clustered at the husband's occupation level, are reported in parentheses. \textsuperscript{*} $p<0.10$; \textsuperscript{**} $p<0.05$; \textsuperscript{***} $p<0.01$.
 \end{minipage}
\end{table}

\clearpage
\begin{table}[H]\centering
\caption{Robustness of ACS Results to Sequential Removal of Demographic Controls}
\label{tab:acs_change_controls1}
\centering
\scalebox{0.8}{
\begin{threeparttable}
\begin{tabular}{llccccc}
\hline
\addlinespace
 & & Control Set 1 & Control Set 2 & Control Set 3 & Control Set 4 & Control Set 5 \\
 & &  & & & & (Baseline) \\
\cline{1-7} \\ 
\multicolumn{7}{l}{Outcome: Log Earnings} \\
\addlinespace
$ Exposed_{o(h(i))} \times Post_{t}$&            &       0.027** &       0.026** &       0.038***&       0.048***&       0.048***\\
                    &            &     (0.010)   &     (0.010)   &     (0.010)   &     (0.010)   &     (0.010)   \\
\addlinespace Mean pre-treatment&            &       9.902   &       9.902   &       9.902   &       9.902   &       9.902   \\
Obs                 &            &     949,443   &     949,443   &     949,443   &     949,443   &     949,443   \\
$R^2$               &            &        0.08   &        0.09   &        0.12   &        0.15   &        0.15   \\
\addlinespace
\multicolumn{7}{l}{Outcome: Usual work hours} \\
\addlinespace
$ Exposed_{o(h(i))} \times Post_{t}$&            &       0.393***&       0.387***&       0.438***&       0.466***&       0.466***\\
                    &            &     (0.093)   &     (0.094)   &     (0.095)   &     (0.096)   &     (0.096)   \\
\addlinespace Mean pre-treatment&            &      36.276   &      36.276   &      36.276   &      36.276   &      36.276   \\
Obs                 &            &     948,946   &     948,946   &     948,946   &     948,946   &     948,946   \\
$R^2$               &            &        0.02   &        0.02   &        0.03   &        0.03   &        0.03   \\
\addlinespace
\multicolumn{7}{l}{Outcome: Weeks Worked} \\
\addlinespace
$ Exposed_{o(h(i))} \times Post_{t}$&            &       0.372***&       0.366***&       0.443***&       0.498***&       0.500***\\
                    &            &     (0.090)   &     (0.089)   &     (0.089)   &     (0.089)   &     (0.089)   \\
\addlinespace Mean pre-treatment&            &      45.635   &      45.635   &      45.635   &      45.635   &      45.635   \\
Obs                 &            &     949,443   &     949,443   &     949,443   &     949,443   &     949,443   \\
$R^2$               &            &        0.01   &        0.02   &        0.03   &        0.03   &        0.03   \\
\addlinespace
\multicolumn{7}{l}{Outcome: Part-time status} \\
\addlinespace
$ Exposed_{o(h(i))} \times Post_{t}$&            &      -0.021***&      -0.021***&      -0.023***&      -0.024***&      -0.024***\\
                    &            &     (0.004)   &     (0.004)   &     (0.004)   &     (0.004)   &     (0.004)   \\
\addlinespace Mean pre-treatment&            &       0.279   &       0.279   &       0.279   &       0.279   &       0.279   \\
Obs                 &            &     948,946   &     948,946   &     948,946   &     948,946   &     948,946   \\
$R^2$               &            &        0.02   &        0.02   &        0.03   &        0.03   &        0.03   \\
\hline \addlinespace
Control: Number of children & & No & Yes & Yes & Yes & Yes \\ 
Control: Women's age (5y sbin) & & No & No & Yes & Yes & Yes \\ 
Control: Women's education level & & No & No & No & Yes & Yes \\ 
Control: Husband's education level & & No & No & No & No & Yes  \\ 
\addlinespace \hline
\end{tabular}

\end{threeparttable}
}
\par\vspace{0.3cm}
\begin{minipage}{\textwidth} \small
\textit{Notes}: This table reports estimates from Equation~\eqref{eq:1} on the 2013--2024 ACS sample of dual-parent households with children of child-rearing age, sequentially removing demographic controls. The baseline specification (Column 5) controls for the wife's age (in five-year bins), number of children, and both spouses' education levels. Columns 1--4 sequentially remove these controls, one at a time, to assess the sensitivity of the results to covariate inclusion. All specifications include state-by-time and husband's-occupation fixed effects. Standard errors, clustered at the husband's occupation level, are reported in parentheses. \textsuperscript{*} $p<0.10$; \textsuperscript{**} $p<0.05$; \textsuperscript{***} $p<0.01$. 
\end{minipage}
\end{table}

\clearpage
\begin{table}[H]\centering
\caption{Robustness of ACS Results to Adding Controls for Children's Age}
\label{tab:acs_change_controls2}
\centering
\scalebox{0.8}{
\begin{threeparttable}
\begin{tabular}{llccccc}
\hline
\addlinespace
 & & Control Set 1 & Control Set 2 & Control Set 3 & Control Set 4 & Control Set 5 \\
 & & (Baseline) & & & & \\
\cline{1-7} \\ 
\multicolumn{7}{l}{Outcome: Log Earnings} \\
\addlinespace
$ Exposed_{o(h(i))} \times Post_{t}$&            &       0.048***&       0.048***&       0.049***&       0.048***&       0.048***\\
                    &            &     (0.010)   &     (0.010)   &     (0.010)   &     (0.010)   &     (0.010)   \\
\addlinespace Mean pre-treatment&            &       9.902   &       9.902   &       9.902   &       9.902   &       9.902   \\
Obs                 &            &     949,443   &     949,443   &     949,443   &     949,443   &     949,443   \\
$R^2$               &            &        0.15   &        0.15   &        0.15   &        0.15   &        0.15   \\
\addlinespace
\multicolumn{7}{l}{Outcome: Usual work hours} \\
\addlinespace
$ Exposed_{o(h(i))} \times Post_{t}$&            &       0.466***&       0.461***&       0.477***&       0.471***&       0.468***\\
                    &            &     (0.096)   &     (0.095)   &     (0.096)   &     (0.095)   &     (0.095)   \\
\addlinespace Mean pre-treatment&            &      36.276   &      36.276   &      36.276   &      36.276   &      36.276   \\
Obs                 &            &     948,946   &     948,946   &     948,946   &     948,946   &     948,946   \\
$R^2$               &            &        0.03   &        0.03   &        0.03   &        0.03   &        0.03   \\
\addlinespace
\multicolumn{7}{l}{Outcome: Weeks Worked} \\
\addlinespace
$ Exposed_{o(h(i))} \times Post_{t}$&            &       0.500***&       0.488***&       0.507***&       0.496***&       0.496***\\
                    &            &     (0.089)   &     (0.088)   &     (0.089)   &     (0.089)   &     (0.089)   \\
\addlinespace Mean pre-treatment&            &      45.635   &      45.635   &      45.635   &      45.635   &      45.635   \\
Obs                 &            &     949,443   &     949,443   &     949,443   &     949,443   &     949,443   \\
$R^2$               &            &        0.03   &        0.04   &        0.03   &        0.04   &        0.04   \\
\addlinespace
\multicolumn{7}{l}{Outcome: Part-time status} \\
\addlinespace
$ Exposed_{o(h(i))} \times Post_{t}$&            &      -0.024***&      -0.024***&      -0.025***&      -0.024***&      -0.024***\\
                    &            &     (0.004)   &     (0.004)   &     (0.004)   &     (0.004)   &     (0.004)   \\
\addlinespace Mean pre-treatment&            &       0.279   &       0.279   &       0.279   &       0.279   &       0.279   \\
Obs                 &            &     948,946   &     948,946   &     948,946   &     948,946   &     948,946   \\
$R^2$               &            &        0.03   &        0.04   &        0.04   &        0.04   &        0.04   \\
\hline \addlinespace
Control: Baseline controls & & Yes & Yes & Yes & Yes & Yes \\ 
Control: Age youngest child & & No & Yes & No & Yes & Yes \\ 
Control: Age oldest child (raw) & & No & No & Yes & Yes & No \\ 
Control: Age oldest child (4y bins) & & No & No & No & No & Yes \\ 
\addlinespace \hline
\end{tabular}

\end{threeparttable}
}
\par\vspace{0.3cm}
\begin{minipage}{\textwidth} \small
\textit{Notes}: This table reports estimates from Equation~\eqref{eq:1} using the 2013--2024 ACS sample of dual-parent households with children of child-rearing age. Column 1 reproduces the baseline specification (controls for the wife's age in five-year bins, number of children, and both spouses' education levels). Columns 2--5 sequentially add controls for the age of the youngest and oldest child, both as continuous variables and in four-year bins. All regressions include state-by-time and husband's-occupation fixed effects. Standard errors, clustered at the husband's occupation level, are reported in parentheses. \textsuperscript{*} $p<0.10$; \textsuperscript{**} $p<0.05$; \textsuperscript{***} $p<0.01$.
\end{minipage}
\end{table}

\clearpage

\begin{table}[H]\centering
\caption{Robustness to Excluding Households with Self-Employed Husbands}
\label{tab:acs_baseline_estimates_no_self_emp}
\centering
\scalebox{0.85}{
\begin{threeparttable}
\begin{tabular}{llcccc}
\hline
\addlinespace
& & Log Earnings & Usual Weekly Hours & Weeks Worked & Part-time \\ 
& &   & Worked & & Status \\
\cline{3-6} \addlinespace
\addlinespace
$ Exposed_{o(h(i))} \times Post_{t}$&            &       0.048***&       0.466***&       0.500***&      -0.024***\\
                    &            &     (0.010)   &     (0.096)   &     (0.089)   &     (0.004)   \\
\addlinespace Mean pre-treatment&            &       9.915   &      36.294   &      45.671   &       0.277   \\
Obs                 &            &     949,443   &     948,946   &     949,443   &     948,946   \\
$R^2$               &            &        0.15   &        0.03   &        0.03   &        0.03   \\
\addlinespace \hline
\end{tabular}

\end{threeparttable}
}
\par\vspace{0.3cm}
\begin{minipage}{\textwidth} \small
\textit{Notes}: This table reports estimates from Equation~\eqref{eq:1} using the 2013--2024 ACS sample of dual-parent households with children of child-rearing age, restricted to households where the husband is classified as an employee (excluding households where the husband is self-employed). For brevity, we report only the post-COVID (2022 onward) interaction coefficient. Treatment is defined based on the husband's occupation-level WFH shock, as in the baseline specification (Subsection~\ref{subsection:treatment_defn}). All regressions include state-by-time and husband's-occupation fixed effects, and control for the wife's age, number of children, and both spouses' education levels. Standard errors, clustered at the husband's occupation level, are reported in parentheses. \textsuperscript{*} $p<0.10$; \textsuperscript{**} $p<0.05$; \textsuperscript{***} $p<0.01$.
\end{minipage}
\end{table}

\clearpage

\begin{table}[H]\centering
\caption{Robustness of Wage Results to Trimming High-Income Households}
\label{tab:acs_trim_income_estimates}
\scalebox{0.9}{
\begin{threeparttable}
\begin{tabular}{llcccc}
\hline
\addlinespace
& &   Log Earnings  & Usual Weekly Hours & Weeks Worked & Part-time \\ 
& &  &   Worked & & Status \\
\cline{3-6} \addlinespace
\multicolumn{6}{l}{\emph{A. No income trim (Baseline)}}\\
\addlinespace
$ Exposed_{o(h(i))} \times Post_{t}$&            &       0.048***&       0.466***&       0.500***&      -0.024***\\
                    &            &     (0.010)   &     (0.096)   &     (0.089)   &     (0.004)   \\
\addlinespace Mean pre-treatment&            &       9.915   &      36.294   &      45.671   &       0.277   \\
Obs                 &            &     949,443   &     948,946   &     949,443   &     948,946   \\
$R^2$               &            &        0.15   &        0.03   &        0.03   &        0.03   \\
\multicolumn{6}{l}{\emph{B. Trim top 1\% income earners}}\\
\addlinespace
$ Exposed_{o(h(i))} \times Post_{t}$&            &       0.044***&       0.467***&       0.498***&      -0.024***\\
                    &            &     (0.010)   &     (0.097)   &     (0.089)   &     (0.004)   \\
\addlinespace Mean pre-treatment&            &       9.897   &      36.229   &      45.641   &       0.278   \\
Obs                 &            &     938,315   &     937,832   &     938,315   &     937,832   \\
$R^2$               &            &        0.14   &        0.03   &        0.03   &        0.03   \\
\multicolumn{6}{l}{\emph{C. Trim top 5\% income earners}}\\
\addlinespace
$ Exposed_{o(h(i))} \times Post_{t}$&            &       0.036***&       0.453***&       0.487***&      -0.023***\\
                    &            &     (0.010)   &     (0.093)   &     (0.090)   &     (0.004)   \\
\addlinespace Mean pre-treatment&            &       9.860   &      36.134   &      45.603   &       0.279   \\
Obs                 &            &     901,176   &     900,732   &     901,176   &     900,732   \\
$R^2$               &            &        0.14   &        0.03   &        0.03   &        0.03   \\
\addlinespace \hline
\end{tabular}

\end{threeparttable}
}
\par\vspace{0.3cm}
\begin{minipage}{\textwidth} \small
\textit{Notes}: This table reports robustness checks on the wife's log wage outcome using the 2013--2024 ACS sample of dual-parent households with children of child-rearing age, assessing sensitivity to households at the top of the household income distribution. Panel A reproduces the baseline (untrimmed) wage specification from \autoref{tab:acs_baseline_annual} for comparison. Subsequent panels re-estimate the specification after progressively trimming the top percentiles of the household income distribution. Treatment is based on the husband's occupation-level WFH shock (Subsection~\ref{subsection:treatment_defn}). For brevity, we report only the post-COVID (2022 onward) interaction coefficient. All regressions include state-by-time and husband's-occupation fixed effects, and control for the wife's age, number of children, and both spouses' education levels. Standard errors, clustered at the husband's occupation level, are reported in parentheses. \textsuperscript{*} $p<0.10$; \textsuperscript{**} $p<0.05$; \textsuperscript{***} $p<0.01$.
\end{minipage}
\end{table}
\clearpage

%\begin{table}[H]\centering
%\caption{Robustness Checks Using Alternative WFH Shocks and Teleworkable %Occupation Overlap}
%\label{tab:acsconsistentbloomdingel}
%\scalebox{0.65}{
%\begin{threeparttable}
%\input{tables/acs_consistent_occupations_bloom_dingel}
%\begin{tablenotes}
%{\em{Notes:}} Panel A defines treated occupations as those where the %increase in the share of remote job postings between 2019 and 2023 %exceeds the median across all Census 2010 4-digit occupations. Panel B %uses the same criterion but limits the sample to occupations classified %as teleworkable by \citet{dingel_how_2020}. Panels C and D repeat the %analyses in Panels A and B, respectively, using broader occupation %definitions based on SOC 2018 3-digit codes. All regressions include %fixed effects for state-by-time, husband’s occupation, and husband’s %industry. Controls include the wife’s age, education, and number of %children, as well as the husband’s education. Standard errors are %clustered at the husband's occupation level.
%\end{tablenotes}
%\end{threeparttable}
%}
%\end{table}
%\clearpage

\begin{table}[H]\centering
\caption{Main Results Using Leave-One Occupation Out at Census 2d Level}
\label{tab:acsleaveoneoutoccupations2d}
\scalebox{0.85}{
\begin{threeparttable}
\begin{tabular}{llcccc}
\hline
\addlinespace
& &  Log Earnings & Usual Weekly Hours & Weeks Worked & Part-time \\ 
& &  & Worked & & Status \\
\cline{3-6} \addlinespace
                    &            &\multicolumn{1}{c}{}&\multicolumn{1}{c}{}&\multicolumn{1}{c}{}&\multicolumn{1}{c}{}\\

Benchmark Estimate  &            &       0.048&       0.463&       0.500&      -0.024\\
Average Bootstrap Estimate&            &       0.048&       0.463&       0.500&      -0.024\\
Average Bootstrap SE&            &       0.010&       0.096&       0.089&       0.004\\
P-Statistic         &            &       0.000&       0.000&       0.000&       0.000\\
\addlinespace \hline
\end{tabular}

\end{threeparttable}
}
\par\vspace{0.3cm}
\begin{minipage}{\textwidth} \small
\textit{Notes}: This table reports leave-one-occupation-out estimates for the wife's labor market outcomes, using the 2013--2024 ACS sample.
 ``Benchmark Estimate'' reports the baseline post-COVID coefficient from Equation~\eqref{eq:1}. ``Average Bootstrap Estimate'' is the mean of the post-COVID coefficient across iterations that each drop one Census 2-digit occupation group at a time and re-estimate Equation~\eqref{eq:1}. ``Average Standard Error'' is the mean of the corresponding standard errors across iterations. The ``P-statistic'' is the share of iteration-specific estimates that are not statistically significant at the 5\% level. All regressions include state-by-time and husband's-occupation fixed effects, and control for the wife's age, education, and number of children, as well as the husband's education. Standard errors in each iteration are clustered at the husband's occupation level.
\end{minipage}
\end{table}
\clearpage

\begin{table}[H]\centering
\caption{Effect of Husbands' WFH Shock on Marriage, Divorce, and Fertility Outcomes}
\label{tab:acs_marriage_divorce}
\scalebox{0.95}{
\begin{threeparttable}
\begin{tabular}{llccc}
\hline
\addlinespace
& &  Married & Divorced & Had a child \\ 
&& in the last 12mm & in the last 12mm & in the last 12mm \\
\cline{3-5} \addlinespace
\addlinespace
$ Exposed_{o(h(i))} \times Post_{t}$&            &       0.001   &       0.000   &      -0.005** \\
                    &            &     (0.001)   &     (0.000)   &     (0.002)   \\
\addlinespace Mean pre-treatment&            &       0.032   &       0.001   &       0.168   \\
Obs                 &            &   1,371,205   &   1,371,038   &   1,371,205   \\
$R^2$               &            &        0.04   &        0.00   &        0.32   \\
\addlinespace \hline
\end{tabular}

\end{threeparttable}
}
\par\vspace{0.3cm}
\begin{minipage}{\textwidth} \small
 \textit{Notes}: This table reports estimates from Equation~\eqref{eq:1} using the pooled 2013--2024 ACS sample of dual-parent households with children of child-rearing age. For brevity, we report only the post-COVID (2022 onward) interaction coefficient. The outcomes -- having married, having divorced, and having had a child -- are indicators for whether the event occurred in the 12 months preceding the interview. Occupation is measured only at the time of the interview, so we implicitly assume individuals remained in the same occupation throughout the preceding 12 months. All regressions control for the wife's age, number of children, and education, as well as the husband's education, and include state-by-time and husband's-occupation fixed effects. Standard errors, clustered at the husband's occupation level, are reported in parentheses. \textsuperscript{*} $p<0.10$; \textsuperscript{**} $p<0.05$; \textsuperscript{***} $p<0.01$. %While this introduces some potential for measurement error, the ACS provides the largest sample available for analyzing these demographic outcomes.
\end{minipage}
\end{table}
\clearpage

\begin{table}[H]\centering
\caption{Effect of Husbands' WFH Shock on Fertility Outcomes}
\label{tab:acs_fertility}
\scalebox{0.95}{
\begin{threeparttable}
\begin{tabular}{llcc}
\hline
\addlinespace
& & Pr(At least one child)     & Number of Children\\
& &  &         \\
\cline{3-4} \addlinespace
\addlinespace
$ Exposed_{o(h(i))} \times Post_{t}$&            &      -0.011***&      -0.026***\\
                    &            &     (0.003)   &     (0.008)   \\
\addlinespace Mean pre-treatment&            &       0.744   &       1.518   \\
Obs                 &            &   3,255,210   &   3,255,210   \\
$R^{2}$             &            &        0.08   &        0.10   \\
\addlinespace \hline
\end{tabular}

\end{threeparttable}
}
\par\vspace{0.3cm}
\begin{minipage}{\textwidth} \small
\textit{Notes}: This table reports estimates from Equation~\eqref{eq:1} using the pooled 2013--2024 ACS sample of dual-parent households with children of child-rearing age. The outcomes are an indicator for having had a child in the 12 months preceding the interview and the number of children currently in the household. Treatment is defined based on the husband's occupation-level WFH exposure (Subsection~\ref{subsection:treatment_defn}). All regressions include controls for the wife's age, number of children, and education, as well as the husband's education, and state-by-time and husband's-occupation fixed effects. Standard errors, clustered at the husband's occupation level, are reported in parentheses. \textsuperscript{*} $p<0.10$; \textsuperscript{**} $p<0.05$; \textsuperscript{***} $p<0.01$.
\end{minipage}
\end{table}
\clearpage

\begin{table}[H]\centering
\caption{Effect of Husbands' WFH Shock on Wives' Labor Market Outcomes Using Reweighting Techniques}
\label{tab:acs_reweighting}
\scalebox{0.95}{
\begin{threeparttable}
\begin{tabular}{llcccc}
\hline
\addlinespace
& &  Log Earnings & Usual Weekly Hours & Weeks Worked & Part-time \\ 
& &  & Worked & & Status \\
\cline{3-6} \addlinespace
\multicolumn{6}{l}{\emph{A. Using DFL Weights}}\\
\addlinespace
$ Exposed_{o(h(i))} \times Post_{t}$&            &       0.048***&       0.470***&       0.499***&      -0.024***\\
                    &            &     (0.010)   &     (0.095)   &     (0.089)   &     (0.004)   \\
\addlinespace Mean pre-treatment&            &       9.902   &      36.276   &      45.635   &       0.279   \\
Obs                 &            &     949,443   &     948,946   &     949,443   &     948,946   \\
$R^{2}$             &            &        0.15   &        0.03   &        0.03   &        0.03   \\
\multicolumn{6}{l}{\emph{B. Using IPW}}\\
\addlinespace
$ Exposed_{o(h(i))} \times Post_{t}$&            &       0.049***&       0.461***&       0.501***&      -0.024***\\
                    &            &     (0.010)   &     (0.097)   &     (0.089)   &     (0.004)   \\
\addlinespace Mean pre-treatment&            &       9.902   &      36.276   &      45.635   &       0.279   \\
Obs                 &            &     949,443   &     948,946   &     949,443   &     948,946   \\
$R^{2}$             &            &        0.15   &        0.03   &        0.03   &        0.03   \\
\addlinespace \hline
\end{tabular}

\end{threeparttable}
}
\par\vspace{0.3cm}
\begin{minipage}{\textwidth} \small
  \textit{Notes}: This table reports estimates from Equation~\eqref{eq:1} using the 2013--2024 ACS sample of dual-parent households with children of child-rearing age, reweighting the post-COVID sample to match the pre-COVID distribution of spouse occupation, wife's age, wife's education, and state. Panel A applies DiNardo-Fortin-Lemieux (DFL) reweighting \citep{dinardo1995labor}; Panel B applies inverse-propensity-score weighting (IPW); see Section~\ref{section:robustness} for construction details. All regressions control for the wife's age, education, and number of children, as well as the husband's education, and include state-by-time and husband's-occupation fixed effects. Standard errors, clustered at the husband's occupation level, are reported in parentheses. \textsuperscript{*} $p<0.10$; \textsuperscript{**} $p<0.05$; \textsuperscript{***} $p<0.01$.
\end{minipage}
\end{table}

% ya esta corriendo
\begin{table}[H]\centering
\caption{Robustness of ACS Results to Alternative Definitions of WFH Shock}
\label{tab:changeshockdef}
\scalebox{0.56}{
\begin{threeparttable}
\begin{tabular}{llcccc}
\hline
\addlinespace
& &  Log Earnings & Usual Weekly Hours & Weeks Worked & Part-time \\ 
& &  & Worked & & Status \\
\cline{3-6} \addlinespace
\multicolumn{6}{l}{\emph{A. Shock: Difference in average 2023-2024 vs 2019 WFH vacancy rates at Census 2010 4d level (Original)}}\\
\addlinespace
$ Exposed_{o(h(i))} \times Post_{t}$&            &       0.048***&       0.466***&       0.500***&      -0.024***\\
                    &            &     (0.010)   &     (0.096)   &     (0.089)   &     (0.004)   \\
\addlinespace Mean pre-treatment&            &       9.915   &      36.294   &      45.671   &       0.277   \\
Obs                 &            &     949,443   &     948,946   &     949,443   &     948,946   \\
$R^2$               &            &        0.15   &        0.03   &        0.03   &        0.03   \\
\multicolumn{6}{l}{\emph{B. Shock: Difference in 3 month moving average 2023-2024 vs 2019 WFH vacancy rates at Census 2010 4d level}}\\
\addlinespace
$ Exposed_{o(h(i))} \times Post_{t}$&            &       0.048***&       0.466***&       0.500***&      -0.024***\\
                    &            &     (0.010)   &     (0.096)   &     (0.089)   &     (0.004)   \\
\addlinespace Mean pre-treatment&            &       9.915   &      36.294   &      45.671   &       0.277   \\
Obs                 &            &     949,443   &     948,946   &     949,443   &     948,946   \\
$R^2$               &            &        0.15   &        0.03   &        0.03   &        0.03   \\
\multicolumn{6}{l}{\emph{C. Shock: Difference in July 2023 vs July 2019 WFH vacancy rates at Census 2010 4d level}}\\
\addlinespace
$ Exposed_{o(h(i))} \times Post_{t}$&            &       0.046***&       0.432***&       0.446***&      -0.022***\\
                    &            &     (0.010)   &     (0.096)   &     (0.090)   &     (0.004)   \\
\addlinespace Mean pre-treatment&            &       9.915   &      36.294   &      45.671   &       0.277   \\
Obs                 &            &     949,443   &     948,946   &     949,443   &     948,946   \\
$R^2$               &            &        0.15   &        0.03   &        0.03   &        0.03   \\
\multicolumn{6}{l}{\emph{D. Shock: WFH teleworkability a-la Dingel and Neumann at Census 2010 4d level}}\\
\addlinespace
$ Exposed_{o(h(i))} \times Post_{t}$&            &       0.047***&       0.413***&       0.396***&      -0.025***\\
                    &            &     (0.010)   &     (0.098)   &     (0.088)   &     (0.004)   \\
\addlinespace Mean pre-treatment&            &       9.919   &      36.303   &      45.709   &       0.277   \\
Obs                 &            &     942,285   &     941,800   &     942,285   &     941,800   \\
$R^2$               &            &        0.15   &        0.03   &        0.03   &        0.03   \\
\multicolumn{6}{l}{\emph{E. Shock: Difference in average 2023-2024 vs 2019 WFH vacancy rates at SOC 3d level}}\\
\addlinespace
$ Exposed_{o(h(i))} \times Post_{t}$&            &       0.051***&       0.492***&       0.484***&      -0.026***\\
                    &            &     (0.011)   &     (0.104)   &     (0.099)   &     (0.005)   \\
\addlinespace Mean pre-treatment&            &       9.916   &      36.296   &      45.686   &       0.277   \\
Obs                 &            &     946,923   &     946,430   &     946,923   &     946,430   \\
$R^2$               &            &        0.15   &        0.03   &        0.03   &        0.03   \\
\multicolumn{6}{l}{\emph{F. Shock: Difference in 3 month moving average 2023-2024 vs 2019 vacancy rates at SOC 3d level}}\\
\addlinespace
$ Exposed_{o(h(i))} \times Post_{t}$&            &       0.054***&       0.507***&       0.513***&      -0.026***\\
                    &            &     (0.011)   &     (0.104)   &     (0.097)   &     (0.005)   \\
\addlinespace Mean pre-treatment&            &       9.916   &      36.296   &      45.686   &       0.277   \\
Obs                 &            &     946,923   &     946,430   &     946,923   &     946,430   \\
$R^2$               &            &        0.15   &        0.03   &        0.03   &        0.03   \\
\multicolumn{6}{l}{\emph{G. Shock: Difference in July 2023 vs July 2019 WFH vacancy rates at SOC 3d level}}\\
\addlinespace
$ Exposed_{o(h(i))} \times Post_{t}$&            &       0.049***&       0.479***&       0.439***&      -0.024***\\
                    &            &     (0.011)   &     (0.106)   &     (0.101)   &     (0.005)   \\
\addlinespace Mean pre-treatment&            &       9.916   &      36.296   &      45.686   &       0.277   \\
Obs                 &            &     946,923   &     946,430   &     946,923   &     946,430   \\
$R^2$               &            &        0.15   &        0.03   &        0.03   &        0.03   \\
\multicolumn{6}{l}{\emph{H. Shock: WFH teleworkability a-la Dingel and Neumann at SOC 3d level}}\\
\addlinespace
$ Exposed_{o(h(i))} \times Post_{t}$&            &       0.053***&       0.451***&       0.435***&      -0.027***\\
                    &            &     (0.011)   &     (0.102)   &     (0.099)   &     (0.004)   \\
\addlinespace Mean pre-treatment&            &       9.919   &      36.303   &      45.709   &       0.277   \\
Obs                 &            &     942,285   &     941,800   &     942,285   &     941,800   \\
$R^2$               &            &        0.15   &        0.03   &        0.03   &        0.03   \\
\addlinespace \hline
\end{tabular}

\end{threeparttable}
}
\par\vspace{0.3cm}
\begin{minipage}{\textwidth}
    \small
\textit{Notes}: This table reports estimates from Equation~\eqref{eq:1} under alternative treatment definitions, using the 2013--2024 ACS sample of dual-parent households with children of child-rearing age. In Panel A, an occupation is treated if the difference between the average 2023--2024 and 2019 remote-job-vacancy shares is above the median across Census 2010 4-digit occupations (our baseline definition). Panel B uses a three-month moving average of the monthly WFH job-postings shares. Panel C defines treatment using the July 2023 versus July 2019 difference in remote/WFH job-vacancy shares. Panel D uses the \citet{dingel_how_2020} teleworkability index, with an occupation treated if its index value is above the median. Panels E--H replicate Panels A--D, respectively, defining occupations at the SOC 2018 3-digit level instead of Census 2010 4-digit. All regressions include state-by-time and husband's-occupation fixed effects. For brevity, we report only the post-COVID (2022 onward) interaction coefficient. All regressions control for the wife's age, education, and number of children, as well as the husband's education. Standard errors, clustered at the husband's occupation level, are reported in parentheses. \textsuperscript{*} $p<0.10$; \textsuperscript{**} $p<0.05$; \textsuperscript{***} $p<0.01$.
\end{minipage}
\end{table}
\clearpage

\begin{table}[H]\centering
\caption{Robustness of Main Results to Using a Continuous Treatment}
\label{tab:changeshockdef_cont}
\scalebox{0.85}{
\begin{threeparttable}
\begin{tabular}{llcccc}
\hline
\addlinespace
& &  Log Earnings & Usual Weekly Hours & Weeks Worked & Part-time \\ 
& &  & Worked & & Status \\
\cline{3-6} \addlinespace
\addlinespace
$ Exposed_{o(h(i))} \times Post_{t}$&            &       0.022***&       0.186***&       0.238***&      -0.010***\\
                    &            &     (0.004)   &     (0.038)   &     (0.040)   &     (0.002)   \\
\addlinespace Mean pre-treatment&            &       9.904   &      36.279   &      45.652   &       0.279   \\
Obs                 &            &     946,923   &     946,430   &     946,923   &     946,430   \\
$R^{2}$             &            &        0.15   &        0.03   &        0.03   &        0.03   \\
\addlinespace \hline
\end{tabular}

\end{threeparttable}
}
\par\vspace{0.3cm}
\begin{minipage}{\textwidth} \small
  \textit{Notes}: This table replicates the baseline ACS estimates from \autoref{tab:acs_baseline_annual}, replacing the binary treatment $\text{High-}\Delta\text{WFH}_{o(h(i))}$ with the continuous, standardized (mean-zero, unit-variance) change in the husband's occupation-level WFH rate constructed from the \citet{hansen2023remote} job postings data. A one-standard-deviation increase in this measure corresponds to an 8 percentage point increase in the occupation-level WFH share. All specifications include the same controls (wife's age, education, and number of children, and husband's education) and fixed effects (state-by-time and husband's occupation) as the baseline specification. Standard errors, clustered at the husband's occupation level, are reported in parentheses. \textsuperscript{*} $p<0.10$; \textsuperscript{**} $p<0.05$; \textsuperscript{***} $p<0.01$.
 %A one–standard-deviation increase in this WFH rate corresponds to a 12 percentage point increase in WFH. For comparison, a one–standard-deviation increase in the job-postings-based WFH measure corresponds to an 8 percentage point increase. 
\end{minipage}
\end{table}
\clearpage

\begin{table}[H]\centering
\caption{Robustness of Main Results to Using a Binary Treatment Based on Change in ACS Uptake of WFH}
\label{tab:changeshockdef_acs}
\scalebox{0.85}{
\begin{threeparttable}
\begin{tabular}{llcccc}
\hline
\addlinespace
& &  Log Earnings & Usual Weekly Hours & Weeks Worked & Part-time \\ 
& &  & Worked & & Status \\
\cline{3-6} \addlinespace
\addlinespace
$ Exposed_{o(h(i))} \times Post_{t} $&            &       0.052***&       0.436***&       0.354***&      -0.024***\\
                    &            &     (0.010)   &     (0.094)   &     (0.096)   &     (0.004)   \\
\addlinespace Mean pre-treatment&            &       9.903   &      36.277   &      45.637   &       0.279   \\
Obs                 &            &     948,549   &     948,052   &     948,549   &     948,052   \\
$R^{2}$             &            &        0.15   &        0.03   &        0.03   &        0.03   \\
\addlinespace \hline
\end{tabular}

\end{threeparttable}
}
\par\vspace{0.3cm}
\begin{minipage}{\textwidth} \small
 \textit{Notes}: This table replicates the baseline ACS estimates from \autoref{tab:acs_baseline_annual}, using an alternative binary treatment based on husbands' self-reported WFH uptake rather than job postings: an occupation is classified as treated if the percentage-point change in the ACS-reported share of husbands working fully from home between the pre- and post-COVID periods is above the median across occupations. All specifications include the same controls (wife's age, education, and number of children, and husband's education) and fixed effects (state-by-time and husband's occupation) as the baseline specification. Standard errors, clustered at the husband's occupation level, are reported in parentheses. \textsuperscript{*} $p<0.10$; \textsuperscript{**} $p<0.05$; \textsuperscript{***} $p<0.01$.
\end{minipage}
\end{table}
\clearpage

\begin{table}[H]\centering
\caption{Effect of Husbands' WFH Shock on Wives Labor Market Outcomes Using a Placebo Shock Timing}
\label{tab:acs_timing_placebo_baseline_annual}
\scalebox{0.85}{
\begin{threeparttable}
\begin{tabular}{llcccc}
\hline
\addlinespace
& &  Log Earnings & Usual Weekly Hours & Weeks Worked & Part-time \\ 
& &  & Worked & & Status \\
\cline{3-6} \addlinespace
\addlinespace
$ Exposed_{o(h(i))} \times Post_{t}$&            &      -0.004   &       0.091   &       0.237** &      -0.000   \\
                    &            &     (0.011)   &     (0.093)   &     (0.118)   &     (0.004)   \\
\addlinespace Mean pre-treatment&            &       9.936   &      36.474   &      45.714   &       0.273   \\
Obs                 &            &     306,390   &     306,237   &     306,390   &     306,237   \\
$R^2$               &            &        0.15   &        0.03   &        0.03   &        0.04   \\
\addlinespace \hline
\end{tabular}

\end{threeparttable}
}
\par\vspace{0.3cm}
\begin{minipage}{\textwidth} \small
\textit{Notes}: This table reports estimates from Equation~\eqref{eq:1} using the 2012--2019 ACS sample of dual-parent households with children of child-rearing age, in a placebo test that artificially shifts the treatment period: the (placebo) pre-period is 2012--2015 and the (placebo) post-period is 2016--2019, so that the true 2020 pandemic shock falls outside the estimation window. Treatment is defined as in the baseline specification (Subsection~\ref{subsection:treatment_defn}). All regressions control for the wife's age, education, and number of children, as well as the husband's education. Specifications include state-by-time and husband's-occupation fixed effects. Standard errors, clustered at the husband's occupation level, are reported in parentheses. \textsuperscript{*} $p<0.10$; \textsuperscript{**} $p<0.05$; \textsuperscript{***} $p<0.01$.
\end{minipage}
\end{table}
\clearpage

\begin{table}[H]\centering
\caption{Effect of Husbands' WFH Shock on Wives' Labor Market Outcomes Using the CPS}
\label{tab:cps_baseline_estimates_combined}
\scalebox{0.75}{
\begin{threeparttable}
\begin{tabular}{llcclcccc}
\hline
\addlinespace
& & \multicolumn{2}{c}{\emph{Previous week}} & & \multicolumn{4}{c}{\emph{Previous year (March CPS)}} \\ \cline{3-4} \cline{6-9}
& & Usual Weekly & Part-time & & Log Earnings & Usual Weekly & Weeks Worked & Part-time \\
& & Work hours   & status    & &              & Work Hours   &              & Status \\
\cline{3-4} \cline{6-9} \addlinespace
\addlinespace
$ Exposed_{o(h(i))} \times Post_{t}$&            &       0.266*  &      -0.012*  &            &       0.097***&       0.676** &       0.883***&      -0.022*  \\
                    &            &     (0.158)   &     (0.006)   &            &     (0.028)   &     (0.285)   &     (0.334)   &     (0.013)   \\
\addlinespace Mean pre-treatment&            &      36.593   &       0.324   &            &       9.939   &      36.250   &      46.472   &       0.260   \\
Obs                 &            &     446,122   &     445,961   &            &      68,298   &      68,298   &      68,298   &      68,298   \\
$R^2$               &            &        0.05   &        0.05   &            &        0.17   &        0.05   &        0.06   &        0.06   \\
\addlinespace \hline
\end{tabular}

\end{threeparttable}
}
\par\vspace{0.3cm}
\begin{minipage}{\textwidth} \small
\textit{Notes}: This table reports estimates from Equation~\eqref{eq:1} based on the 2013--2026 CPS sample of dual-parent households with children of child-rearing age. Variables measured over the previous week are drawn from the CPS basic monthly files, whereas variables measured over the previous year (e.g., annual earnings) are drawn from the March CPS/ASEC supplement. Treatment is defined as in the ACS baseline specification (Subsection~\ref{subsection:treatment_defn}). All specifications include controls for the wife's age, education, and number of children, as well as the husband's education, and state-by-year and husband's-occupation fixed effects. Standard errors, clustered at the husband's occupation level, are reported in parentheses. \textsuperscript{*} $p<0.10$; \textsuperscript{**} $p<0.05$; \textsuperscript{***} $p<0.01$.
\end{minipage}
\end{table}
\clearpage

\begin{table}[H]\centering
\caption{Effect of Husbands' WFH Shock on Time Spent in Labor and Household Supply - Share of the Day (Placebo Timing)}
\label{tab:mechplacebotiming}
\scalebox{0.55}{
\begin{threeparttable}
\begin{tabular}{llcccccclcc}
\hline
\addlinespace
& & \multicolumn{6}{c}{\emph{Time Spent in Labor Production}} & & \multicolumn{2}{c}{\emph{Time Spent for Personal Care and Leisure}} \\ \cline{3-8} \cline{10-11}
& & Working & Main Work & Main Work & Main Work        & Secondary Work        & Commuting &   &  Personal Care & Leisure      \\
& &             &                   & From Home & Outside Home &                                       &                       &       &                                &                      \\ 
\cline{3-8} \cline{10-11} \addlinespace
\multicolumn{11}{l}{\large{\textbf{Wives:}}} \\
\addlinespace
$ Exposed_{o(h(i))} \times Post_{t}$&            &      -0.003   &      -0.004   &      -0.005   &       0.001   &       0.001   &       0.000   &            &      -0.008   &      -0.005   \\
                    &            &     (0.017)   &     (0.017)   &     (0.006)   &     (0.016)   &     (0.001)   &     (0.002)   &            &     (0.008)   &     (0.010)   \\
\addlinespace \addlinespace
\addlinespace Mean pre-treatment&            &       0.152   &       0.150   &       0.018   &       0.132   &       0.001   &       0.010   &            &       0.374   &       0.118   \\
Obs                 &            &      14,324   &      14,324   &      14,324   &      14,324   &      14,324   &      14,324   &            &      14,324   &      14,324   \\
$R^2$               &            &        0.76   &        0.76   &        0.77   &        0.75   &        0.80   &        0.71   &            &        0.73   &        0.76   \\
\addlinespace
\addlinespace
\multicolumn{11}{l}{\large{\textbf{Husbands:}}} \\
\addlinespace
$ Exposed_{o(h(i))} \times Post_{t}$&            &      -0.005   &      -0.011   &       0.004   &      -0.015   &       0.002   &      -0.001   &            &      -0.002   &       0.016   \\
                    &            &     (0.016)   &     (0.016)   &     (0.008)   &     (0.018)   &     (0.004)   &     (0.004)   &            &     (0.007)   &     (0.010)   \\
\addlinespace \addlinespace
\addlinespace Mean pre-treatment&            &       0.321   &       0.314   &       0.028   &       0.286   &       0.005   &       0.027   &            &       0.346   &       0.120   \\
Obs                 &            &      13,745   &      13,745   &      13,745   &      13,745   &      13,745   &      13,745   &            &      13,745   &      13,745   \\
$R^2$               &            &        0.79   &        0.79   &        0.78   &        0.78   &        0.73   &        0.73   &            &        0.77   &        0.78   \\
\end{tabular}
\begin{tabular}{llccccccccl} \addlinespace
& & \multicolumn{8}{c}{\emph{Time Spent in Household Production}} &  \\ \cline{3-11} 
& & Housework & Cooking & Shopping & Primary     & Secondary            & Total                & Total                         &  Prop. Secondary CC &         \\
& &           &                 &          & Childcaring & CC while working & Secondary CC     & Childcaring       &  in Total CC &                    \\ 
\cline{3-11} \addlinespace
\multicolumn{11}{l}{\large{\textbf{Wives:}}} \\
$ Exposed_{o(h(i))} \times Post_{t}$&            &      -0.009   &       0.000   &       0.001   &       0.009   &       0.020   &      -0.019   &      -0.010   &      -0.004   &            \\
                    &            &     (0.010)   &     (0.006)   &     (0.002)   &     (0.009)   &     (0.040)   &     (0.017)   &     (0.020)   &     (0.026)   &            \\
\addlinespace \addlinespace
\addlinespace Mean pre-treatment&            &       0.104   &       0.049   &       0.016   &       0.108   &       0.128   &       0.242   &       0.350   &       0.657   &            \\
Obs                 &            &      14,324   &      14,324   &      14,324   &      14,324   &       7,770   &      14,324   &      14,324   &      14,111   &            \\
$R^2$               &            &        0.79   &        0.74   &        0.75   &        0.74   &        0.85   &        0.76   &        0.76   &        0.76   &            \\
\addlinespace
\addlinespace
\multicolumn{11}{l}{\large{\textbf{Husbands:}}} \\
$ Exposed_{o(h(i))} \times Post_{t}$&            &      -0.011*  &      -0.002   &       0.003   &       0.001   &       0.027   &       0.003   &       0.004   &      -0.027   &            \\
                    &            &     (0.006)   &     (0.003)   &     (0.002)   &     (0.005)   &     (0.018)   &     (0.014)   &     (0.015)   &     (0.039)   &            \\
\addlinespace
\addlinespace Mean pre-treatment&            &       0.041   &       0.014   &       0.008   &       0.045   &       0.036   &       0.127   &       0.172   &       0.681   &            \\
Obs                 &            &      13,745   &      13,745   &      13,745   &      13,745   &      12,162   &      13,745   &      13,745   &      12,543   &            \\
$R^2$               &            &        0.78   &        0.74   &        0.80   &        0.76   &        0.76   &        0.77   &        0.77   &        0.78   &            \\
\addlinespace \hline
\end{tabular}

\end{threeparttable}
}
\par\vspace{0.3cm}
\begin{minipage}{\textwidth}
    \small
 \textit{Notes}: This table replicates \autoref{tab:mechatus1} using a placebo test in which the onset of the (placebo) post-treatment period is artificially moved to March 2016, analogous to the 2012--2015/2016--2019 split used in \autoref{tab:acs_timing_placebo_baseline_annual}: the sample is restricted to pre-2020 ATUS diary days, with the (placebo) pre-period spanning 2013--February 2016 and the (placebo) post-period spanning March 2016--2019. The outcome definitions (share of the 24-hour day spent in each activity, organized into labor-production, personal-care/leisure, and household-production panels), fixed effects, controls, and clustering are otherwise identical to those in \autoref{tab:mechatus1}; see that table's notes for full definitions. Standard errors, clustered at the husband's occupation level, are reported in parentheses. \textsuperscript{*} $p<0.10$; \textsuperscript{**} $p<0.05$; \textsuperscript{***} $p<0.01$.
\end{minipage}
\end{table}

\clearpage
 
\begin{table}[H]\centering
\caption{Effect of Husbands' WFH Occupation Shocks on Childcare Time and Child-Related Activities}
\label{tab:sipp_mechanism_chcare_cross}
\scalebox{0.65}{
\begin{threeparttable}
\begin{tabular}{llccccccc}
\hline
\addlinespace
& & Childcare by& Childcare by & Childcare by  & Childcare by  & Childcare by  & Childcare & Log Childcare \\
& & Wife while & Spouse while  & Grandparent   & Relative              & Non-relative  & by Daycare& Expenditure       \\ 
& & Working    & Wife Works    &                               &                               &                               &                       &                                                      \\ 
\cline{3-9} \addlinespace
\addlinespace
$ Exposed_{o(h(i))} \times Post_{t}$&            &       0.007   &       0.069** &       0.005   &       0.023   &      -0.013   &      -0.064** &      -0.199** \\
                    &            &     (0.032)   &     (0.033)   &     (0.028)   &     (0.029)   &     (0.020)   &     (0.025)   &     (0.100)   \\
\addlinespace Mean pre-treatment&            &       0.282   &       0.470   &       0.409   &       0.450   &       0.188   &       0.250   &       4.662   \\
Obs                 &            &     121,927   &     121,927   &     176,958   &     176,958   &     176,958   &     176,958   &      64,547   \\
$R^2$               &            &        0.16   &        0.12   &        0.13   &        0.12   &        0.11   &        0.16   &        0.29   \\
\addlinespace \hline
\end{tabular}

\end{threeparttable}
}
\par\vspace{0.3cm}
\begin{minipage}{\textwidth}
    \small
 \textit{Notes}: This table reports estimates from Equation~\eqref{eq:1} using the 2014--2024 SIPP sample of dual-parent households with children of child-rearing age. The dependent variables are drawn from the SIPP childcare topical modules, which ask households with minor children whether, in a typical week, each of the following was involved for at least one day in looking after the children: the wife, the husband (conditional on the wife working), a grandparent, another relative, or a paid daycare provider; we also report a dummy for whether the household pays for daycare and log self-reported childcare expenditure. For brevity, we report only the post-COVID (2022 onward) interaction coefficient. All regressions include state-by-time and husband's-occupation fixed effects, and control for the wife's age, education, and number of children, as well as the husband's education. Standard errors, clustered at the husband's occupation level, are reported in parentheses. \textsuperscript{*} $p<0.10$; \textsuperscript{**} $p<0.05$; \textsuperscript{***} $p<0.01$.
\end{minipage}
\end{table}
\clearpage

\begin{table}[H]\centering
\caption{Effect of Husbands' WFH Shock on Differential Migration Towards Favorable Geographic Labor Market Counties for Women} % habla pe gil
\label{tab:acs_migrating_origdest_labor_chars}
\scalebox{0.7}{
\begin{threeparttable}
\begin{tabular}{llcccccc}
\hline
\addlinespace
 & & \multicolumn{6}{c}{$\Delta$ Destination-Origin} \\
& & Log Earnings & Usual Weekly Hours & Weeks Worked & Part-time & Commuting & WFH Uptake \\ 
& &  & Worked &  & Status &  &  \\ 
\cline{3-8} \addlinespace
\addlinespace
$ Exposed_{o(h(i))} \times Post_{t}$&            &       0.001   &       0.014   &       0.031   &      -0.001   &       0.022   &       0.001   \\
                    &            &     (0.004)   &     (0.021)   &     (0.024)   &     (0.001)   &     (0.060)   &     (0.001)   \\
\addlinespace Mean pre-treatment&            &      -0.003   &      -0.006   &      -0.003   &       0.000   &      -0.033   &       0.000   \\
Obs                 &            &      86,943   &      86,943   &      86,943   &      86,943   &      86,936   &      86,943   \\
$R^{2}$             &            &        0.03   &        0.03   &        0.03   &        0.03   &        0.03   &        0.03   \\
\addlinespace \hline
\end{tabular}

\end{threeparttable}
}
\par\vspace{0.3cm}
\begin{minipage}{\textwidth}
    \small
 \textit{Notes}: This table reports estimates from Equation~\eqref{eq:1} using the pooled 2013--2023 ACS sample of dual-parent households with children of child-rearing age, restricted to households that migrated across counties in the year preceding the interview. The dependent variables are the difference (destination minus origin county) in average labor market characteristics for women: usual weekly hours worked, weeks worked, part-time status, WFH uptake, and commuting time, plus earnings; county-level averages are computed among women in the ACS. Treatment is defined based on the husband's occupation-level WFH exposure (Subsection~\ref{subsection:treatment_defn}). All regressions include controls for the wife's age, number of children, and education, as well as the husband's education, and state-by-time and husband's-occupation fixed effects. Standard errors, clustered at the husband's occupation level, are reported in parentheses. \textsuperscript{*} $p<0.10$; \textsuperscript{**} $p<0.05$; \textsuperscript{***} $p<0.01$.
\end{minipage}
\end{table}

\clearpage

\section{Figures}
\renewcommand\thefigure{\thesection.\arabic{figure}}

\setcounter{figure}{0}

\begin{figure}[H] 
    \caption{Distribution of WFH Exposure Using Job Postings Data}\label{fig:shock_distribution}
    \includegraphics[width=0.9\linewidth]{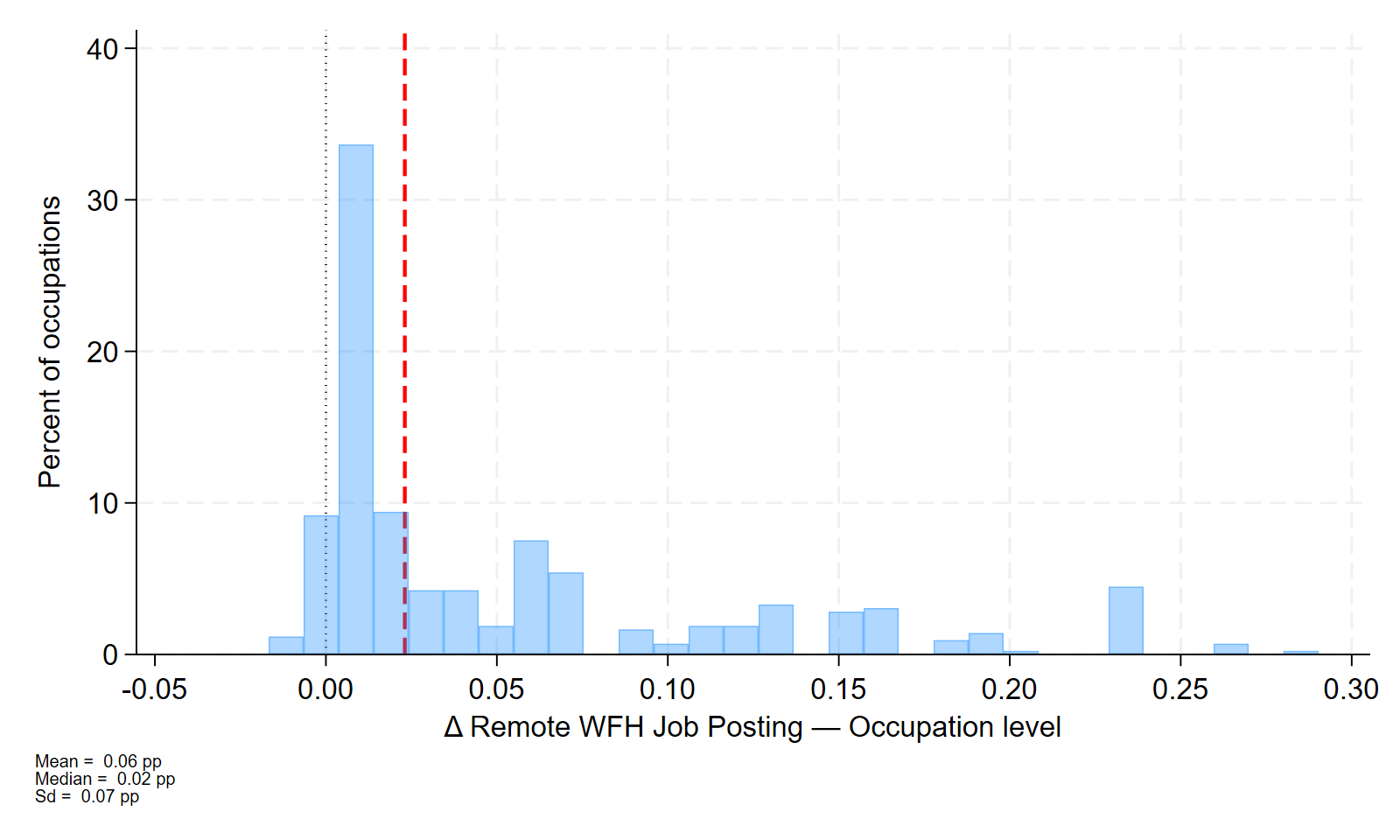}
    \centering
    \figurenotes{Author's elaboration using WFH job postings data from \citet{hansen2023remote}. The figure shows the distribution, across occupations, of the difference between the share of hybrid and remote-work job postings in the post-pandemic period (2023--2024) and the pre-pandemic period (2019); occupation codes are converted from the original SOC 3-digit classification used in the job postings data to the Census 2010 occupation codes used in the ACS. Occupations with a value at or above the sample median are classified as treated; those below are classified as control (see Subsection~\ref{subsection:treatment_defn}).}
\end{figure}

\begin{figure}[H]
    \caption{Average of 5 Normalized Job Characteristics (O*Net)}\label{fig:onet2025_five}
    \begin{tabular}{c}
        (a) Pre  \\
        \includegraphics[width=0.8\textwidth]{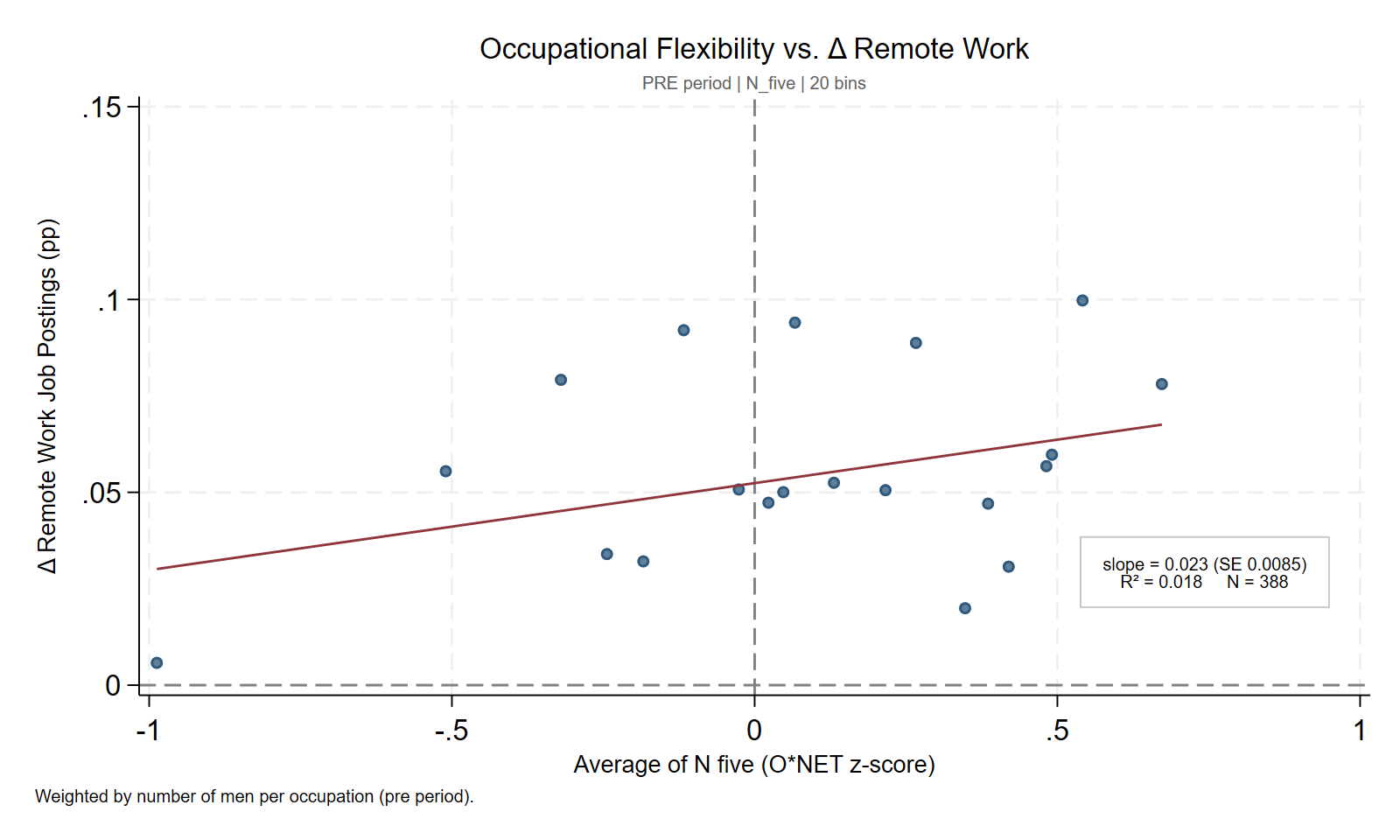} \\
        (b) Post \\ \includegraphics[width=0.8\textwidth]{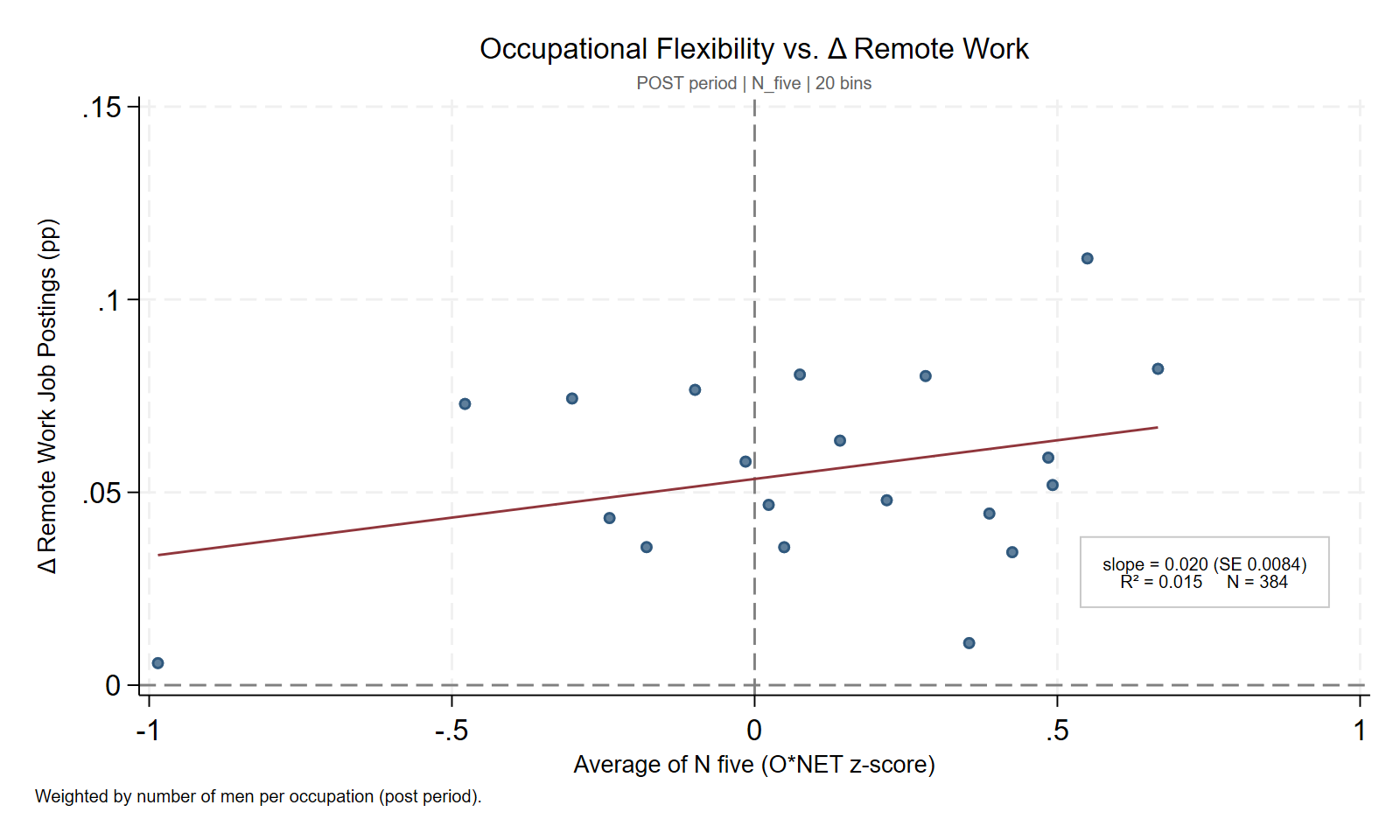} 
    \end{tabular}
    \centering
    \begin{minipage}{\textwidth}
    \small
   \textit{Notes}: This figure plots, by treatment-intensity bin, the average of five normalized O*NET 2025 job characteristics across Census 2010 occupations, weighted by ACS male-employment counts. Following \citet{goldin_grand_2014}, the index averages \emph{Time pressure}, \emph{Contact with others}, \emph{Establishing and maintaining interpersonal relationships}, \emph{Structured work}, and \emph{Freedom to make decisions}; higher values indicate greater job inflexibility (see \autoref{tab:desc_stat_onet} for construction details). Panel (a) uses pre-COVID (ACS 2019) employment weights; panel (b) uses post-COVID (ACS 2023) weights. The relationship between our occupation-level treatment intensity (Subsection~\ref{subsection:treatment_defn}) and this composite flexibility measure is relatively flat.
    \end{minipage}
\end{figure}

\begin{figure}[H]
    \caption{Average of 7 Normalized Job Characteristics (O*Net)}\label{fig:onet2025_seven}
    \begin{tabular}{c}
        (a) Pre  \\
        \includegraphics[width=0.7\textwidth]{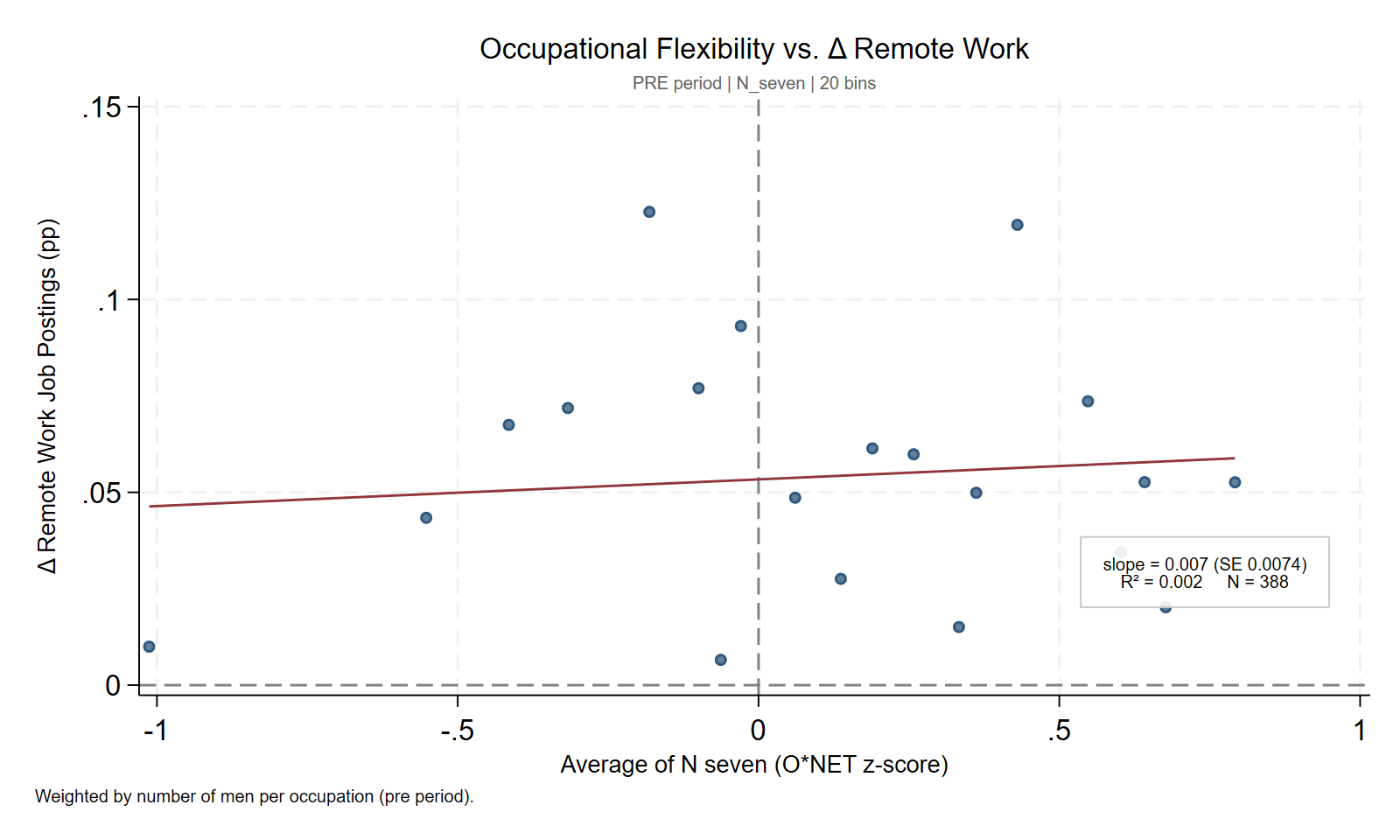} \\
        (b) Post \\ \includegraphics[width=0.7\textwidth]{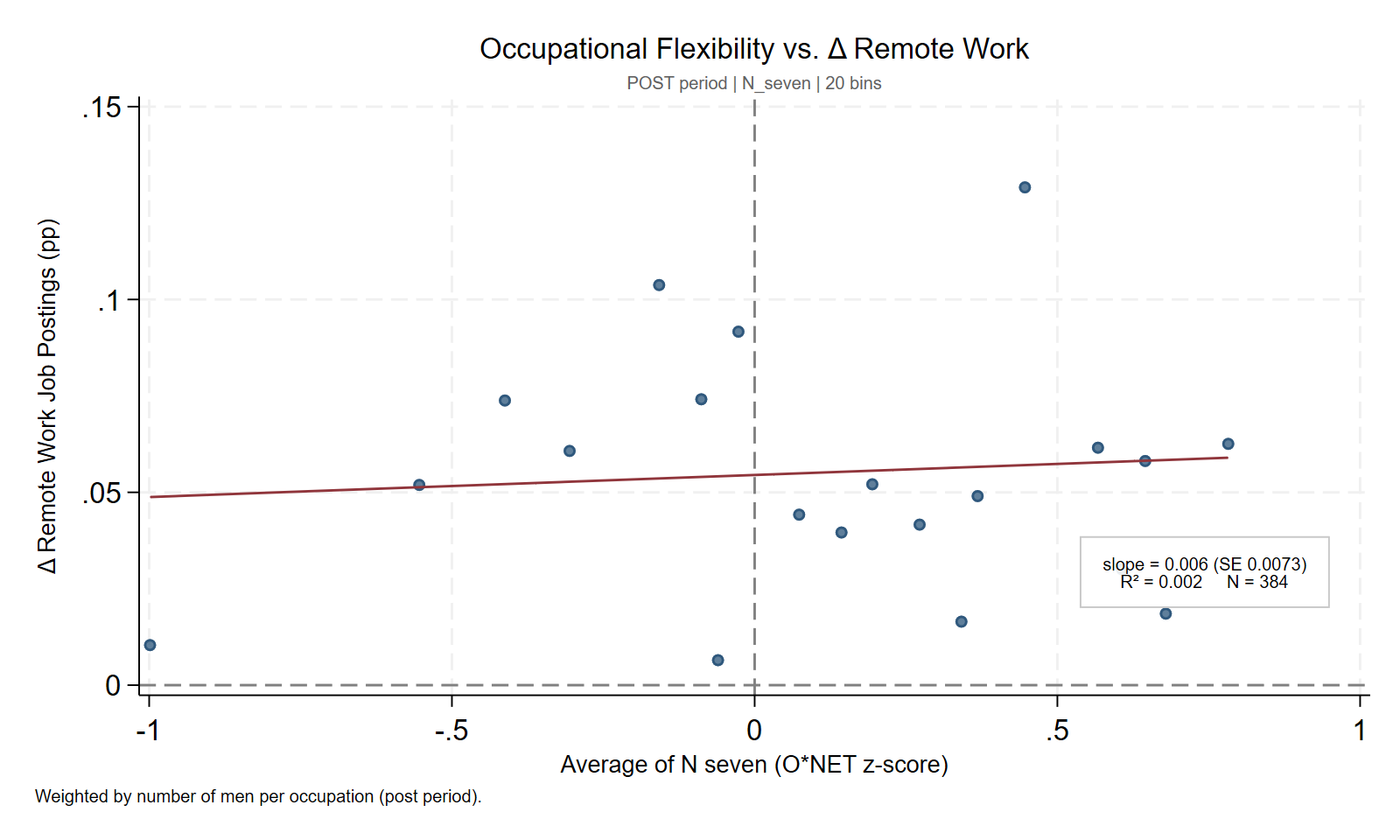} 
    \end{tabular}
    \centering
    \begin{minipage}{\textwidth}
    \small
   \textit{Notes}: This figure extends \autoref{fig:onet2025_five} by adding two O*NET characteristics -- \emph{Face-to-face discussions} and \emph{Frequency of decision making} -- to construct a seven-characteristic index of job inflexibility, following \citet{goldin_grand_2014}; higher values indicate more structured, less substitutable jobs (see \autoref{tab:desc_stat_onet}). Values are weighted by ACS male employment at the occupation level. Panel (a) uses 2019 (pre-pandemic) weights; panel (b) uses 2023 (post-pandemic) weights. As with the five-characteristic index, our occupation-level treatment status (Subsection~\ref{subsection:treatment_defn}) is not systematically related to this flexibility measure.
    \end{minipage}
\end{figure}

\begin{figure}[H]
    \caption{Effects of Husbands' WFH Shock on Own Share of the Day Spent in Labor Production}\label{fig:dynamic_laborprod_men_ATUS} 
    \begin{tabular}{c}
        %Working
        \emph{(a) Working (ATUS)} \\
        \includegraphics[width=0.5\linewidth]{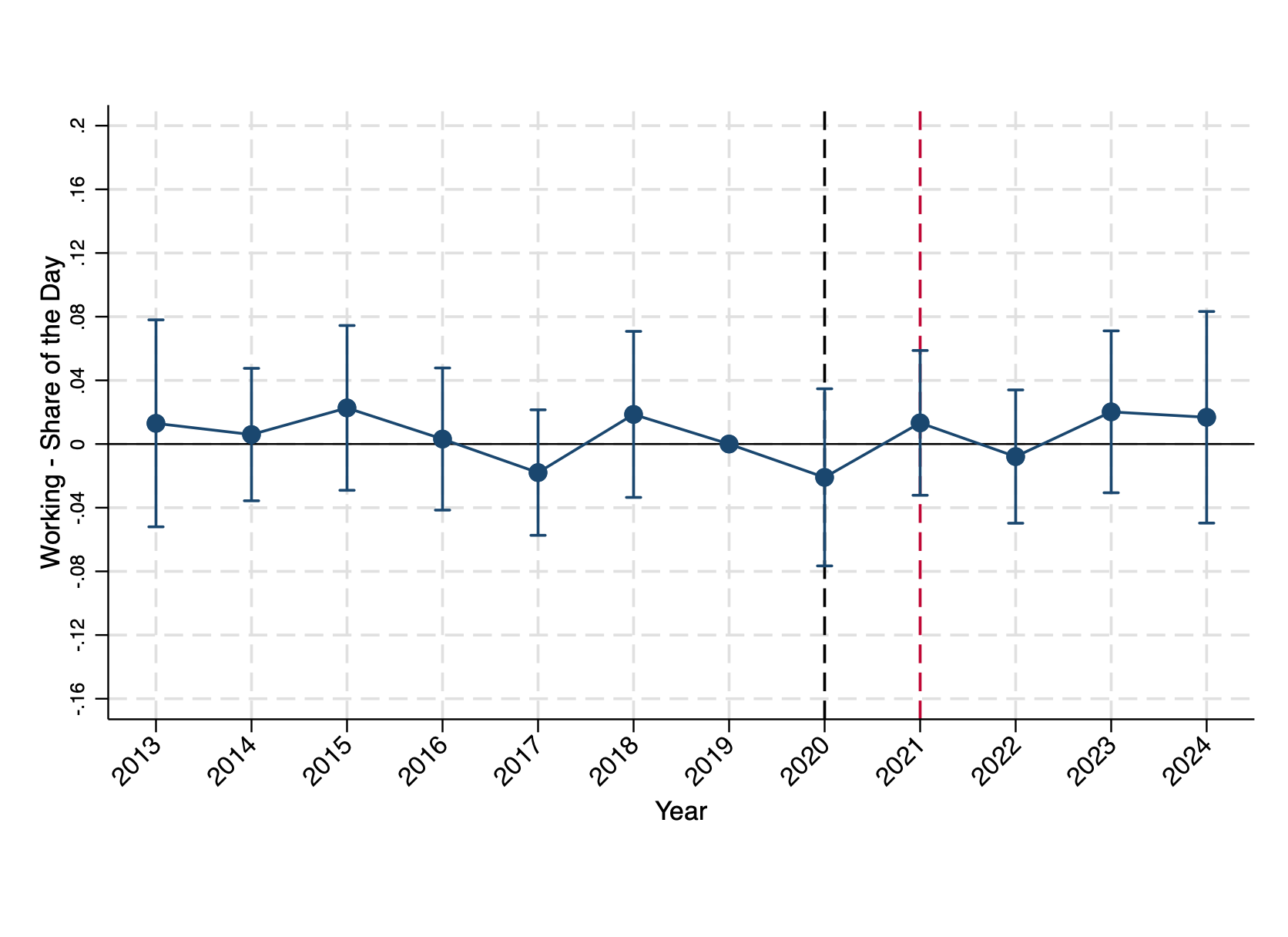}
        \\
        \emph{(b) Mainwork From Home (ATUS)}  \\
        \includegraphics[width=0.5\linewidth]{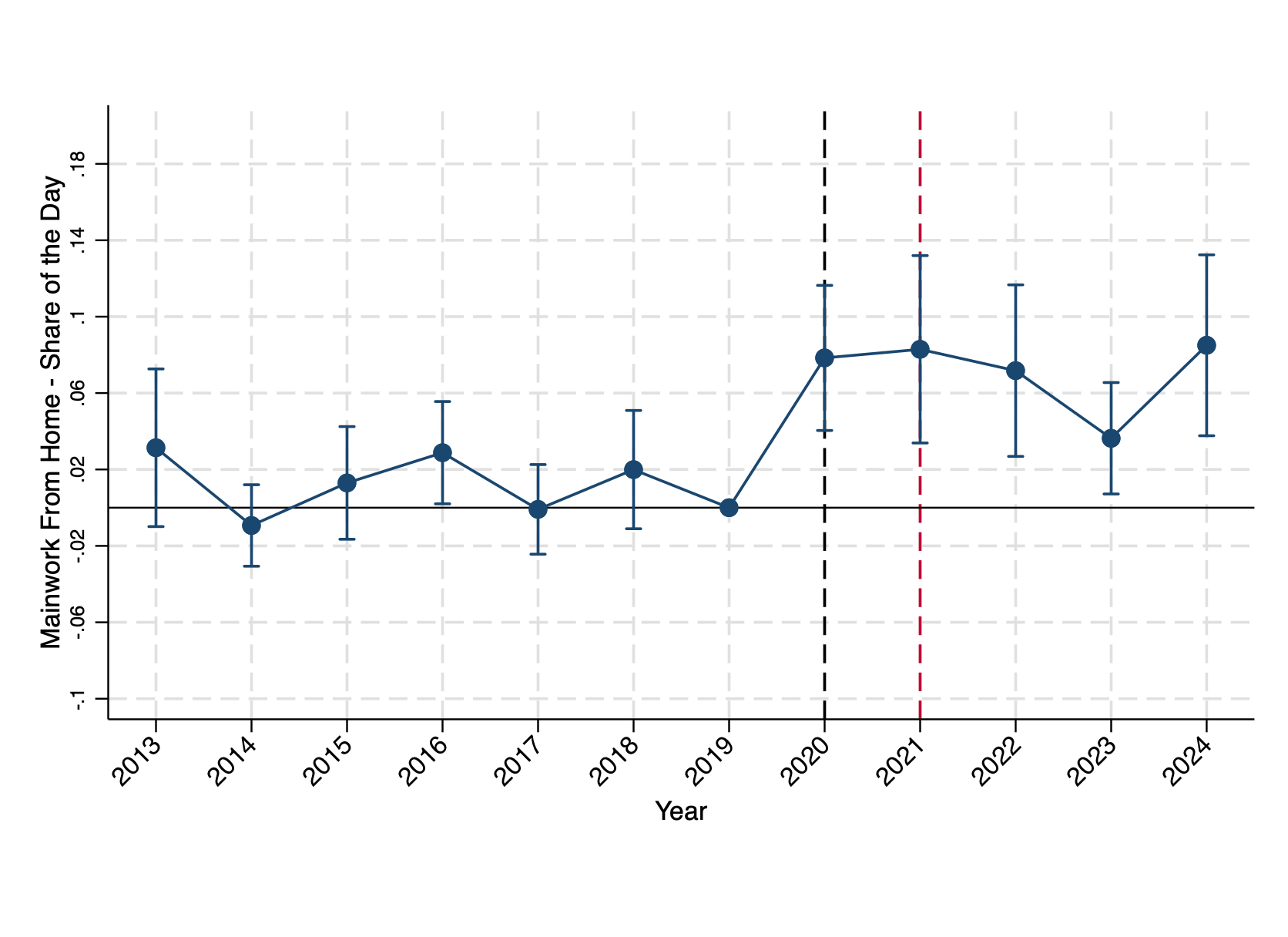} \\
        %Mainwork outside home
        \emph{(c) Mainwork Outside Home (ATUS)} \\
        \includegraphics[width=0.5\linewidth]{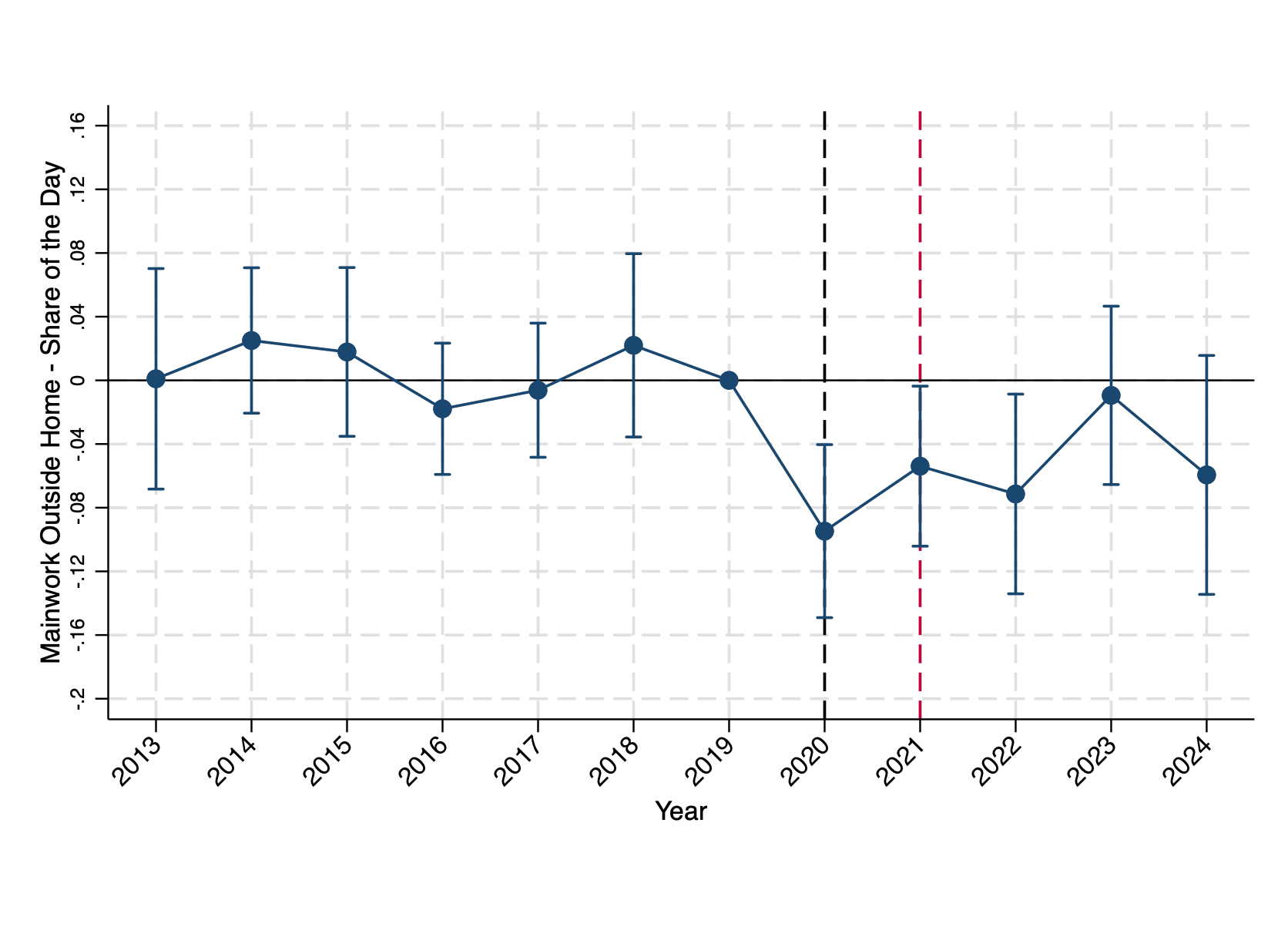}
    \end{tabular}
    \centering
    \vspace{0.5em}
    \begin{minipage}{\textwidth} \small
     \textit{Notes}: This figure plots yearly coefficients from the dynamic difference-in-differences specification in Equation~\eqref{eq:2}, estimated on the women-respondent ATUS sample (wives' own time-use diaries) of dual-parent households with children of child-rearing age, comparing wives whose husbands were in occupations with large WFH shocks to those whose husbands were in occupations with smaller changes (Subsection~\ref{subsection:treatment_defn}). The outcome is the wife's share of the 24-hour day spent working (overall). All regressions include state-by-time and husband's-occupation fixed effects, and control for the wife's age, number of children, and education levels of both spouses. Confidence intervals are shown at the 95\% level and standard errors are clustered at the husband's occupation level.
    \end{minipage}
\end{figure}
\clearpage

%\begin{figure}[H]  \includegraphics[width=0.9\linewidth]{figs/evolution_bloom_wfh_ind.png} \centering
% \caption{Evolution of WFH by industry (2 digits).}\label{fig:evolution_WFH_ind}
%\figurenotes{Author's elaboration using WFH job postings data from \citet{hansen2023remote}. Classification based on NAICS 2022 2-digits.}
%\end{figure}

\begin{figure}[H]
    \caption{Effects of Husbands' WFH Shock on Wives' Share of the Day Working}\label{fig:atus_dynamic_DiD_working_share_wives} 
    \begin{tabular}{c}
        %Working
        \includegraphics[width=0.85\linewidth]{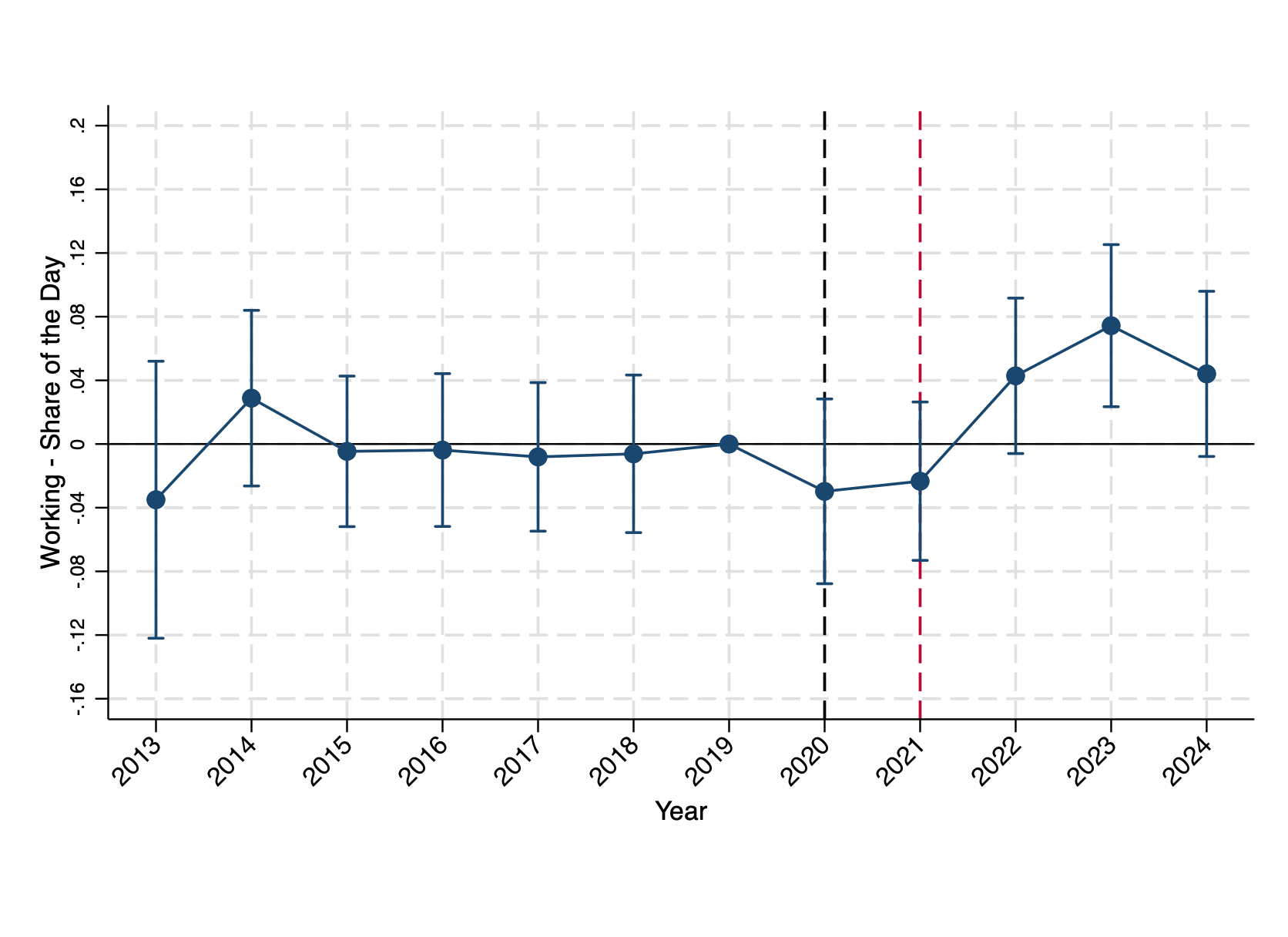}
    \end{tabular}
    \centering
    \vspace{0.5em}
    \begin{minipage}{\textwidth} \small
    \textit{Notes}: We use our main ATUS sample from 2013 to 2024. We plot coefficients from our dynamic difference-in-differences specification \eqref{eq:2} comparing men who were in occupations with large WFH shocks to those in occupations with smaller changes. The outcome is the female spouse's share of day working. All regressions include state-by-time and husband-occupation fixed effects, and control for wife’s age, number of children, and education levels of both spouses. Confidence intervals are shown at the 95\% level and standard errors are clustered at the husband’s occupation level.
    \end{minipage}
\end{figure}
\clearpage

\begin{figure}[H]
    \caption{Adjusted Labor Market Trends by Treatment Group}\label{fig:ddid_main_by_group}
    \begin{tabular}{cc}
        (a) Log Earnings & (b) Usual Weekly Work Hours \\
        \includegraphics[width=.45\linewidth]{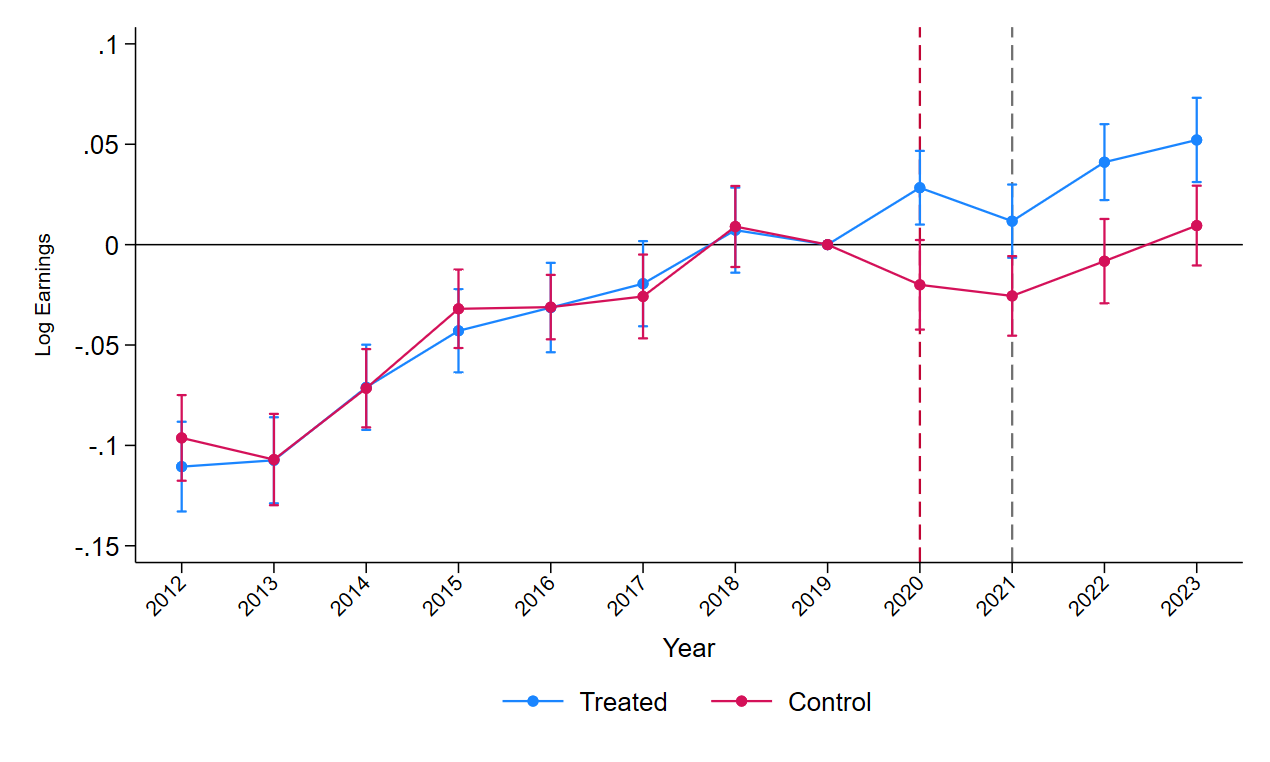} &
        \includegraphics[width=.45\linewidth]{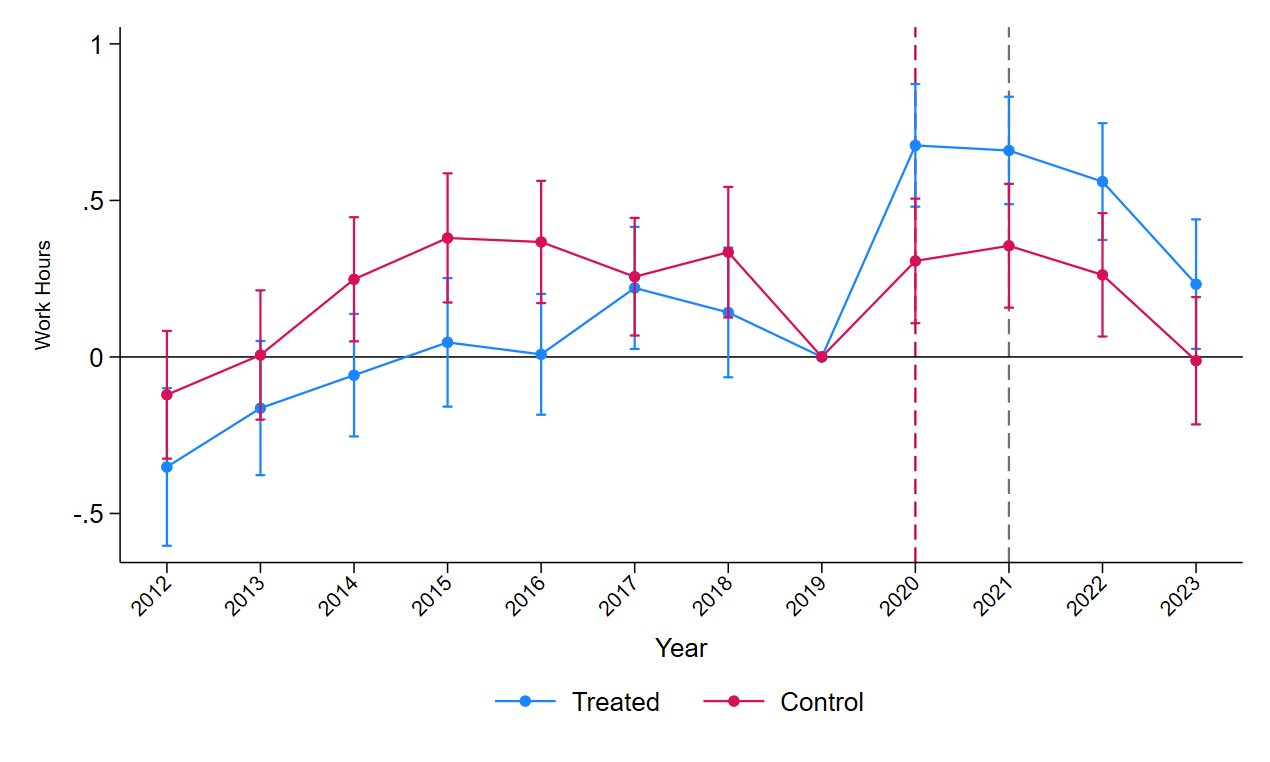} \\
          (c) Weeks Worked & (d) Part-time Status \\
        \includegraphics[width=.45\linewidth]{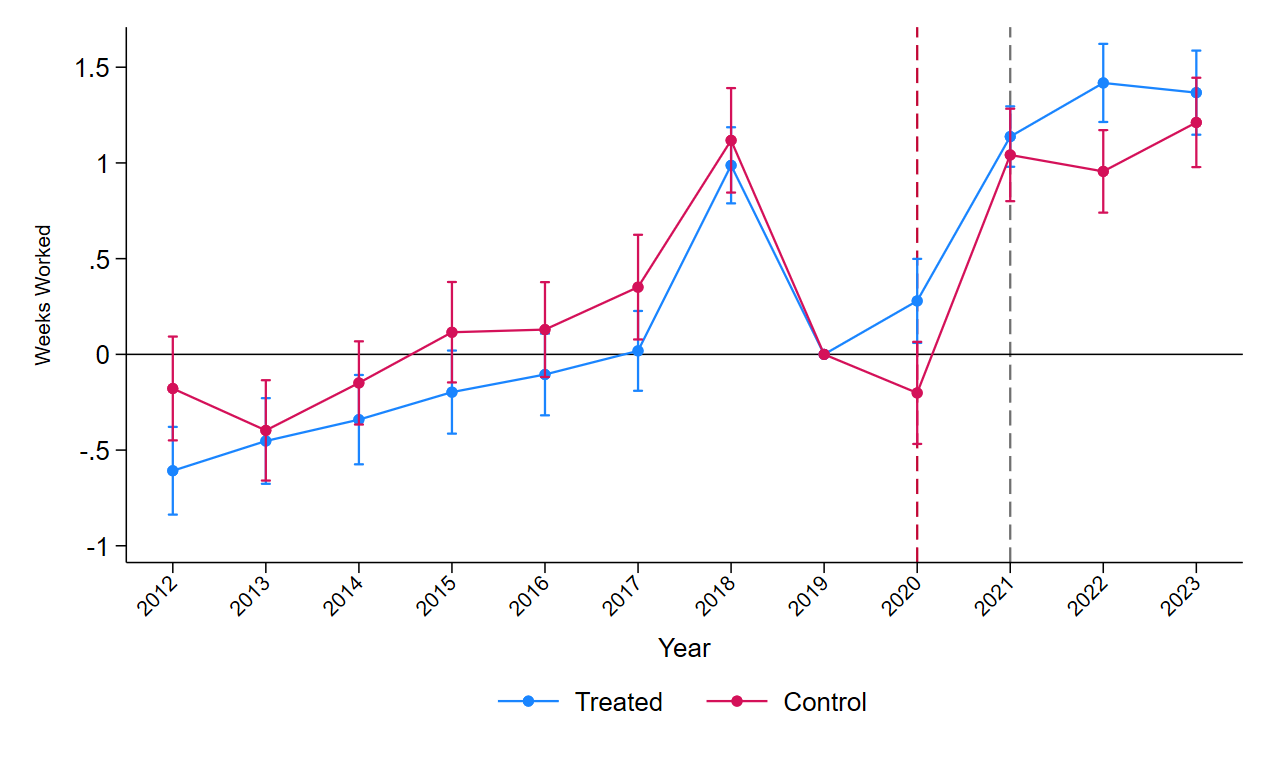} &
        \includegraphics[width=.45\linewidth]{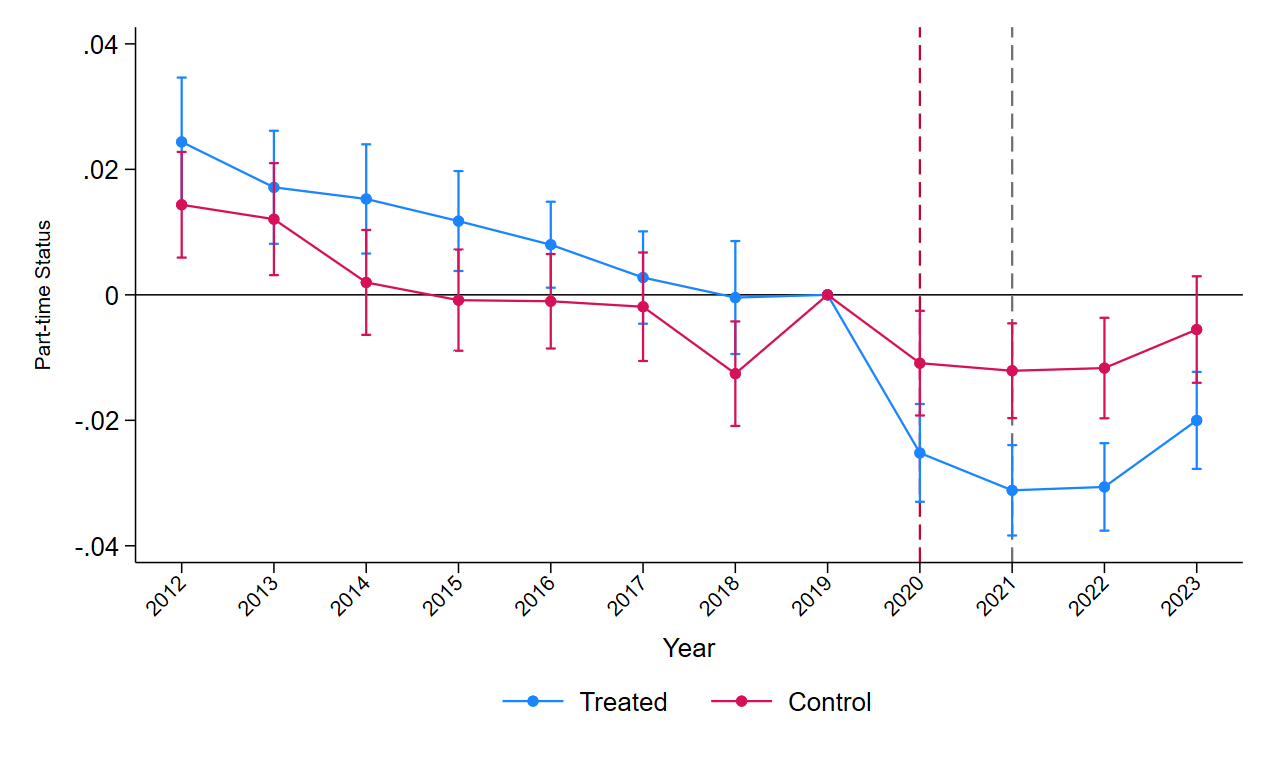} 
    \end{tabular}
    \centering
    \vspace{0.5em}
    \begin{minipage}{\textwidth} \small
    \textit{Notes}: Each panel plots adjusted yearly trends in the wife's labor market outcomes -- log annual earnings (a), usual weekly hours worked (b), weeks worked (c), and part-time status (d) -- for the treated (blue line) and control (red line) groups separately (Subsection~\ref{subsection:treatment_defn}), using the 2013--2024 ACS sample of dual-parent households with children of child-rearing age. The plotted values are predicted values from separate regressions estimated by group:
    \[
    y_{i(H),s,t} = f_s + f_{o(H)} + \sum_{\substack{y=2012 \\ y \neq 2019}}^{2023} \beta^y \cdot 1(\text{year}(t) = y) + X_{i(H),t} \delta + \epsilon_{i(H),s,t},
    \]
    where outcomes are residualized on state ($f_s$) and husband's-occupation ($f_{o(H)}$) fixed effects and controls for the wife's age, number of children, and both spouses' education ($X_{i(H),t}$). The omitted year is 2019.
    \end{minipage}
\end{figure}
\clearpage

\begin{figure}[H]
    \caption{Dynamic Effects of Husbands' WFH Shock on Wives' Employment and Log Wages}\label{fig:ddid_emp_wages}
    \begin{tabular}{c}
        (a) Annual Employment \\
        \includegraphics[width=.75\linewidth]{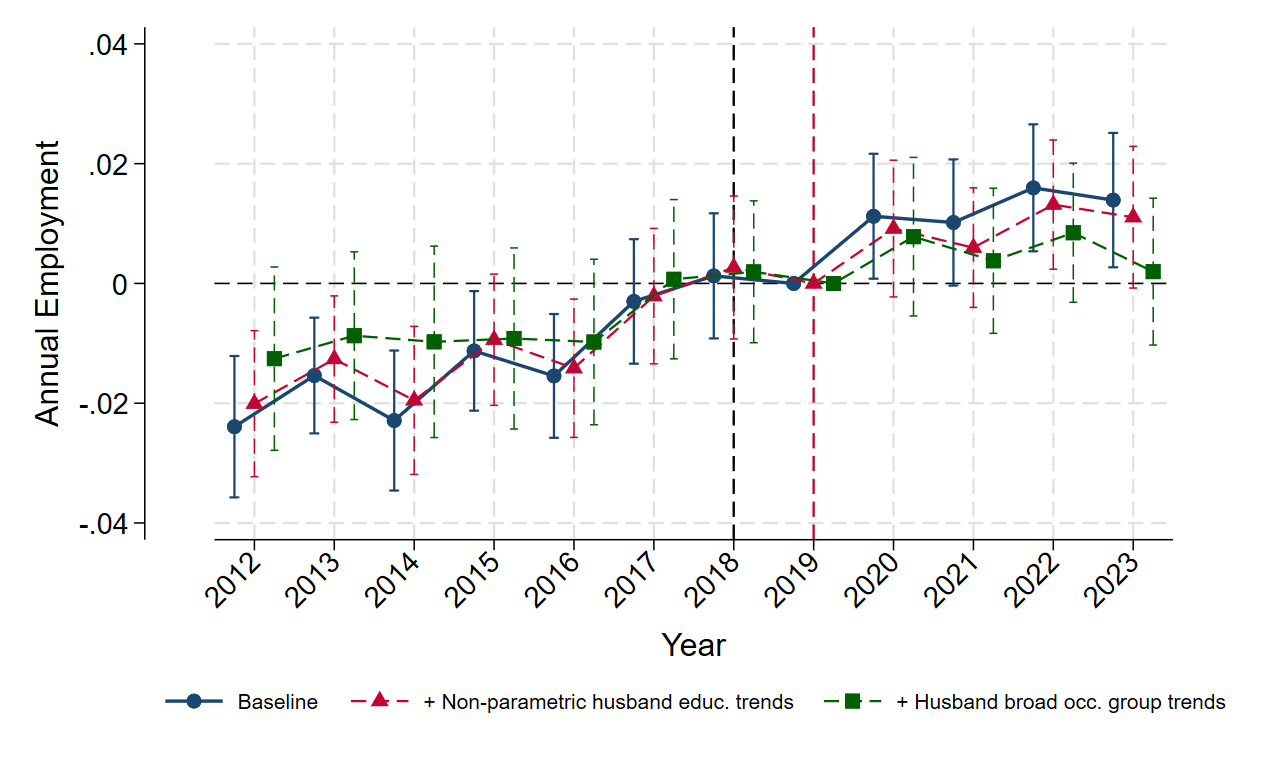} \\
        (b) Log Wages \\ \\
        \includegraphics[width=.75\linewidth]{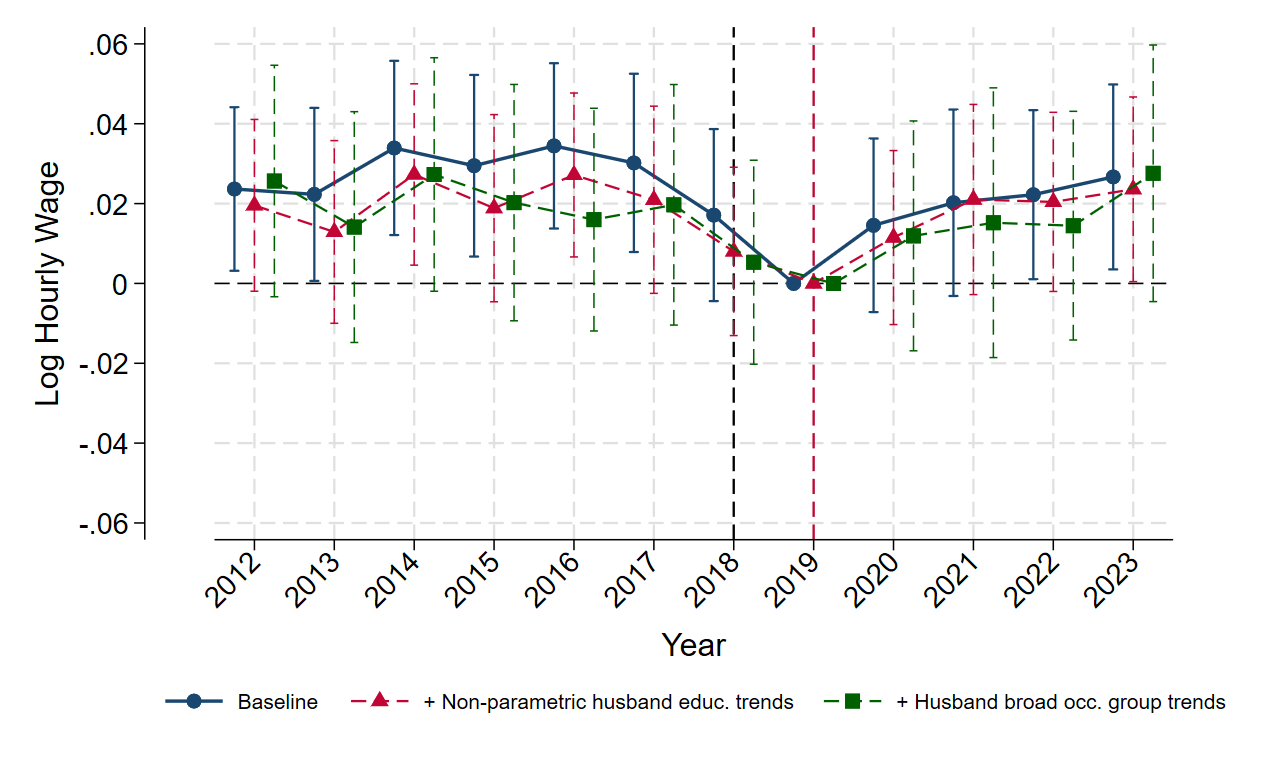}  
    \end{tabular}
    \centering
    \vspace{0.5em}
    \begin{minipage}{\textwidth} \small
   \textit{Notes}: Each panel plots yearly coefficients from the dynamic difference-in-differences specification in Equation~\eqref{eq:2}, estimated on the 2013--2024 ACS sample of dual-parent households with children of child-rearing age, for the wife's annual employment status (panel a) and log hourly wage conditional on employment (panel b). Each specification additionally controls non-parametrically for husband's-education-by-year and husband's-broad-occupation-group-by-year fixed effects, as discussed in Section~\ref{section:robustness}. The omitted year is 2019. All regressions include state-by-time and husband's-occupation fixed effects, and control for the wife's age, number of children, and education levels of both spouses. Confidence intervals are shown at the 95\% level and are based on standard errors clustered at the husband's occupation level.
    \end{minipage}
\end{figure}

\clearpage

\begin{figure}[H]\centering
    \caption{Heterogeneous Effects of Husbands' WFH Shock on Wives' Labor Market Outcomes by Spouses' Education Level}\label{fig:acs_het_edlevel}
    \begin{tabular}{c}
        A. By Women Education Level \\ 
        \includegraphics[width=0.7\linewidth]{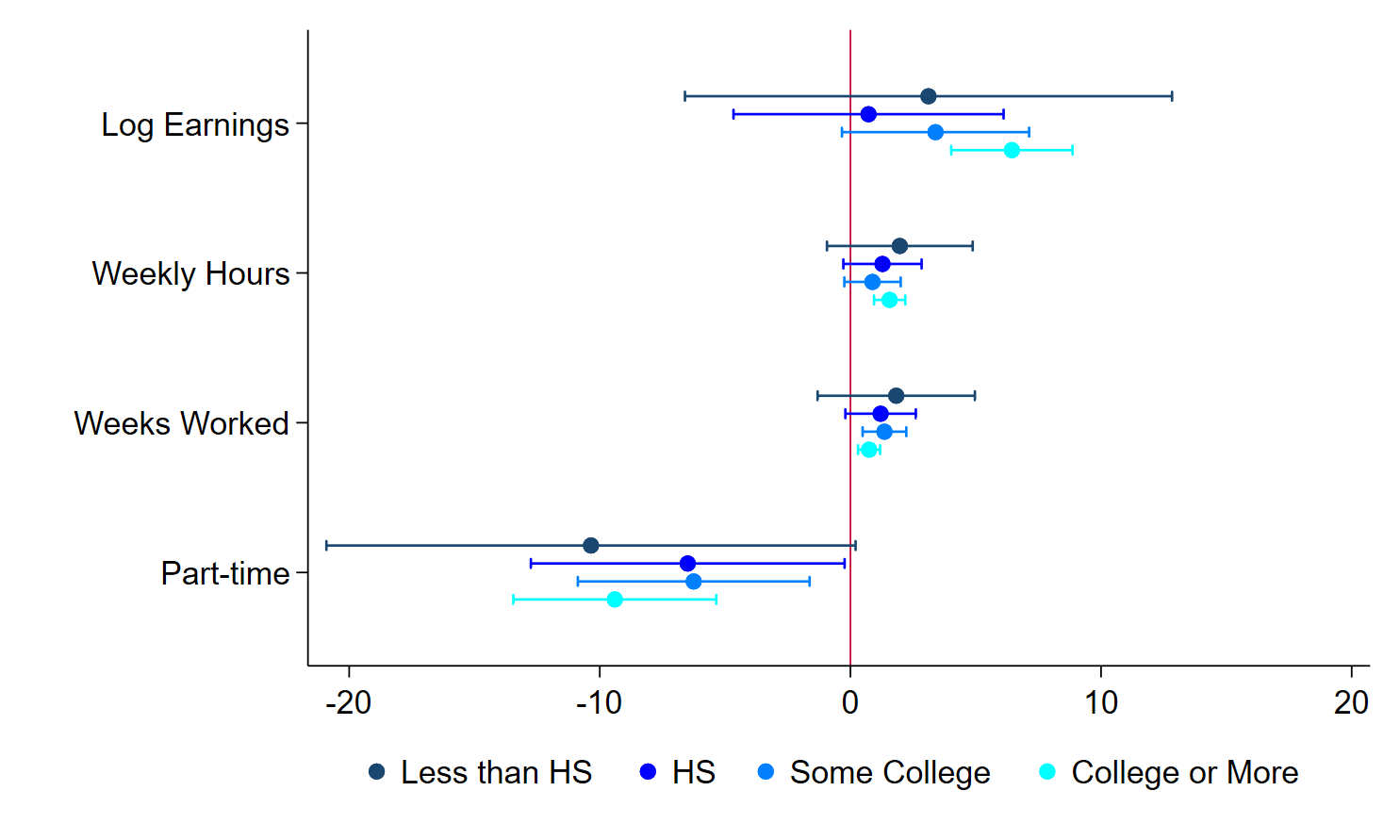} \\
        B. By Male Spouse Education Level \\ \includegraphics[width=0.7\linewidth]{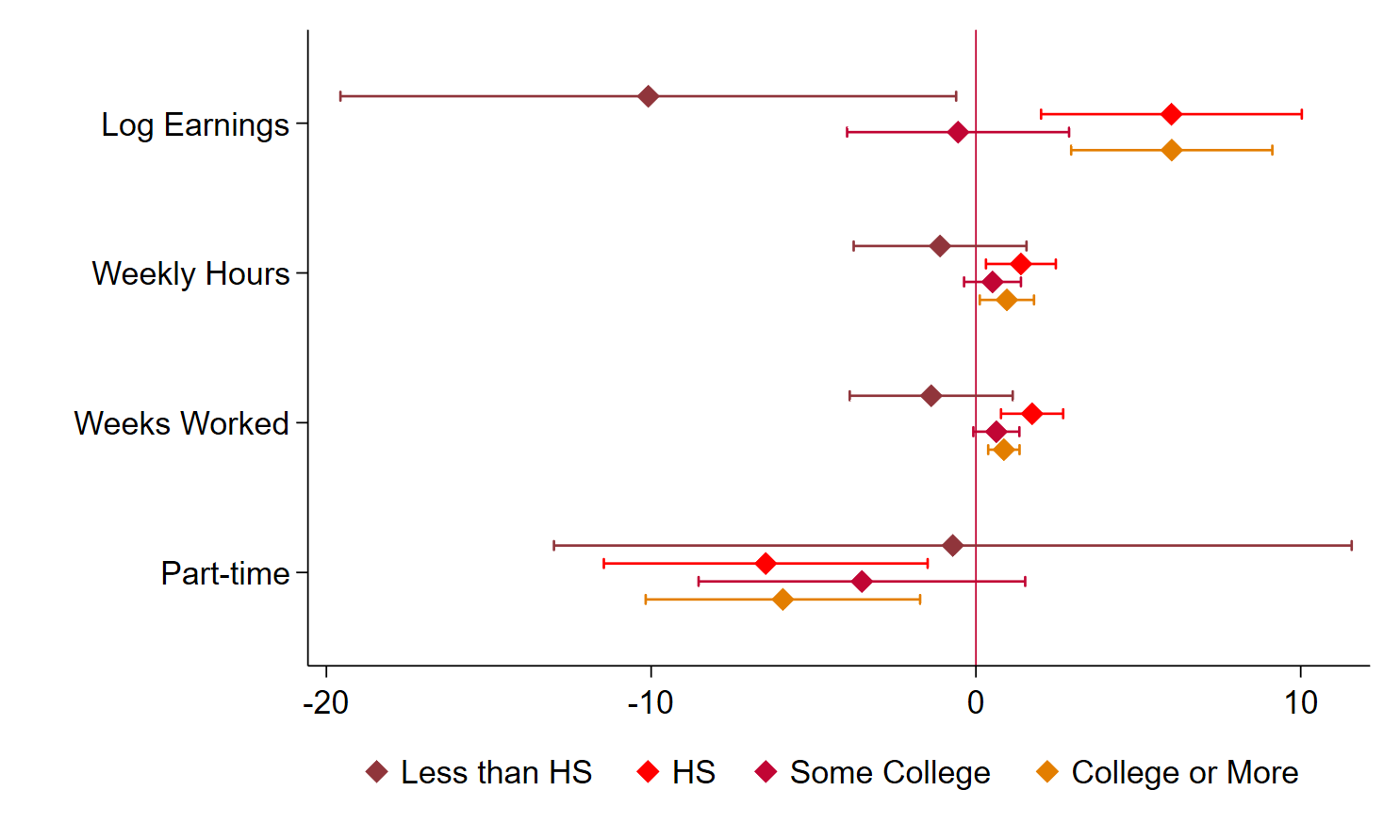} 
    \end{tabular}
    \begin{minipage}{\textwidth} \small
         \textit{Notes}: This figure plots standardized coefficients from separate estimations of Equation~\eqref{eq:1} on subsamples of the 2013--2024 ACS sample of dual-parent households with children of child-rearing age. In panel A, the sample is split into four subsets by the wife's education: less than high school (HS), completed HS, some college, and completed college or above. In panel B, the sample is instead split by the husband's education, using the same four categories. For each subsample and outcome (listed on the $y$-axis), we plot the post-COVID interaction coefficient, $\text{High-}\Delta\text{WFH}_{o(h(i))} \times Post_t$, from Equation~\eqref{eq:1}. Coefficients for weekly hours, weeks worked, and part-time status are standardized by dividing by the subsample's pre-treatment mean; the coefficient for log earnings is multiplied by 100 (percentage-point scale). All regressions include state-by-time and husband's-occupation fixed effects, and control for the wife's age, number of children, and education levels of both spouses (excluding the split variable). Confidence intervals are shown at the 95\% level and standard errors are clustered at the husband's occupation level.
    \end{minipage}
\end{figure}
\clearpage

\begin{figure}[H]
    \caption{Effect of Husbands' WFH Shock on Wives' Labor Market Outcomes, by Age of Youngest Child}\label{fig:acs_dynamic_DiD_ageyc}
    \begin{tabular}{cc}
        (a) Log Earnings  &  (b) Weekly Work Hours \\
        \includegraphics[width=0.45\linewidth]{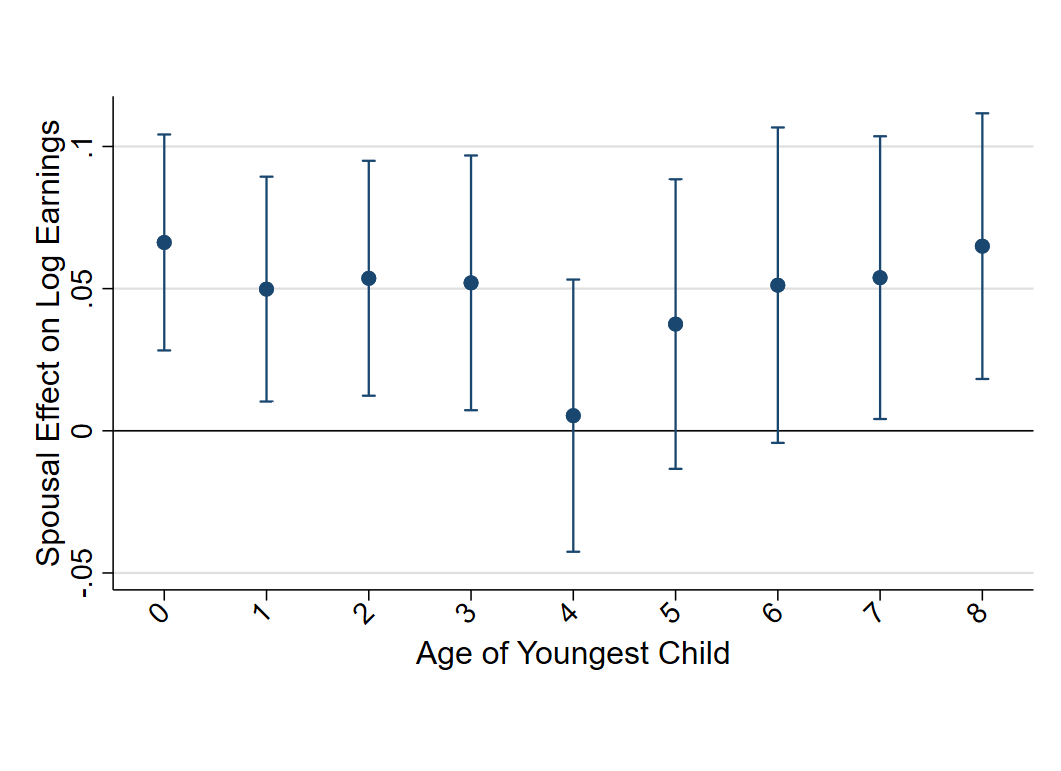} & \includegraphics[width=0.45\linewidth]{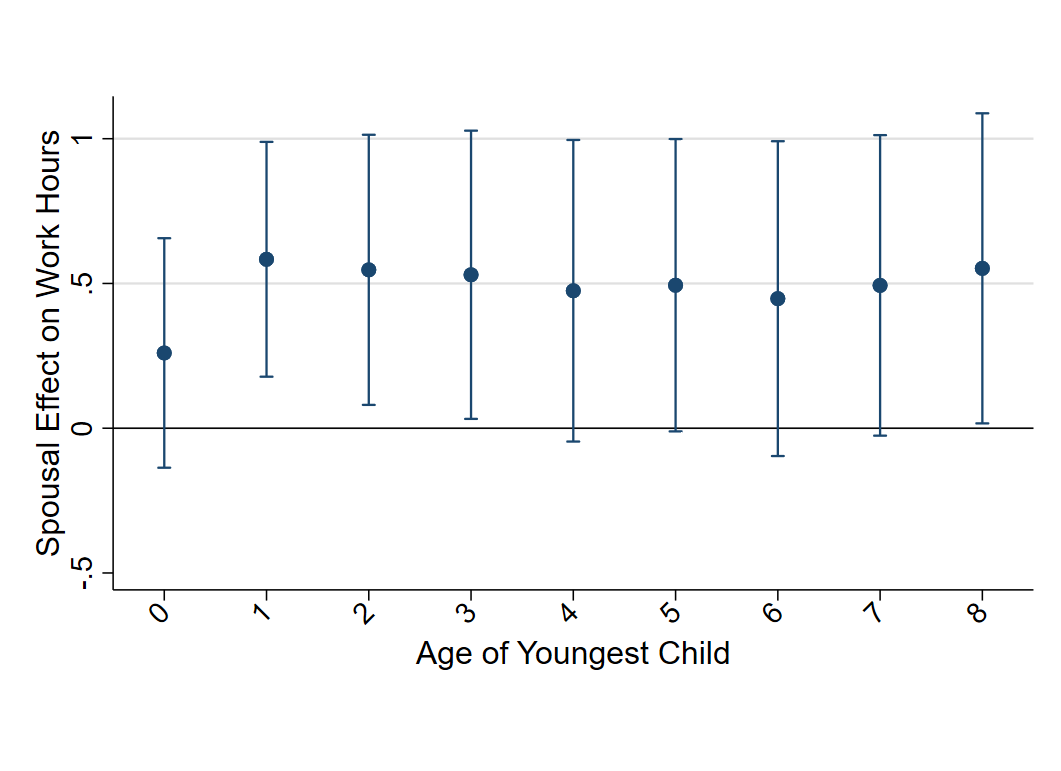} \\
        (c) Weeks Worked & (d) Part-time Status \\ \includegraphics[width=0.45\linewidth]{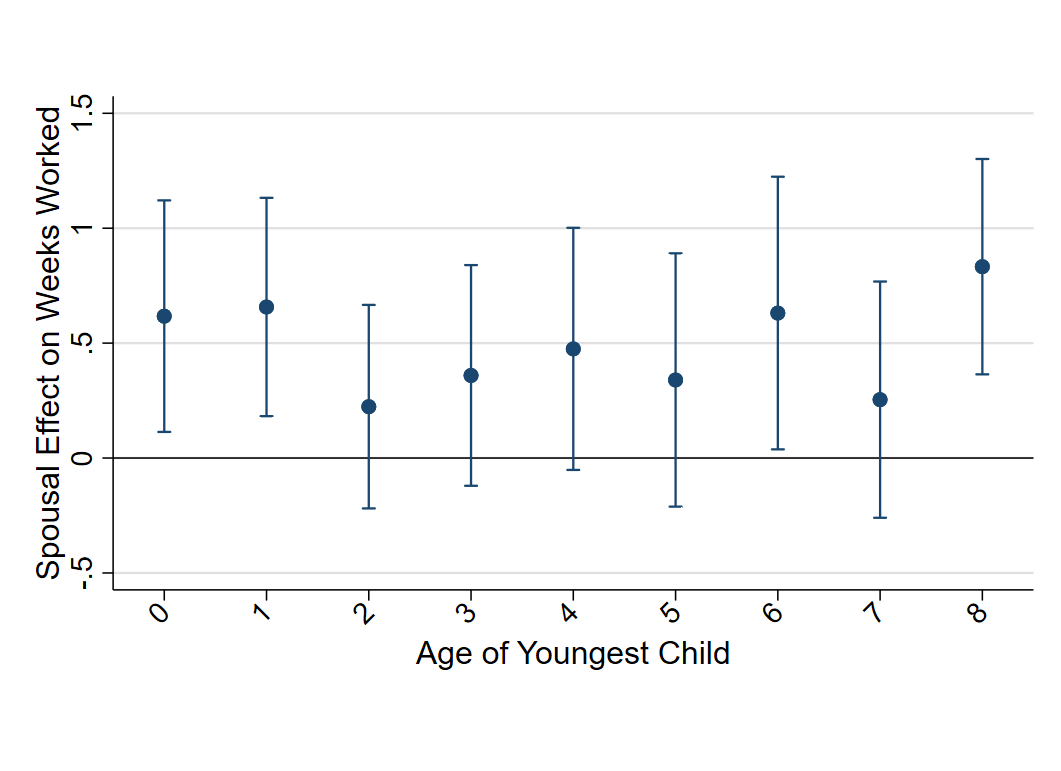} &         \includegraphics[width=0.45\linewidth]{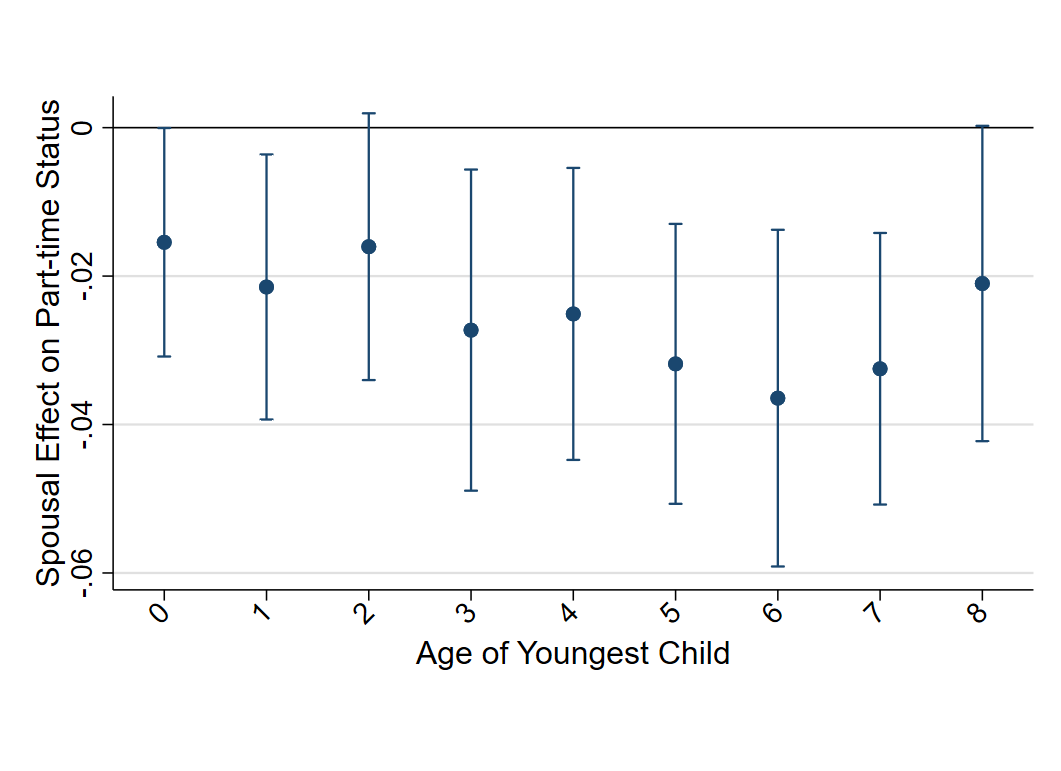}
        \end{tabular}
    \centering
    \vspace{0.5em}
    \begin{minipage}{\textwidth} \small
    \textit{Notes}: Each panel plots the post-COVID interaction coefficient from Equation~\eqref{eq:1}, $\text{High-}\Delta\text{WFH}_{o(h(i))} \times Post_t$, and its 95\% confidence interval, estimated separately on nine subsamples of the 2013--2024 ACS sample of dual-parent households, split by the age of the household's youngest child (0 to 8, one subsample per age in years). All regressions include state-by-time and husband's-occupation fixed effects, and control for the wife's age, number of children, and education levels of both spouses. Standard errors are clustered at the husband's occupation level.
    \end{minipage}
\end{figure}
\clearpage

\begin{figure}[H]
 \caption{Dynamic Effect of Husbands' WFH Shock on Wives' WFH Uptake and Commuting
 }\label{fig:markettimeuseI}
    \begin{tabular}{cc}
    \multicolumn{2}{c}{\emph{(a) Women's WFH Uptake}} \\
    \multicolumn{2}{c}{\includegraphics[width=0.8\linewidth] {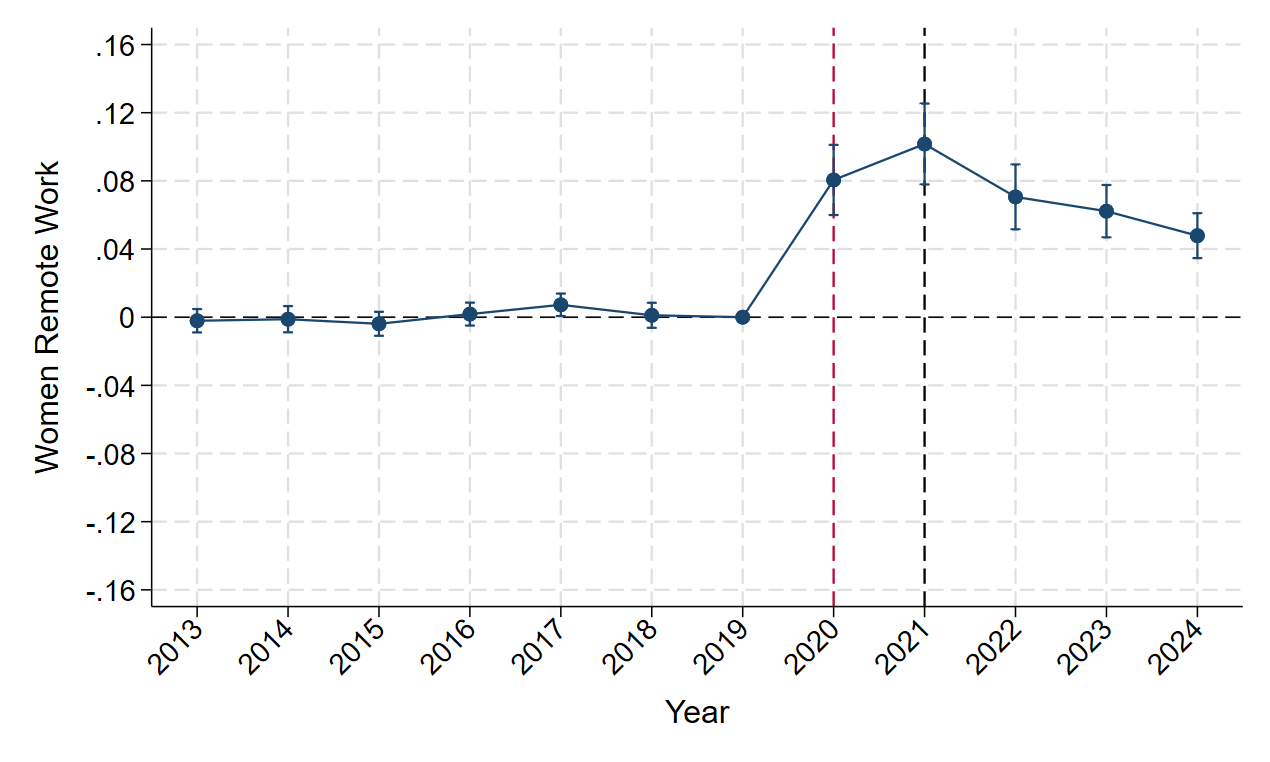}} \\
    \multicolumn{2}{c}{\emph{(b) Women's Commuting Time}} \\
    \multicolumn{2}{c}{\includegraphics[width=0.8\linewidth]{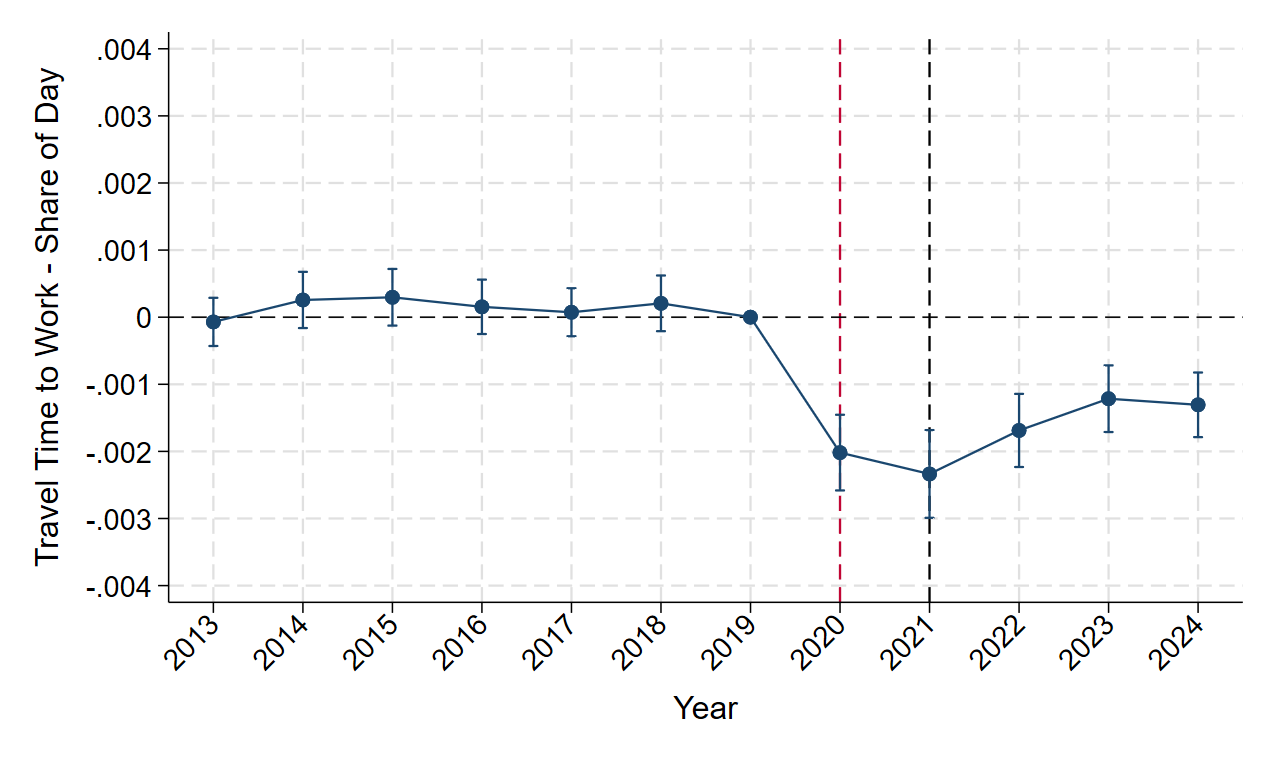}} \\
    \end{tabular}
     \vspace{0.5em}
     \centering
    \vspace{0.5em}
    \begin{minipage}{0.99\linewidth} \small
     \textit{Notes}: This figure plots yearly coefficients from the dynamic difference-in-differences specification in Equation~\eqref{eq:2}, estimated on the 2013--2024 ACS sample of dual-parent households with children of child-rearing age, comparing wives whose husbands were in occupations with large WFH shocks to those whose husbands were in occupations with smaller changes (Subsection~\ref{subsection:treatment_defn}). Panel (a) plots the effect on the wife's own probability of working fully from home (self-reported, week prior to interview). Panel (b) plots the effect on the wife's commuting time as a share of the day. We include state-by-time and husband's-occupation fixed effects, and control for the wife's age, number of children, and education levels of both spouses. Confidence intervals are shown at the 95\% level and standard errors are clustered at the husband's occupation level.
    \end{minipage}
\end{figure}
\clearpage

%\begin{figure}[H]
%    \caption{Event Study for Decomposition of Spousal Spillovers vs Direct Correlated Effects (Occupation Version) }\label{fig:acs_dynamic_DiD_own_occ}
%    \begin{tabular}{cc}
%        (a) Log Earnings  &  (b) Weekly Work Hours \\
%        \includegraphics[width=0.45\linewidth]{figs/acs_dynamic_own_occ_logearnings.png} & \includegraphics[width=0.45\linewidth]{figs/acs_dynamic_own_occ_workhrs.png} \\
%        (c) Weeks Worked & (d) Part-time Status \\ \includegraphics[width=0.45\linewidth]{figs/acs_dynamic_own_occ_wkswork.png} &         \includegraphics[width=0.45\linewidth]{figs/acs_dynamic_own_occ_part_time.png} \\
 %       (e) WFH Uptake & (f) Commuting Time \\ \includegraphics[width=0.45\linewidth]{figs/acs_dynamic_own_occ_wfh_only.png} &         \includegraphics[width=0.45\linewidth]{figs/acs_dynamic_own_occ_trantime.png}
 %       \end{tabular}
 %   \centering
 %   \vspace{0.5em}
 %   \begin{minipage}{\textwidth} \small
 %   \textit{Notes}: For each main outcome, we estimate an event study version of our spillover decomposition analysis in (\ref{eq:2}) and plot the main yearly coefficients of interest along with 95\% CI: treatment from male spouse's occupation (blue circles), treatment from women college degree (red triangles), and their interaction (green squares). The coefficient associated with male spouse occupation captures effect of spousal WFH shock on women that have college degrees where bachelors have not seen large increases in WFH uptake.
  %  \end{minipage}
%\end{figure}
%\clearpage

\begin{figure}[H]
    \caption{Assortative Matching by Spousal WFH Exposure, Pre- and Post-COVID}\label{fig:assortative_matching}
    \begin{tabular}{c}
        (a) Pre-COVID \\
        \includegraphics[width=0.75\linewidth]{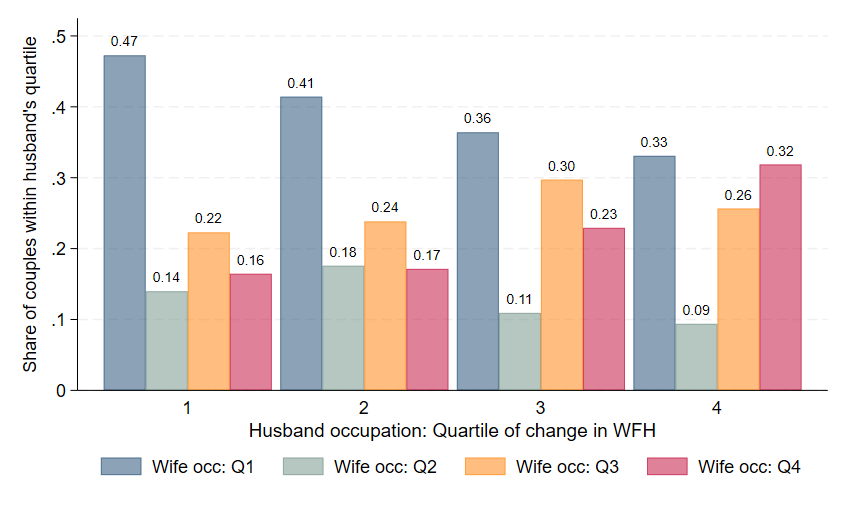} 
        \\ (b) Post-COVID  \\ 
        \includegraphics[width=0.75\linewidth]{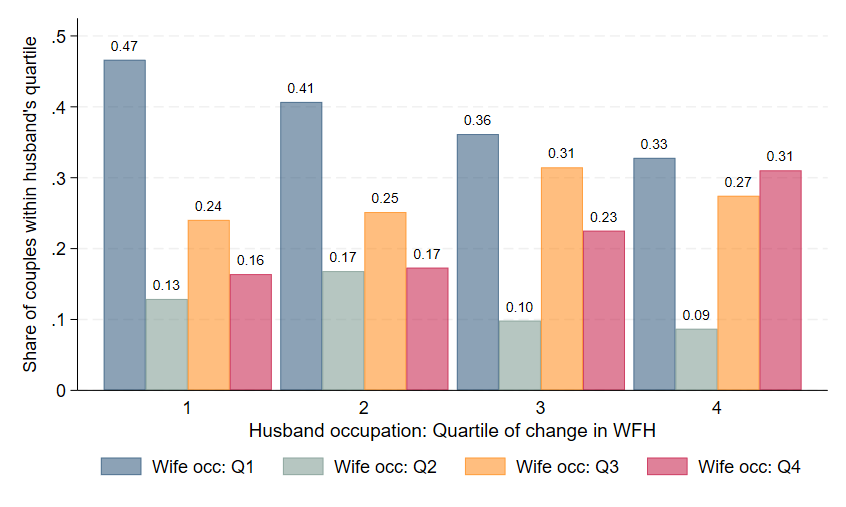}
    \end{tabular}
    \centering
    \vspace{0.5em}
    \begin{minipage}{\textwidth} \small
   \textit{Notes}: Each panel shows the distribution of wives across their own WFH-exposure quartiles, conditional on their husband's occupation-level WFH-exposure quartile, separately for the pre-COVID (panel a, before 2020) and post-COVID (panel b, 2020 onward) periods, using the pooled 2013--2024 ACS sample of dual-parent households with children of child-rearing age. Quartiles are based on the occupation-level change in WFH probability described in Subsection~\ref{subsection:treatment_defn}. The figure assesses whether assortative matching by WFH exposure changed differentially following the pandemic; a stable pattern across panels supports the identifying assumption used in Appendix~\ref{app:metrics}.
    \end{minipage}
\end{figure}
\clearpage

\begin{figure}[H]
    \caption{Changes in Fertility Outcomes by Husbands' WFH Shock}\label{fig:desc_fertility}
    \begin{tabular}{c}
        (a) Probability of Having a Child  \\
        \includegraphics[width=0.75\linewidth]{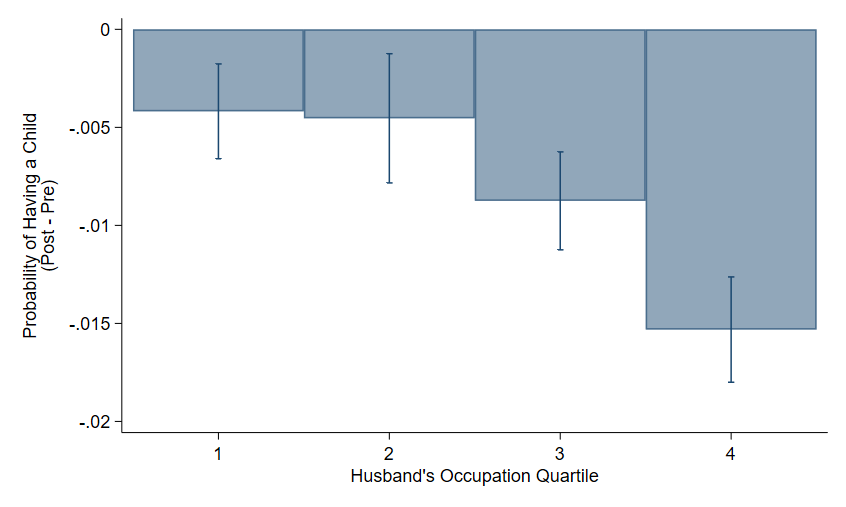} \\ (b) Number of Children \\ 
        \includegraphics[width=0.75\linewidth]{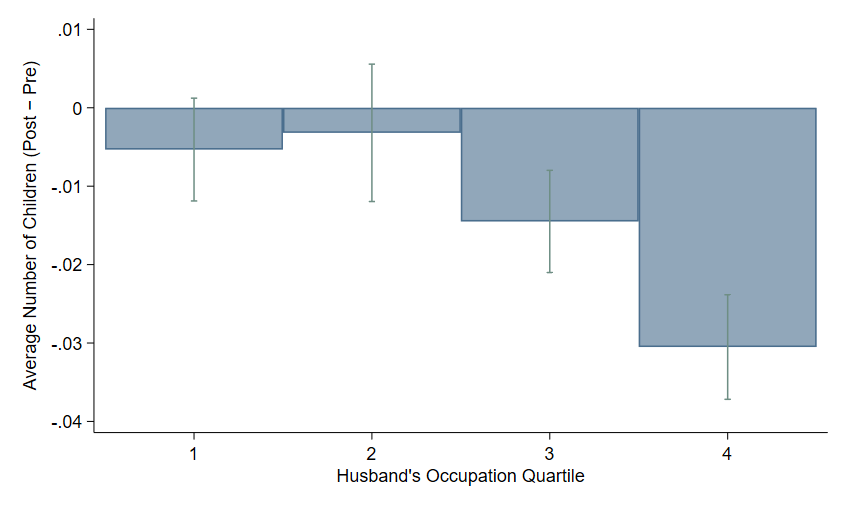}
    \end{tabular}
    \centering
    \vspace{0.5em}
    \begin{minipage}{\textwidth}   \small 
    \textit{Notes}: Each panel plots the change in a fertility-related outcome from the pre-COVID (before 2020) to the post-COVID (2020 onward) period, across quartiles of the husband's occupation-level WFH-exposure change (Subsection~\ref{subsection:treatment_defn}). Panel (a) plots the change in the probability of having had a child in the preceding 12 months; panel (b) plots the change in the average number of children in the household. The figure uses the pooled 2013--2024 ACS sample of dual-parent households and assesses whether fertility patterns shifted differentially across exposure groups in the post-pandemic period; corresponding regression estimates are reported in \autoref{tab:acs_fertility}.
    \end{minipage}
\end{figure}

\clearpage

\begin{figure}[H]
    \caption{Kernel Density Plot Showing Common Support for Reweighting}\label{fig:overlap}    
    \includegraphics[width=0.95\linewidth]{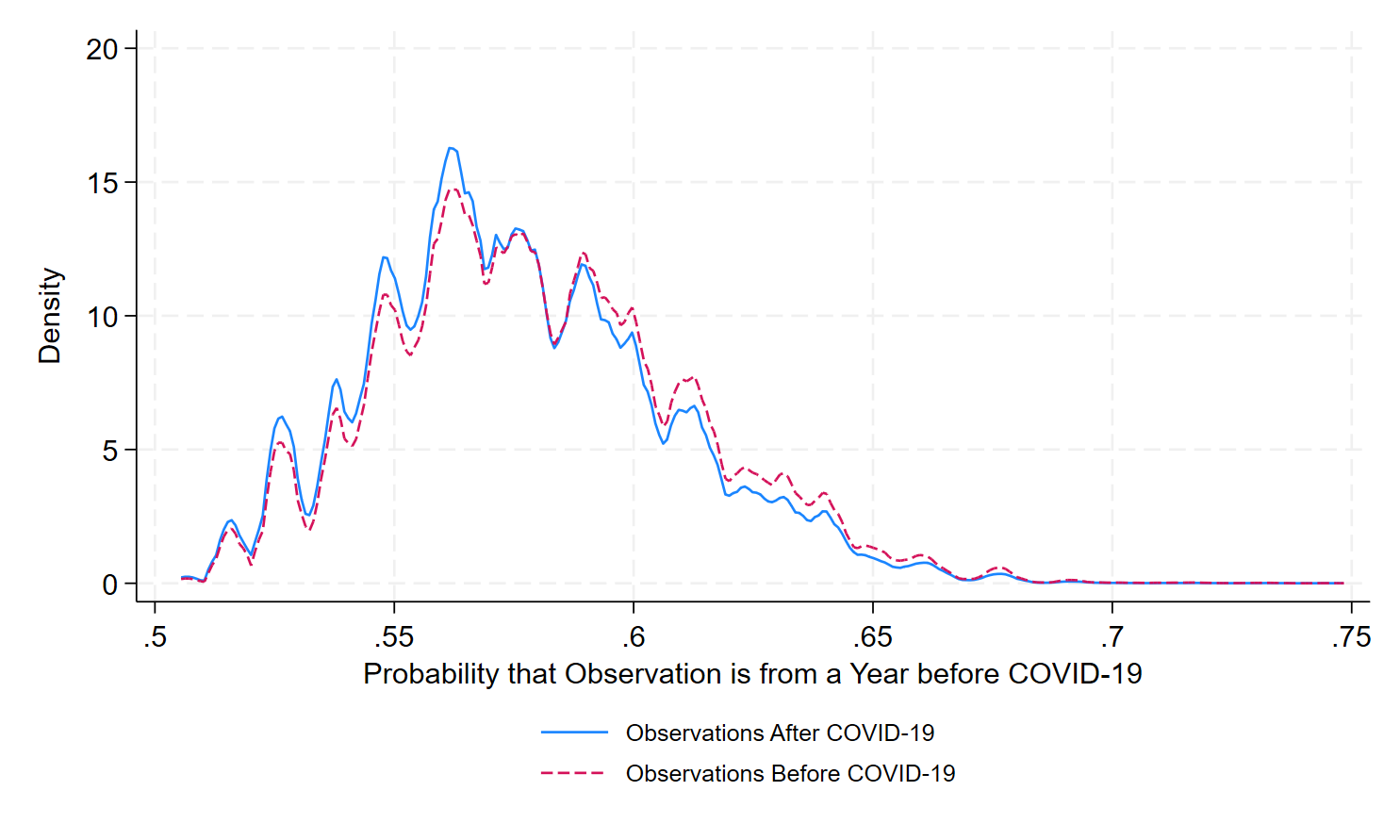}
    \centering
    \begin{minipage}{\textwidth} \small
    \textit{Notes}: This figure uses the 2013--2024 ACS sample of dual-parent households with children of child-rearing age. It plots the kernel density of the estimated probability that an observation is from the pre-COVID period, obtained from a logit model with covariates for husband's occupation, wife's age, wife's education, and state, separately for observations actually drawn from the pre- and post-COVID periods. The predicted values come from:
    \[
    \Pr(\text{Pre-COVID}) = f_s + f_{o(H)} + f_{e} + f_{a}  + \varepsilon,
    \]
    where \( f_s \), \( f_{o(H)} \), \( f_e \), and \( f_a \) are fixed effects for state, husband's occupation, wife's education, and wife's age bin, respectively. Substantial overlap between the two densities indicates good common support in covariate distributions before and after COVID, supporting the reweighting strategies used in \autoref{tab:acs_reweighting} -- DiNardo-Fortin-Lemieux reweighting \citep{dinardo1995labor} and inverse-propensity-score weighting.
    \end{minipage}
\end{figure}

\clearpage

\begin{figure}[H]
    \caption{Dynamic Effect of Husbands' WFH Shock on Having Migrated Last Year}\label{fig:acs_dynamic_DiD_mig_dummy}    
    \includegraphics[width=0.9\linewidth]{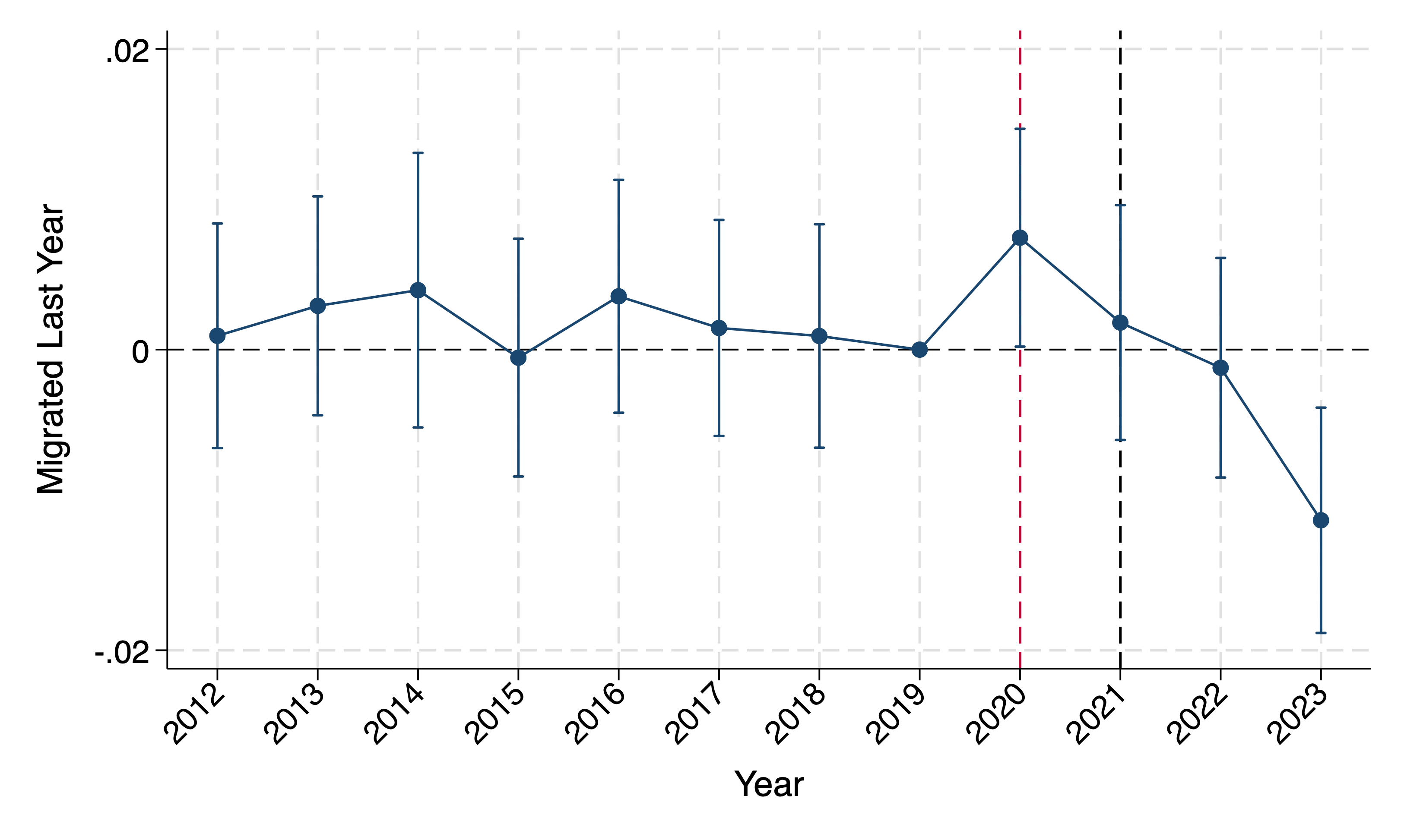}
    \centering
    \begin{minipage}{\textwidth} \small
   \textit{Notes:} This figure plots yearly coefficients from the dynamic difference-in-differences specification in Equation~\eqref{eq:2}, estimated on the 2013--2024 ACS sample of dual-parent households with children of child-rearing age. The outcome is an indicator for having migrated across counties in the 12 months preceding the interview. Treatment compares husbands in occupations with large WFH shocks to those in occupations with smaller changes (Subsection~\ref{subsection:treatment_defn}). The omitted year is 2019. All regressions include state-by-time and husband's-occupation fixed effects, and control for the wife's age, number of children, and education levels of both spouses. Confidence intervals are shown at the 95\% level and are based on standard errors clustered at the husband's occupation level.
    \end{minipage}
\end{figure}

\clearpage

\section{Data Appendix} \label{app:data_construction}

\setcounter{table}{0}
\setcounter{figure}{0}

\paragraph{American Community Survey (ACS)}
We use ACS \citep{ipums_usa_2025} samples from 2013 to 2024, focusing on retrospective measures covering the preceding 12 months. Since occupation is recorded at the time of the survey while our outcomes pertain to the preceding 12 months, we implicitly assume that each individual's occupation at the time of the survey is consistent with their occupation over the prior 12 months. This assumption is inconsequential as long as any occupational transitions occurring within the prior 12 months do not result in individuals switching between the treatment and control groups. It follows that the effective coverage period for our outcomes spans 2012--2023.

\paragraph{WFH Job Postings Data \citep{hansen2023remote}}
Our preferred measure of WFH exposure is derived from job postings data on remote and hybrid work arrangements from 2019 to 2024. These data were constructed by a team of researchers in collaboration with Lightcast, using a large language model (LLM) trained to classify job advertisements based on whether the position allows remote work at least one day per week. The resulting measures are available at the month-year--occupation level and at the month-year--industry level.\footnote{Occupations are classified using the Standard Occupational Classification 2018 (SOC) at 3 digits, and industry codes are classified using North American Industry Classification System (NAICS) 2022 at 3 digits.} We use this data to build the WFH probability changes by occupation described in Section~\ref{subsection:treatment_defn}. Although this measure is constructed from scraped online job postings, \ref{fig:evolution_WFH} shows that the Hansen data tracks remarkably closely the evolution of actual WFH uptake at the occupation level as measured in the ACS yearly data.

\paragraph{Teleworkability index} As part of robustness checks, in  Section \ref{section:robustness} we also use the WFH teleworkability index by \citet{dingel_how_2020} as an alternative measure of WFH feasibility.

%\begin{figure}[H]
%   \caption{Evolution of Men's Full Remote Work Uptake by Occupation}\label{fig:evolution_WFH_acs}
%    \includegraphics[width=0.9\linewidth]{figs/acs_spwfh_only_rates.png}
%    \centering
%    \begin{minipage}{\textwidth} \small
%    \textit{Notes:} Author's elaboration using our main sample of ACS male parents described in Section \ref{sec:data}. We compute the share of men working in select broad occupation groups that also report to have worked from home in the past week.
%    \end{minipage}
%\end{figure}

\paragraph{American Time Use Survey (ATUS)}
We use the ATUS \citep{atus_ipums_2023} samples from 2013 to 2024. The ATUS draws its sample from households completing their final CPS interview, ensuring consistency in demographic and labor market variables across surveys. Each ATUS respondent reports detailed time use for a designated 24-hour period, allowing us to examine outcomes such as time spent on childcare, labor and household production. We can also assess time spent on secondary activities, such as childcaring while doing another activity. We restrict the sample to respondents who completed the time-use diary on a weekday and non-holiday day in order to capture household dynamics on a typical workday. Importantly, the ATUS interviews only one respondent per household. By merging the ATUS with the CPS using unique person identifiers, we obtain time-use data for either the wife or the husband, but never both spouses within the same household.

\paragraph{ATUS Representativeness}
As the ATUS sample is a selected subsample of CPS respondents, we compare its characteristics to those of the full CPS sample to assess potential differences in composition. Appendix \autoref{tab:descstatsmechs} shows modest differences in treatment exposure and socioeconomic characteristics, with ATUS households appearing slightly wealthier and more educated. 
Importantly,
%the estimated effects on wives' labor-market outcomes are directionally consistent with those obtained in the ACS samples (see \autoref{tab:atuscpsout}), and
the increase in women's time spent working following the WFH shock remains statistically significant and persistent within the ATUS sample alone (see Appendix \autoref{tab:mechatus1}). These patterns suggest that sample composition does not drive the results.
As the ATUS interviews only one individual per household, we never observe the time use of both spouses simultaneously. This limitation requires the assumption that, conditional on gender, ATUS respondents are representative of those without time-use data. Under this assumption, the sample of male respondents identifies husbands' time use, while the sample of female respondents identifies wives' time use.

\begin{table}[H]\centering
\caption{Mean and SD from CPS Compared to Women-Respondent and Men-Respondent in the ATUS Sample}
\label{tab:descstatsmechs}
\scalebox{0.6}{
\begin{threeparttable}
\begin{tabular}{lcccccc}  \hline \\ & \multicolumn{2}{c}{CPS} & \multicolumn{4}{c}{ATUS} \\ & \multicolumn{2}{c}{Sample} & \multicolumn{2}{c}{Women-Resp Sample} & \multicolumn{2}{c}{Men-Resp Sample} \\ & Mean & SD & Mean & SD & Mean & SD \\
\hline \addlinespace
\vspace{0.1cm} \\ \textit{Panel A: Husband Characteristics} & & & & \\ \vspace{0.1cm} \\&            &            &            &            &            &            \\
Husband In Shocked Occupation (Treatment)&        0.50&        0.50&        0.54&        0.50&        0.59&        0.49\\
Husband Employed (Last Year)&        0.91&        0.29&        0.92&        0.27&        0.94&        0.24\\
Husband Log Earnings&       10.57&        0.84&       10.63&        0.85&       10.66&        0.86\\
Husband Log Hourly Wage&        2.95&        0.74&        3.00&        0.75&        3.04&        0.75\\
Husband Usual Work Hours&       43.64&        9.46&       44.07&        9.44&       43.87&        9.10\\
Husband Weeks Worked&       50.15&        6.71&       50.21&        6.66&       50.12&        6.89\\
Husband Part-time (Last Year)&        0.03&        0.18&        0.03&        0.17&        0.04&        0.19\\
Husband - Below HS Level&        0.08&        0.28&        0.07&        0.25&        0.05&        0.22\\
Husband - HS Level  &        0.23&        0.42&        0.19&        0.39&        0.16&        0.37\\
Husband - Above HS Level&        0.69&        0.46&        0.75&        0.43&        0.79&        0.41\\
Husband Age         &       37.53&        6.88&       37.76&        6.49&       37.85&        6.60\\
\vspace{0.1cm} \\ \textit{Panel B: Wife Characteristics} & & & & \\ \vspace{0.1cm} \\&            &            &            &            &            &            \\
Wife Employed (Last Year)&        0.66&        0.47&        0.66&        0.48&        0.70&        0.46\\
Wife Log Earnings   &       10.03&        1.11&       10.10&        1.17&       10.08&        1.13\\
Wife Log Hourly Wage&        2.75&        0.77&        2.86&        0.75&        2.82&        0.79\\
Wife Usual Work Hours&       37.00&       10.64&       37.01&       11.13&       36.68&       10.85\\
Wife Weeks Worked   &       46.73&       11.55&       46.95&       11.24&       47.22&       10.58\\
Wife Part-time (Last Year)&        0.24&        0.42&        0.27&        0.44&        0.24&        0.43\\
Wife - Below HS Level&        0.07&        0.25&        0.05&        0.22&        0.04&        0.20\\
Wife - HS Level     &        0.18&        0.38&        0.13&        0.33&        0.11&        0.32\\
Wife - Above HS Level&        0.76&        0.43&        0.82&        0.38&        0.85&        0.36\\
Wife Age            &       35.08&        6.00&       35.31&        5.54&       35.59&        5.63\\
HH Log Earnings     &       10.91&        0.85&       10.98&        0.85&       11.03&        0.82\\
Number of own children in household&        2.08&        0.89&        2.02&        0.86&        2.02&        0.85\\
Age of Youngest Child&        3.48&        2.55&        3.44&        2.56&        3.45&        2.56\\
Age of Oldest Child &        7.52&        5.09&        6.95&        4.53&        6.96&        4.62\\
Youngest Child 0-4  &        0.64&        0.48&        0.65&        0.48&        0.64&        0.48\\
Youngest Child 5-8  &        0.36&        0.48&        0.35&        0.48&        0.36&        0.48\\
\addlinespace
\addlinespace \hline \end{tabular}

\end{threeparttable}
}
\par\vspace{0.3cm}
\begin{minipage}{\textwidth}
    \small
 \textit{Notes}: The table reports mean and standard deviations between the different samples: the benchmark CPS sample from the main findings and the ATUS samples. There are two ATUS samples: women respondents sample and men respondents sample. The former is when wives are those that report their time diary of their activities, while the latter is when husbands report them. Panel A captures variables related to husband characteristics, Panel B is related to wives characteristics and Panel C are related to household variables. 
\end{minipage}
\end{table}

\paragraph{Current Population Survey (CPS)}
We replicate our main findings with public-use microdata from the CPS \citep{ipums_cps_2024} spanning the years 2013--2026. For the CPS, we integrate data from both the basic monthly files and the Annual Social and Economic Supplement (ASEC), also known as the March files.\footnote{March files are available just until 2025, whereas we were able to download basic monthly files up until April 2026 (excluding March 2026).} While weekly measures of employment are available in both the basic monthly and March files, annual employment/labor force participation and income variables are available exclusively in the March files.\footnote{See Appendix B of \cite{kleven2024eitc} for a detailed discussion of differences between the CPS basic monthly and March CPS files, and between weekly and annual variables. In contrast with ACS, annual variables refer to the \emph{previous calendar year}.}

\paragraph{Survey of Income and Program Participation (SIPP)}
We use data from the 2014--2024 SIPP surveys. We exploit the SIPP's childcare topical modules to answer questions on changes in childcare expenditures and responsibility allocations.

\clearpage
\section{Discussion of Identification Strategy} \label{app:metrics}

Our empirical design estimates the causal effect of an increase in the husband's WFH 
probability on the wife's labor market outcomes. In particular, we estimate a 
\emph{household reduced-form effect} of the husband's WFH uptake on the wife's outcomes.

Since we do not control for the wife's own WFH exposure in the baseline specification, 
our baseline estimates necessarily combine the \emph{spillover} effect arising from the 
husband's WFH exposure with the \emph{direct} effect operating through the wife's own 
WFH exposure, which in the model is generated by assortative matching, as discussed in 
Section~\ref{sec:empirics}.

To illustrate the implications, consider a simplified version of our baseline 
specification:
\begin{equation}\label{eq:1app}
y_{it} = \alpha + \beta_0 D^{o(h(i))} + \gamma D_t + \beta_h D^{o(h(i))} \times D_t 
+ u_{it},
\end{equation}
where $D^{o(h(i))}=1$ if the husband of woman $i$ is employed in a treated occupation 
(i.e., one experiencing a large WFH shock), and $D_t=1$ in post-COVID years.

Suppose that unobserved variation in the wife's own WFH exposure enters the error term 
as:
\begin{equation}
u_{it} = 
\beta_1 D^{o(i)}
+ \beta_w D^{o(i)} \times D_t
+ \beta_2 D^{o(i)} \times D^{o(h(i))}
+ \beta_{hw} D^{o(i)} \times D^{o(h(i))} \times D_t
+ \varepsilon_{it},
\end{equation}
where $D^{o(i)} = 1$ if the wife herself is employed in a treated occupation.

Even under the assumption that the cross-occupation interaction terms $\beta_2$ and 
$\beta_{hw}$ are zero, the difference-in-differences estimator $\hat{\beta}_h$ 
from~\eqref{eq:1app} identifies:
\begin{equation}
\begin{split}
\hat{\beta}_h
=&\; \beta_h
+ \beta_1 \cdot 
\Big[
\underbrace{
P(D^{o(i)} = 1 \mid D^{o(h)} = 1, \text{Post})
- P(D^{o(i)} = 1 \mid D^{o(h)} = 1, \text{Pre})
}_{\Delta P(D^{o(i)} = 1 \mid D^{o(h)} = 1)}
\\[0.25em]
&\hspace{3.1em}
-
\underbrace{
\big(
P(D^{o(i)} = 1 \mid D^{o(h)} = 0, \text{Post})
- P(D^{o(i)} = 1 \mid D^{o(h)} = 0, \text{Pre})
\big)
}_{\Delta P(D^{o(i)} = 1 \mid D^{o(h)} = 0)}
\Big]
\\[0.5em]
&\; -\; \beta_w \cdot 
\Big(
P(D^{o(i)} = 1 \mid D^{o(h)} = 1, \text{Post})
- P(D^{o(i)} = 1 \mid D^{o(h)} = 0, \text{Post})
\Big).
\end{split}
\end{equation}

Even under the assumption that assortative matching remains stable before and after 
COVID --- which finds support in \autoref{fig:assortative_matching} --- that is,
\[
\Delta P(D^{o(i)} = 1 \mid D^{o(h)} = 1) 
= 
\Delta P(D^{o(i)} = 1 \mid D^{o(h)} = 0)
= 0,
\]
the estimator $\hat{\beta}_h$ will still load on the direct effect $\beta_w$ through 
differences in wives' occupational exposure across treated and untreated households in 
the post-COVID period.

Section~\ref{section:robustness} presents several exercises exploiting both spouses' 
occupations and/or college education, designed to separate these two components and 
quantify the relative contributions of the \emph{spillover} and \emph{direct} effects. 
%A key data limitation, however, is that in many surveys -- including the ACS -- occupational information is observed only for individuals currently employed. This constraint implies two challenges. First, we cannot systematically observe a wife’s own occupation when she is not employed, preventing us from measuring her WFH exposure during non-employment spells. 

\clearpage

%===============================================================================
% APPENDIX: Quantifying the change in the gender gap
%===============================================================================
\section{Quantifying the Change in the Gender Gap}
\label{app:gender_gap_accounting}

Appendix \autoref{tab:acs_gender_gap} computes the variation in the gender gap due to husbands' WFH exposure. The quantification follows \cite{kleven2019children}. We first run Equation \eqref{eq:1} for both genders separately and get the post-treatment estimated effects ($\hat{\beta}^g$), as reported in \autoref{tab:acs_baseline_annual} and in Appendix \autoref{tab:acs_men_annual}.

Let $\widehat y^{\,g}_{ist}$ denote the fitted value
of outcome $y$ for spouse of gender $g$ and household $i$, in state $s$ and time $t$. The treatment average level is the average fitted value over this group, $\ell^{g}_{\mathrm{tr}}$.
The counterfactual level is the prediction for the same households with the treatment switched off, which is equivalent to subtracting the post-treatment effect $\widehat\beta^{g}$ to the treatment average level:
\begin{equation}
\ell^{g}_{\mathrm{cf}} \;=\; \ell^{g}_{\mathrm{tr}} - \widehat\beta^{g}.
\label{eq:acc_cf}
\end{equation}

We define the gender gap as the ratio of the women's to the men's level. For outcomes measured in levels (usual weekly hours, weeks worked, and part-time status), for each $m \in (cf,tr)$,
\begin{equation}
G_{m} = \frac{\ell^{W}_{m}}{\ell^{M}_{m}},
\label{eq:acc_gap_levels}
\end{equation}
Earnings enter the regression in logs; which means the gap in earnings levels is therefore the exponential of the difference in log levels.
\begin{equation}
G_{m} = e^{\bigl(\ell^{W}_{m}-\ell^{M}_{m}\bigr)}
\label{eq:acc_gap_log}
\end{equation}

Finally, we estimate the change in the gender gap attributable to the husband's remote work shock, considering as baseline the counterfactual gender gap.
\begin{equation}
\Delta G \;=\; \frac{G_{\mathrm{tr}}}{G_{\mathrm{cf}}} - 1.
\label{eq:acc_dgap}
\end{equation}
For earnings, this reduces to $\Delta G = e^{(\widehat\beta^{W}-\widehat\beta^{M})}-1 \approx \widehat\beta^{W}-\widehat\beta^{M}$ and is independent of the counterfactual level. A positive $\Delta G$ for earnings, hours, or weeks, or negative in the case of part-time status, indicates that wives' outcomes rise relative to husbands'. In other words, that the gap narrows.

\clearpage

\end{document}